\pdfoutput=1
\documentclass[aps,prd,nofootinbib,twocolumn,superscriptaddress,letterpaper,amsmath,amssymb,preprintnumbers]{revtex4}

\usepackage{graphicx}  % needed for figures
\usepackage{dcolumn}   % needed for some tables
\usepackage{bm}        % for math
\usepackage{amssymb}   % for math
\usepackage{feynmf}%%%{feynmp}
\usepackage{slashed}
\usepackage{multirow}
\usepackage{mathrsfs}
\def\gsim{\lower0.5ex\hbox{$\:\buildrel >\over\sim\:$}}
\def\lsim{\lower0.5ex\hbox{$\:\buildrel <\over\sim\:$}}

\def \n{\noindent}

\let\n=\nu
\let\t=\tau

\newcommand{\be}{\begin{equation}}
\newcommand{\ee}{\end{equation}}
\newcommand{\bea}{\begin{eqnarray}}
\newcommand{\eea}{\end{eqnarray}}

\newcommand{\nbox}{{\,\lower0.9pt\vbox{\hrule \hbox{\vrule height 0.2 cm
\hskip 0.2 cm \vrule height 0.2 cm}\hrule}\,}}

\def\sub#1{_{\lower.25ex\hbox{$\scriptstyle#1$}}}

\def\to{\rightarrow}
\def\slash{\not\!}

\newskip\zatskip \zatskip=0pt plus0pt minus0pt
\def\matth{\mathsurround=0pt}
\def\lsim{\mathrel{\mathpalette\atversim<}}
\def\gsim{\mathrel{\mathpalette\atversim>}}
\def\sigv{\ifmmode \langle\sigma v\rangle\else $\langle\sigma v\rangle$\fi}
\newskip\zatskip \zatskip=0pt plus0pt minus0pt
\def\matth{\mathsurround=0pt}
\def\lsim{\mathrel{\mathpalette\atversim<}}
\def\gsim{\mathrel{\mathpalette\atversim>}}
\def\atversim#1#2{\lower0.7ex\vbox{\baselineskip\zatskip\lineskip\zatskip
  \lineskiplimit
  0pt\ialign{$\matth#1\hfil##\hfil$\crcr#2\crcr\sim\crcr}}}

\def\missET {{\not\!\! E_T}}

\begin{document}

\thispagestyle{empty}
\vspace*{-3.5cm}

\vspace{0.5in}

%\begin{flushright}
%\today\\
%\end{flushright}
%\vspace{0.5in}
\title{Mono-Higgs: a new collider probe of dark matter}

\begin{center}
\begin{abstract}
We explore the LHC phenomenology of dark matter (DM) pair production in association with a 125 GeV Higgs boson.  This signature, dubbed `mono-Higgs,' appears as a single Higgs boson plus missing energy from DM particles escaping the detector.  We perform an LHC  background study for mono-Higgs signals at $\sqrt{s} = 8$ and $14$ TeV for four Higgs boson decay channels: $\gamma\gamma$, $b \bar b$, and $ZZ^* \to 4\ell$, $\ell\ell j j$.  We estimate the LHC sensitivities to a variety of new physics scenarios within the frameworks of both effective operators and simplified models.  For all these scenarios, the $\gamma\gamma$ channel provides the best sensitivity, whereas the $b\bar b$ channel suffers from a large $t \bar t$ background.  Mono-Higgs is unlike other mono-$X$ searches ($X$=jet, photon, etc.), since the Higgs boson is unlikely to be radiated as initial state radiation, and therefore probes the underlying DM vertex directly.
\end{abstract}
\end{center}

\author{Linda Carpenter}
\affiliation{Department of Physics and Astronomy, Ohio State University, OH}

\author{Anthony DiFranzo}
\affiliation{Department of Physics and Astronomy, University of
  California, Irvine, CA 92697}

\author{Michael Mulhearn}
\affiliation{Department of Physics, University of
  California, Davis, CA 95616}

\author{Chase Shimmin}
\affiliation{Department of Physics and Astronomy, University of
  California, Irvine, CA 92697}

\author{Sean Tulin}
\affiliation{Department of Physics and Astronomy, University of Michigan, MI}

\author{Daniel Whiteson}
\affiliation{Department of Physics and Astronomy, University of
  California, Irvine, CA 92697}

\preprint{MCTP-13-42}
%\preprint{UCI-HEP-TR-2012-XX}
%\pacs{}
\maketitle

% introduction
%\linenumbers

\section{Introduction}

Although most of the matter in the Universe is dark matter (DM), its underlying particle nature remains unknown and cannot be explained within the Standard Model (SM).  Many DM candidates have been proposed, largely motivated in connection with new physics at the electroweak symmetry breaking scale~\cite{Jungman:1995df,Bertone:2004pz}.  Weak-scale DM also naturally accounts for the observed relic density via thermal freeze-out~\cite{Scherrer:1985zt}.  With the discovery of the Higgs boson~\cite{Aad:2012tfa,Chatrchyan:2012ufa}, a new window to DM has opened.  If DM is indeed associated with the scale of electroweak symmetry breaking, Higgs-boson-related signatures in colliders are a natural place to search for it.

Invisible Higgs boson decays provide one well-known avenue for exploring possible DM-Higgs-boson couplings, provided such decays are kinematically allowed.  Null results from searches at the Large Hadron Collider (LHC) for an invisibly decaying Higgs boson produced in association with a $Z$ boson, combined with current Higgs boson data, already provide a model-independent constraint on the Higgs invisible branching ratio of ${\mathcal B}_{\rm inv} < 38\%$ at $95\%$ CL~\cite{Belanger:2013kya}; see Ref.~\cite{Aad:2013oja} for results in $Wh$.  On the other hand, invisible Higgs boson decays are not sensitive to DM with mass above $m_h/2 \approx 60$ GeV.  Therefore, it is clearly worthwhile to investigate other Higgs-boson-related collider observables.

DM production at colliders is characterized by missing transverse energy ($\missET$) from DM particles escaping the detector and recoiling against a visible final state $X$.  Recent mono-$X$ studies at the LHC have searched for a variety of different $X+ \missET$ signals, such as where $X$ is a hadronic jet ($j$)~\cite{Chatrchyan:2012me,Aad:2011xw}, photon ($\gamma$)~\cite{Aad:2012fw,Chatrchyan:2012tea}, or $W/Z$ boson~\cite{Carpenter:2012rg,Aad:2013oja}.   The discovery of the Higgs boson opens a new collider probe of dark matter.  This paper explores the theoretical and experimental aspects of this new LHC signature of dark matter: DM pair production in association with a Higgs boson, $h\chi\chi$, dubbed `mono-Higgs', giving a detector signature of $h+\missET$. We consider mono-Higgs signals in four final state channels for $h$: $b\bar b$, $\gamma\gamma$, and $ZZ^* \to 4 \ell$ and $ZZ^*\rightarrow \ell\ell j j$.

There is an important difference between mono-Higgs and other mono-$X$ searches.  In proton-proton collisions, a $j$/$\gamma$/$W$/$Z$ can be emitted directly from a light quark as initial state radiation (ISR) through the usual SM gauge interactions, or it may be emitted as part of the new effective vertex coupling DM to the SM.  In contrast, since Higgs boson ISR is highly suppressed due to the small coupling of the Higgs boson to quarks, a mono-Higgs is preferentially emitted as part of the effective vertex itself.  In a sense, a positive mono-Higgs signal would probe directly the structure of the effective DM-SM coupling.

Mono-$X$ studies have largely followed two general paths.  In the effective field theory (EFT) approach, one introduces different non-renormalizable operators that generate $X + \missET$ without specifying the underlying ultraviolet (UV) physics.  Since the operators are non-renormalizable, they are suppressed by powers of $1/\Lambda$, where $\Lambda$ is the effective mass scale of UV particles that are integrated out.  Alternatively, in the simplified models approach, one considers an explicit model where the UV particles are kept as degrees of freedom in the theory.  Although the EFT approach is more model-independent, it cannot be used reliably when the typical parton energies in the events are comparable to $\Lambda$~\cite{DelNobile:2013sia}, and additionally it is blind to possible constraints on the UV physics generating its operators (e.g., dijet resonance searches).  Simplified models avoid these short-comings, but at the expense of being more model-dependent.  The two approaches are therefore quite complementary and in the present work, we consider both.

The remainder of our work is outlined as follows.  In Sec.~\ref{sec:models}, we construct both EFT operators and simplified models for generating mono-Higgs signatures at the LHC.  Our simplified models consist of DM particles coupled to the SM through an $s$-channel mediator that is either a $Z^\prime$ vector boson or a scalar singlet $S$.  In Sec.~\ref{sec:colliders}, we assess the sensitivity of LHC experiments to mono-Higgs signals at the 8 TeV and 14 TeV LHC, with 20 fb$^{-1}$ and 300 fb$^{-1}$ respectively, in four Higgs boson decay channels ($b \bar b$, $\gamma\gamma$, $4\ell$, $\ell\ell j j$), including both new physics and SM backgrounds.  In Sec.~\ref{sec:conclude}, we conclude.

\section{New physics operators and models}
\label{sec:models}

We describe new physics interactions between DM and the Higgs boson that may lead to mono-Higgs signals at the LHC.  In all cases, the DM particle is denoted by $\chi$ and may be a fermion or scalar.  We also assume $\chi$ is a gauge singlet under $SU(3)_C \times SU(2)_L \times U(1)_Y$.

\begin{figure}
\includegraphics{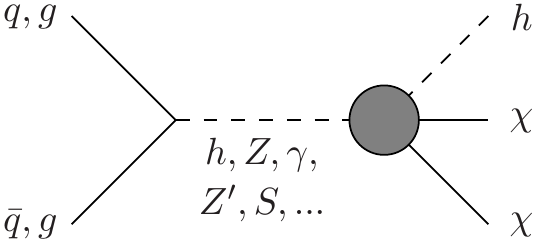}\\
\caption{Schematic diagram for mono-Higgs production in $pp$ collisions mediated by electroweak bosons ($h,Z,\gamma$) or new mediator particles such as a $Z^\prime$ or scalar singlet $S$.  The gray circle denotes an effective interaction between DM, the Higgs boson, and other states.}
\label{fig:effh}
\end{figure}

First, we consider operators within an EFT framework where $\chi$ is the only new degree of freedom beyond the SM.  Next, we consider simplified models with an $s$-channel mediator coupling DM to the SM.   For both cases, Fig.~\ref{fig:effh} illustrates schematically the basic Feynman diagram for producing $h + \missET$ (although not {\it all} models considered here fit within this topology).  Quarks or gluons from $pp$ collisions produce an intermediate state (e.g., an electroweak boson or a new mediator particle) that couples to $h \chi \chi$.

At the end of this section, we identify several benchmark scenarios (both EFT operators and simplified models) that we consider in our mono-Higgs study, see  Table~\ref{tab:bench}.

\subsection{Effective operator models}

%We consider models of the $hh\chi\bar{\chi}$ vertex of two classes. First, we build effective operators which describe this interaction without knowledge of the ultraviolet details of the interaction, see Fig~\ref{fig:effh}. These operators are useful approximations if the mediating interaction involves heavy new particles which can be integrated out.

The simplest operators involve direct couplings between DM particles and the Higgs boson through the Higgs portal $|H|^2$~\cite{McDonald:1993ex,Burgess:2000yq,Patt:2006fw,Kim:2006af,MarchRussell:2008yu,Low:2011kp,LopezHonorez:2012kv}.  For scalar DM, we have a renormalizable interaction at dimension-4:
\be \label{eq:dim4}
\lambda |H|^2 \chi^2 \, ,
\ee
where $\chi$ is a real scalar and $\lambda$ is a coupling constant.  For (Dirac) fermion DM, we have two operators at dimension-5:
\be \label{eq:dim5}
\frac{1}{\Lambda} |H|^2 \bar \chi \chi  \, , \quad  \frac{1}{\Lambda} |H|^2 \bar \chi i \gamma_5 \chi \, ,
\ee
suppressed by a mass scale $\Lambda$.  Mono-Higgs can arise via $gg \to h^* \to h \chi \chi$ through these operators.  However, it is important to note that these interactions lead to invisible Higgs boson decay for $m_\chi < m_h/2$.  Treating each operator independently, the partial widths in each case are
\begin{subequations}
\begin{align}
\Gamma(h \to \chi \chi) &= \frac{\lambda^2 v^2}{4 \pi m_h} & & {\rm scalar} \; \chi \\
\Gamma(h \to \chi \bar \chi) &= \frac{v^2 m_h}{8\pi \Lambda^2} & & {\rm fermion} \; \chi
\end{align}
\end{subequations}
neglecting $\mathcal{O}(m_\chi^2/m_h^2)$ terms, where $v \approx 246$ GeV is the Higgs vacuum expectation value.  If invisible decays are kinetimatically open, it is required that $\lambda \lesssim 0.016$ ($\Lambda \gtrsim 10$ TeV) for scalar (fermion) DM to satisfy ${\mathcal B}_{\rm inv} < 38\%$ obtained in Ref.~\cite{Belanger:2013kya}.  In this case, since the couplings must be so suppressed, the leading mono-Higgs signals from DM are from di-Higgs production where one of the Higgs bosons decays invisibly, as we show below.  On the other hand, if $m_\chi \gtrsim m_h$, invisible Higgs boson decay is kinematically blocked and the DM-Higgs couplings can be much larger.

At dimension-6, there arise several operators that give mono-Higgs signals through an effective $h$-$Z$-DM coupling.  For scalar DM, we have
\be \label{eq:O6Zscalar}
\frac{1}{\Lambda^2} \chi^\dagger i \smash{ \overset{\leftrightarrow}{\partial^\mu}} \chi  H^\dagger i D_\mu H
\ee
while for fermionic DM we have
\be \label{eq:O6Zferm}
\frac{1}{\Lambda^2} \bar\chi \gamma^\mu \chi  H^\dagger i D_\mu H \, , \quad \frac{1}{\Lambda^2} \bar\chi \gamma^\mu\gamma_5 \chi  H^\dagger i D_\mu H \, .
\ee
When the Higgs acquires its vev, the Higgs bilinear becomes
\be \label{eq:hhbilin}
\frac{1}{\Lambda^2} H^\dagger i D_\mu H \; \to \; - \frac{g_2 v^2}{4 c_W\Lambda^2} Z_\mu \Big( 1 + \frac{h}{v} \Big)^2 \, ,
\ee
where $g_2$ is the $SU(2)_L$ gauge coupling and $c_W \equiv \cos\theta_W$ is the cosine of the weak mixing angle.  Thus, these operators generate mono-Higgs signals via $q \bar q \to Z^* \to h \chi \chi$.  However, for $m_\chi < m_Z/2$, these operators are strongly constrained by the invisible $Z$ width.  The partial width for scalar DM is
\be
\Gamma(Z\to \chi \chi^\dagger) = \frac{g_2^2 v^4 m_Z}{768\pi c_W^2 \Lambda^4} \quad {\rm scalar} \; \chi \, ,
\ee
neglecting $\mathcal{O}(m_\chi^2/m_Z^2)$ terms.  For fermionic DM, the partial width is larger by a factor of four for either of the operators in Eq.~\eqref{eq:O6Zferm}.  Requiring $\Gamma_Z^{\rm inv} \lesssim 3$ MeV~\cite{Beringer:1900zz} imposes that $\Lambda \gtrsim 400$ GeV ($550$ GeV) for scalar (fermion) DM if such decays are kinematically open.

At higher dimension, there are many different operators to consider for coupling $h\chi\chi$ to additional SM fields.  Here we focus in particular on operators arising at dimension-8 that couple DM particles and the Higgs field with electroweak field strength tensors~\cite{Chen:2013gya}.  (Such operators have been considered recently in connection with indirect detection signals~\cite{Chen:2013gya,Fedderke:2013pbc}.)  For fermionic DM, there are many such operators, e.g.,
\begin{subequations} \label{eq:O8all}
\begin{align}
&\frac{1}{\Lambda^4} \bar{\chi} \gamma^{\mu} \chi  B_{\mu \nu} H^{\dagger} D^{\nu}H \, , & &\frac{1}{\Lambda^4} \bar{\chi} \gamma^{\mu} \chi  W^a_{\mu \nu} H^{\dagger}t^a D^{\nu} H \\
&\frac{1}{\Lambda^4} \bar{\chi} \sigma^{\mu \nu}  \chi  B_{\mu \nu}H^{\dagger} H \, , &  &\frac{1}{\Lambda^4} \bar{\chi} \sigma^{\mu \nu}  \chi  W^a_{\mu \nu}H^{\dagger} t^a H \,
\end{align}
\end{subequations}
where $W^a_{\mu\nu}$ and $B_{\mu\nu}$ are the $SU(2)_L$ and $U(1)_Y$ field strength tensors, respectively.  Additional operators arise where $\bar\chi \gamma^\mu \chi$ can be replaced by the axial current $\bar\chi \gamma^\mu\gamma_5 \chi$, or the field strength tensors are replaced with their duals.  For illustrative purposes, we investigate the mono-Higgs signals from one operator
\be \label{eq:O8}
\frac{1}{\Lambda^4} \bar\chi \gamma^\mu \chi B_{\mu\nu} H^\dagger D^\nu H \, .
\ee
This operator leads to $h + \missET$ via $q\bar q \to Z^*/\gamma^* \to h \chi \chi$.  It is noteworthy that the Feynman rule for this process involves derivative couplings, i.e., $\partial_\mu Z_\nu \partial^\nu h$.  Consequently, compared to our other effective operators, this one leads to a harder $\missET$ spectrum and has by far the best kinematic acceptance efficiency, as we show below.  We also note that the operators in \eqref{eq:O8all} also induce mono-$W/Z/\gamma$ signals, as required by gauge invariance, when both Higgs fields $H$ are replaced by $v$.  For a single operator, the ratio between mono-$h/W/Z/\gamma$ is fixed, and therefore constraints on each channel are relevant.  In the presence of a signal, on the other hand, all channels are complementary in disentangling the underlying operator(s).

\subsection{Simplified models}

Beyond the EFT framework, it is useful to consider simple, concrete models for how DM may couple to the visible sector.  Simplified models provide a helpful bridge between bottom-up EFT studies and realistic DM models motivated by top-down physics~\cite{Alves:2011wf}.  %On the bottom-up side, studies of simplified models give insight about which effective operators may typically be dominant within the EFT and whether the EFT is valid at the energy and sensitivity reach explored at the LHC.  On the top-down side, simplified models can represent a wide class of models that give similar phenomenological signatures in terms of a few key parameters, without specifying the additional details of the ultraviolet theory that may not be relevant for a given observable.
Here, we explore a few representative scenarios where the dark and visible sectors are coupled through a new massive mediator particle.  Mono-Higgs signals are a prediction of these scenarios since in general the mediator may couple to the Higgs boson.

\subsubsection{Vector mediator models ($Z^\prime$)}

A $Z^\prime$ vector boson is a well-motivated feature of many new physics scenarios, arising either as a remnant of embedding the SM gauge symmetry within a larger rank group or as part of a hidden sector that may be sequestered from the SM (see e.g.~\cite{Langacker:2008yv} and references therein).   The $Z^\prime$ has an added appeal for DM since the corresponding $U(1)^\prime$ gauge symmetry ensures DM stability, even if the symmetry is spontaneously broken.\footnote{It is required that the $U(1)^\prime$ is broken by $n>1$ units, where the $\chi$ field carries $n = 1$ unit of $U(1)^\prime$ charge.  This breaks the $U(1)^\prime$ down to a $\mathbb{Z}_n$ discrete symmetry.}  Although how the $Z^\prime$ couples to SM particles is highly model-dependent, we focus here on simple scenarios that are representative of both extended gauge models and hidden sector models.  For practical purposes, this distinction affects whether the $Z^\prime$-quark vertex is a gauge-strength coupling or is suppressed by a small mixing angle, which in turn impacts DM production at the LHC.

One gauge extension of the SM is to suppose that baryon number ($B$) is gauged, with the $Z^\prime$ being the gauge boson of $U(1)_B$~\cite{Carone:1994aa}.  The consistency of such theories often implies the existence of new stable baryonic states that  are neutral under the SM gauge symmetry, providing excellent DM candidates~\cite{Agashe:2004ci,FileviezPerez:2010gw}.  Taking the DM particle $\chi$ to carry baryon number $B_\chi$, the $Z^\prime$-quark-DM part of the Lagrangian is
\be \label{ZprimeDM}
\mathscr{L} \supset  g_q  \bar q \gamma^\mu q  Z^\prime_\mu +\left\{ \begin{array}{cc}
i  g_\chi  \chi^\dagger \smash{ \overset{\leftrightarrow}{\partial^\mu}}  \chi  Z^\prime_\mu + g_\chi^2 |\chi|^2 Z^\prime_\mu Z^{\prime\mu} & {\rm scalar}  \\  g_\chi  \bar\chi \gamma^\mu \chi Z_\mu^\prime& {\rm fermion} \end{array} \right.
\ee
depending on whether $\chi$ is a scalar or fermion.  The $Z^\prime$ couplings to quarks and DM are related to the $U(1)_B$ gauge coupling $g_B$ by $g_q = g_B/3$ and $g_\chi = B_\chi g_B$, respectively.  This scenario is an example of a leptophobic $Z^\prime$ model, and many precision constraints are evaded since the $Z^\prime$ does not couple to leptons~\cite{delAguila:1985cb}.

\begin{figure}
\includegraphics{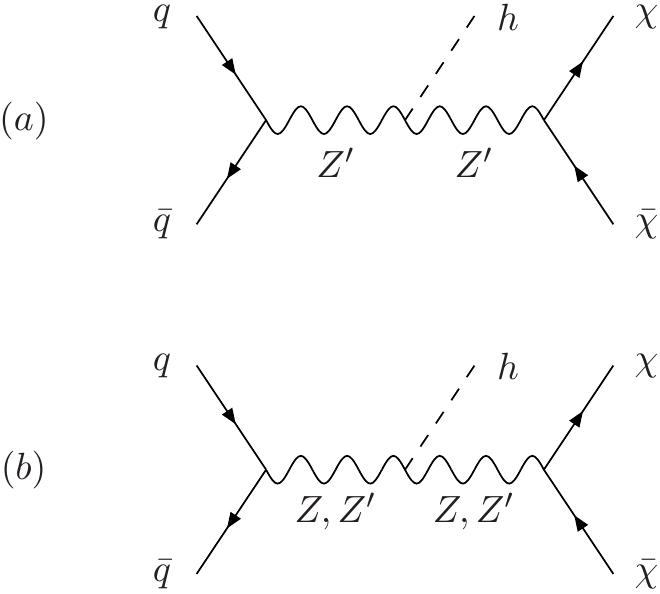}\\
\caption{Diagram showing collider production mode in a simplified model including a $Z'$ boson which decays to $\chi\bar{\chi}$.}
\label{fig:effhZpxx}
\end{figure}

To investigate mono-Higgs signals, we ask whether the $Z^\prime$ is coupled to the Higgs boson $h$.  To generate the $Z^\prime$ mass, the minimal possibility is to introduce a ``baryonic Higgs'' scalar to spontaneously break $U(1)_B$.  Analogous to the SM, there remains a physical baryonic Higgs particle, denoted $h_B$, with a coupling $h_B Z^\prime Z^\prime$.  This coupling comes from the $Z^\prime$ mass term
\be\label{ZBmassterm}
\mathscr{L} \supset \frac{1}{2} m_{Z^\prime}^2 \left( 1 + \frac{h_B}{v_B} \right)^2 Z^\prime_\mu Z^{\prime\mu} \, ,
\ee
where $v_B$ is the baryonic Higgs vev.
Generically $h_B$ will mix with the SM Higgs boson, giving rise to an interaction of the form
\be \label{ZprimeHiggs}
\mathscr{L} \supset - g_{h Z^\prime Z^\prime}  h Z^\prime_\mu Z^{\prime\mu} \, , \qquad g_{h Z^\prime Z^\prime} = \frac{m_{Z^\prime}^2 \sin{\theta}}{v_B}
\ee
where $\theta$ is the $h$-$h_B$ mixing angle.  Combining Eqs.~\eqref{ZprimeDM} and \eqref{ZprimeHiggs} allows for mono-Higgs signals at the LHC, shown in Fig.~\ref{fig:effhZpxx}(a).  At energies below $m_{Z^\prime}$, the relevant effective operators for fermionic DM are
\be \label{U1Beft}
\mathscr{L}_{\rm eff} = - \frac{g_q g_\chi }{m_{Z^\prime}^2} \bar{q} \gamma^\mu q \bar\chi \gamma_\mu \chi \Big( 1 + \frac{g_{h Z^\prime Z^\prime} }{m_{Z^\prime}^2} h \Big) \, ,
\ee
and similarly for scalar DM.  The first term in Eq.~\eqref{U1Beft} is relevant for mono-$j/\gamma/W/Z$ signals (through ISR), while the second term gives rise to mono-Higgs.  It is clear that mono-Higgs, depending on a different combination of underlying parameters, offers a complementary handle for DM studies.

An alternate framework for the $Z^\prime$ is that of a hidden sector (see e.g.~\cite{Chang:2006fp,Pospelov:2007mp,Feldman:2007wj,Feng:2008mu,Gopalakrishna:2008dv}).  In this case, we suppose that DM remains charged under the $U(1)^\prime$, while all SM states are neutral.  The Lagrangian we consider is
\be \label{eq:LintZH}
\mathscr{L} \supset \frac{g_2}{2 c_W} J_{\rm NC}^\mu Z_\mu + g_\chi \bar \chi \gamma^\mu \chi Z_\mu^\prime \, ,
\ee
where $J^\mu_{\rm NC}$ is the usual SM neutral current coupled to the $Z$, and the $Z^\prime$ is coupled to fermionic DM.  Although the two sectors appear decoupled, small couplings can arise through mixing~\cite{Holdom:1985ag,Babu:1997st,Gopalakrishna:2008dv}.  One simple possibility is that the $Z^\prime$ has a mass mixing term with the $Z$.  In this case, one diagonalizes the $Z,Z^\prime$ system by a rotation
\be
Z \to c_\theta Z - s_\theta Z^\prime , \;\; Z^\prime \to c_\theta Z^\prime + s_\theta Z \, ,
\ee
where $\theta$ is the $Z$-$Z^\prime$ mixing angle, and $s_\theta \equiv \sin\theta$ and $c_\theta \equiv \cos\theta$.  Thus, the physical $Z,Z^\prime$ states are linear combinations of the gauge eigenstates, and each one inherits the couplings of the other from Eq.~\eqref{eq:LintZH}.  We note that such mixing gives a contribution to the $\rho$ parameter of $\delta \rho = \sin^2\theta(m_{Z^\prime}^2/m_Z^2 - 1)$~\cite{Babu:1997st}.  Current precision electroweak global fits exclude $|\delta \rho | \gtrsim 10^{-3}$~\cite{Beringer:1900zz}, although any tension is also affected by new physics entering other observables in the global fit.

Mono-Higgs signals arise through diagrams shown in Fig.~\ref{fig:effhZpxx}(b).  The $h Z Z^\prime$ vertex arises as a consequence of the fact $Z$-$Z^\prime$ mixing violates $SU(2)_L$ and is given by
\be
\mathscr{L} \supset \frac{m_Z^2 s_\theta}{v} h Z^\prime_\mu Z^\mu \, .
\ee

\subsubsection{Scalar mediator models}

\begin{figure}
\includegraphics{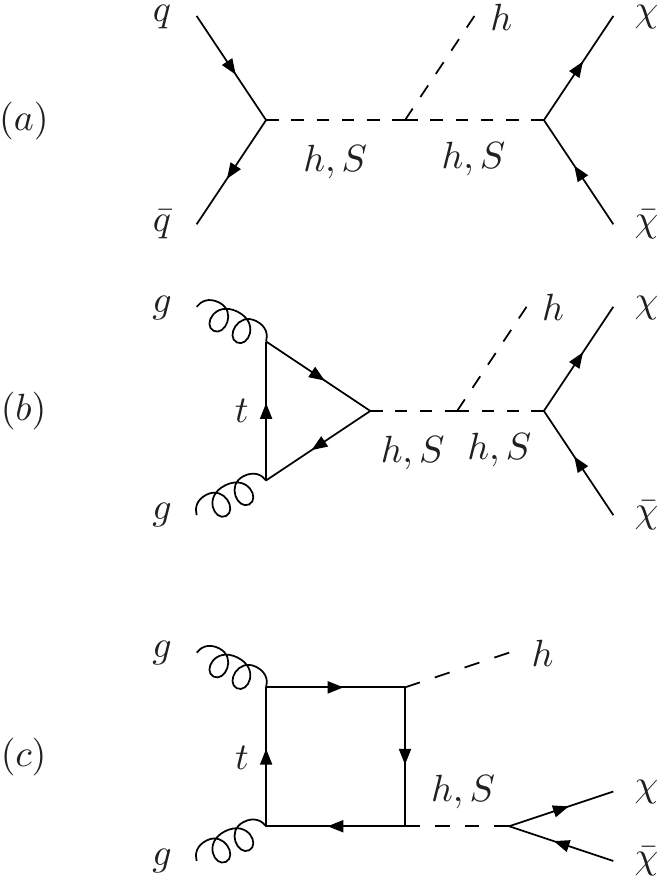}\\
\caption{Diagram showing collider production mode in a simplified model including a $Z'$ boson which decays to $\chi\bar{\chi}$.}
\label{fig:effhSxx}
\end{figure}

New scalar particles may provide a portal into the dark sector~\cite{MarchRussell:2008yu}.  The simplest possibility is to introduce a real scalar singlet, denoted $S$, with a Yukawa coupling to DM
\be
\mathscr{L} \supset - y_\chi \bar\chi \chi S \, .
\ee
By virtue of gauge invariance, $S$ may couple to the SM (at the renormalizable level) only through the Higgs field~\cite{O'Connell:2006wi}.  The relevant terms in the scalar potential are
\begin{align}
&V \supset a |H|^2 S + b |H|^2 S^2 + \lambda_h |H|^4 \notag \\
& \;\;\longrightarrow \tfrac{1}{2} a (h +  v)^2 S + \tfrac{1}{2}b (h +  v)^2 S^2 + \frac{\lambda_h}{4} (h +  v)^4 ,
\label{singlethiggsmix}
\end{align}
where $a,b$ are new physics couplings and $\lambda_h$ is the usual Higgs quartic.  The second line in Eq.~\eqref{singlethiggsmix} follows once the Higgs field acquires a vev, thereby leading to a mixing term $a v$ in the $h$-$S$ mass matrix.  (Without loss of generality, the vev of $S$ can be taken to be zero through a field shift~\cite{O'Connell:2006wi}.)   The two scalar system is diagonalized by a field rotation
\be \label{hSrotation}
h \to c_\theta h + s_\theta S \, , \quad S \to c_\theta S - s_\theta h
\ee
where the mixing angle $\theta$ is defined by $\sin 2\theta = 2av/(m_S^2 - m_h^2)$, with $s_\theta \equiv \sin \theta$ and $c_\theta \equiv \cos\theta$.  After the field rotation, the quark and DM Yukawa terms become
\be \label{LintScalar2}
\mathscr{L} \supset - y_\chi \bar\chi \chi (  c_\theta S - s_\theta h ) - \frac{m_q}{v} \bar q q (c_\theta h + s_\theta S )  \, .
\ee
The mixing angle is constrained by current Higgs data, which is consistent with $\cos\theta = 1$ within $\mathcal{O}(10\%)$ uncertainties~\cite{Belanger:2013kya,Falkowski:2013dza,Djouadi:2013qya,Giardino:2013bma,Ellis:2013lra}, thereby requiring $\sin\theta \lesssim 0.4$.

Mono-Higgs signals in this model arise through processes shown in Fig.~\ref{fig:effhSxx}(a,b).  These processes depend on the $h^2 S$ and $h S^2$ cubic terms in Eq.~\eqref{singlethiggsmix}.  At leading order in $\sin\theta$, these terms are
\be
V_{\rm cubic} \approx \frac{\sin\theta}{v} ( 2 m_h^2 + m_S^2) h^2 S  + b v h S^2 + ...
\ee
where we have expressed $a$ and $\lambda_h$ in terms of $\sin\theta$ and $m_h^2$, respectively.  We note that the $h^2 S$ term is fixed (at leading order in $\sin\theta$) once the mass eigenvalues $m_h, m_S$ and mixing angle are specified.  However, the $h S^2$ is not fixed and remains a free parameter depending on $b$.  Alternately, a Higgs can be radiated directly from the $t$ quark in the production loop, shown in Fig.~\ref{fig:effhSxx}(c).  In our study, we include the $gghS$ box contribution through an effective Lagrangian
\be
\mathscr{L}_{\rm eff} = - \frac{\alpha_s \sin 2\theta}{24 \pi v^2 }  G^a_{\mu\nu} G^{a \mu\nu} h S \, ,
\ee
which we have evaluated in the large $m_t$ limit.  Although this will likely overestimate our $h + \missET$ signal~\cite{Haisch:2012kf}, we defer an evaluate of the true box form factor to future study.

\subsection{Benchmark Models}

For the purposes of our collider study to follow, we consider several illustrative benchmark scenarios for both EFT operators and simplified models.  These models are summarized in Table \ref{tab:bench}.  For the $Z^\prime$ models, we henceforth denote the $Z^\prime$ coupled to baryon number as $Z^\prime_B$ and the hidden sector $Z^\prime$ mixed with the $Z$ as $Z^\prime_H$.  Otherwise, the parameters and interactions are as described above.

\begin{table}
\caption{Summary of benchmark models for $h + \missET$ signals.}
\label{tab:bench}
\begin{tabular*}{\columnwidth}{l @{\extracolsep{\fill} } l l}
\hline\hline
\multicolumn{3}{c}{Effective operators} \\
\hline
$ |\chi|^2 |H|^2$ & $\lambda = 0.01 $ \\
& $\lambda = 1 $\\
\hline
$\bar\chi \chi |H|^2$ & $\Lambda = 100$ GeV \\
& $\Lambda = 10$ TeV \\
\hline
$\bar\chi i \gamma_5 \chi |H|^2$ & $\Lambda = 100$ GeV \\
& $\Lambda = 10$ TeV \\
\hline
$\chi^\dagger \partial^\mu \chi  H^\dagger D_\mu H$ & $\Lambda = 300$ GeV  \\
\hline
$\bar \chi \gamma^\mu \chi  B_{\mu\nu} H^\dagger D_\nu H$ & $\Lambda = 100$ GeV  \\
\hline\hline
\end{tabular*}
\begin{tabular*}{\columnwidth}{l @{\extracolsep{\fill} } l}
\multicolumn{2}{c}{Simplified models with $s$-channel mediator}\\
\hline
$Z^\prime_B$ & $m_{Z^\prime} = 100$ GeV, $g_\chi \!=\! g_B \! = \!1$, $g_{h Z^\prime Z^\prime}/m_{Z^\prime} = 0.3$\\
& $m_{Z^\prime} = 1000$ GeV, $g_\chi \!=\! g_B \! = \!1$, $g_{h Z^\prime Z^\prime}/m_{Z^\prime} = 0.3$\\
\hline
$Z^\prime_H$ & $m_{Z^\prime} = 100$ GeV, $g_\chi = 1$, $\sin\theta = 0.1$\\
 & $m_{Z^\prime} = 1000$ GeV, $g_\chi = 1$, $\sin\theta = 0.1$\\
\hline
Scalar $S$ & $m_{S} = 100$ GeV, $y_\chi = 1$, $\sin\theta = 0.3$, $b=3$\\
 & $m_{S} = 1000$ GeV, $y_\chi = 1$, $\sin\theta = 0.3$, $b=3$ \\
\hline\hline
\end{tabular*}
\end{table}

\section{Collider Sensitivity}
\label{sec:colliders}

In this section we estimate the sensitivity of the LHC to mono-Higgs production with $pp$ collisions at $\sqrt{s}=8$ TeV and 14 TeV with $\mathcal{L}=20$
fb$^{-1}$ and $\mathcal{L}=300$
fb$^{-1}$, respectively.

Signal events are generated in {\sc madgraph}5~\cite{madgraph5}, with
showering and hadronization by {\sc pythia}~\cite{pythia} and detector
simulation with {\sc delphes}~\cite{delphes} assuming pileup conditions of
$\mu=20$ and $\mu=50$ for $\sqrt{s}=8,14$ TeV, respectively.

\begin{figure}
\includegraphics[width=1.65in]{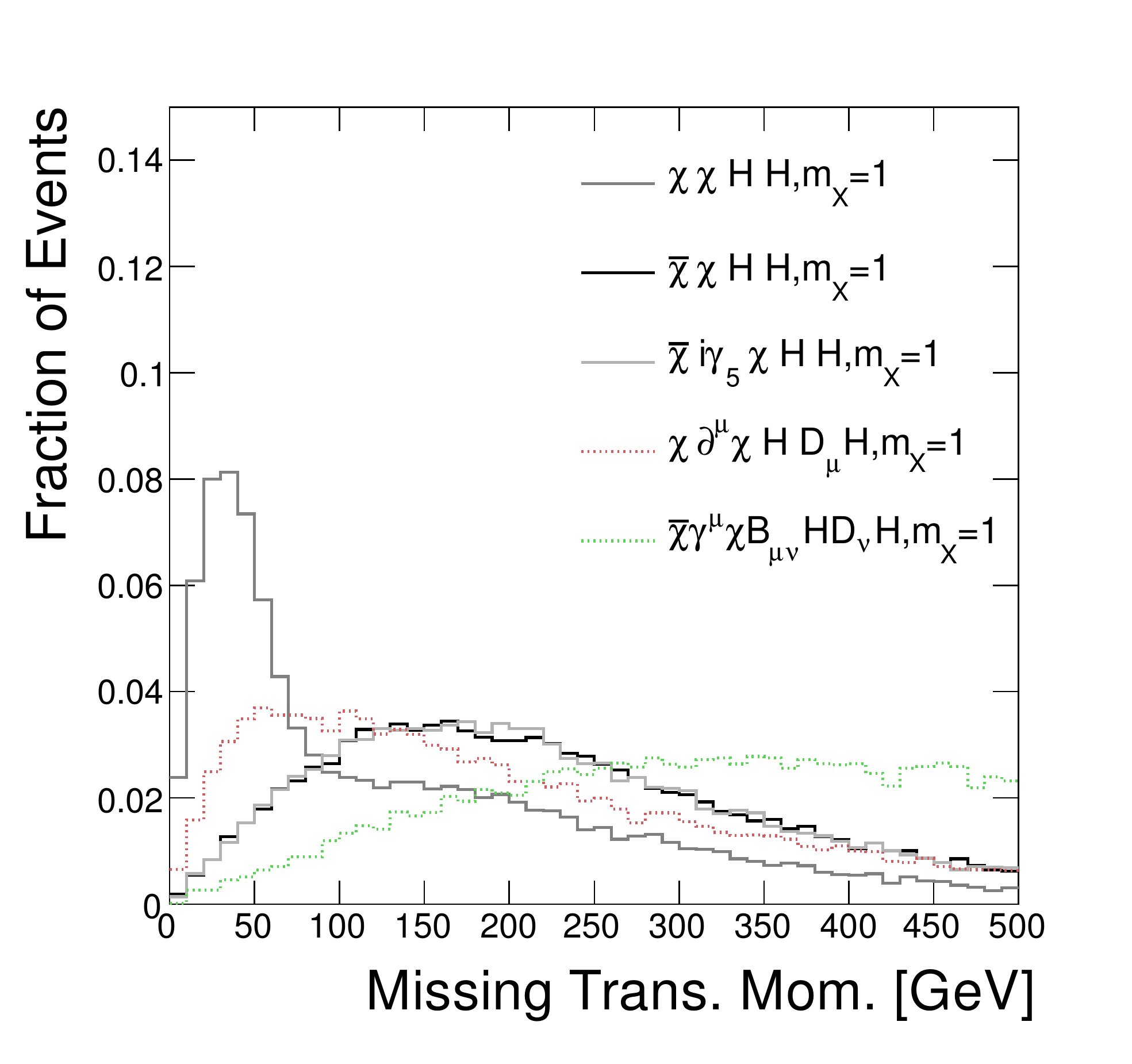}
\includegraphics[width=1.65in]{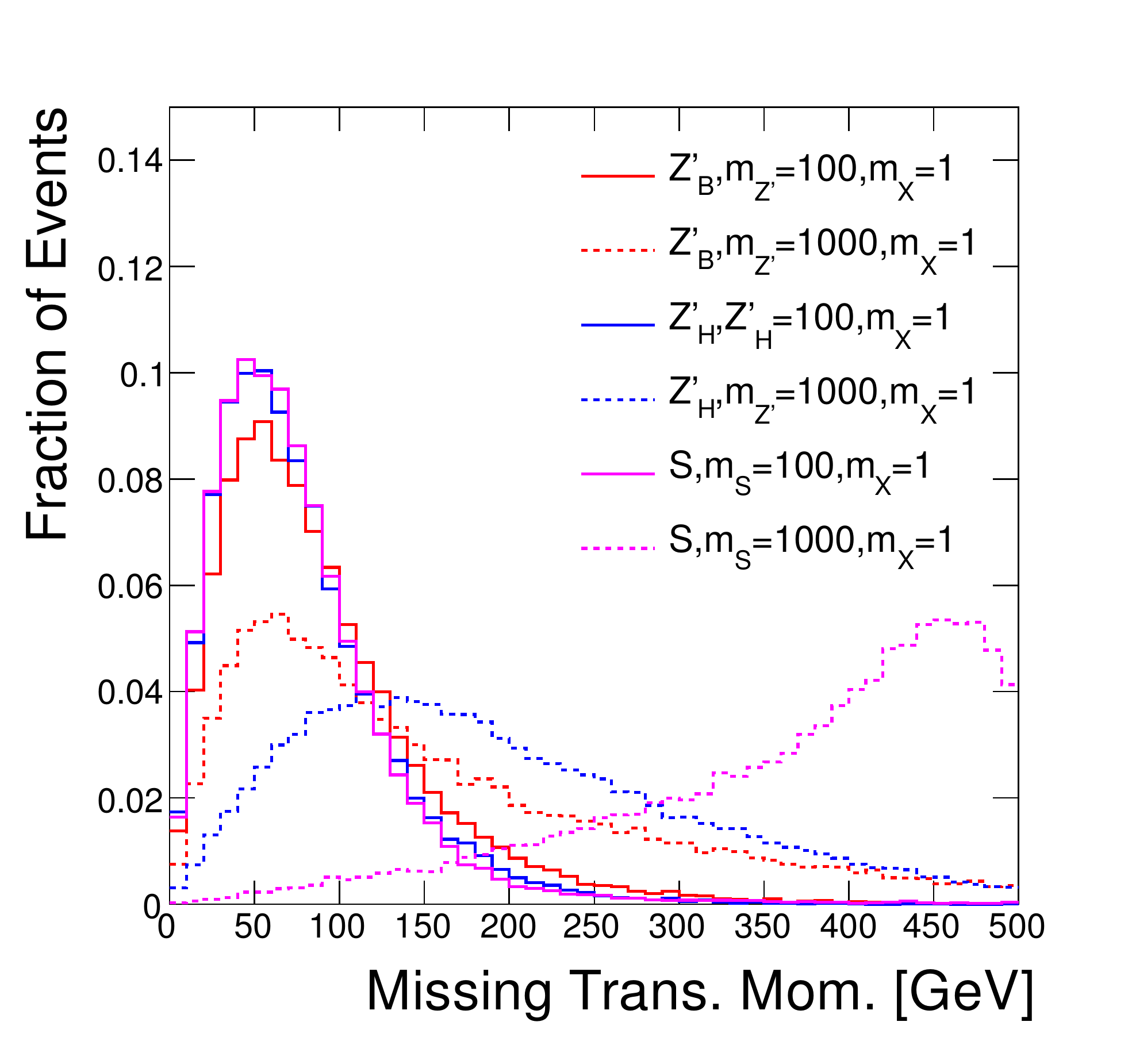}\\
\includegraphics[width=1.65in]{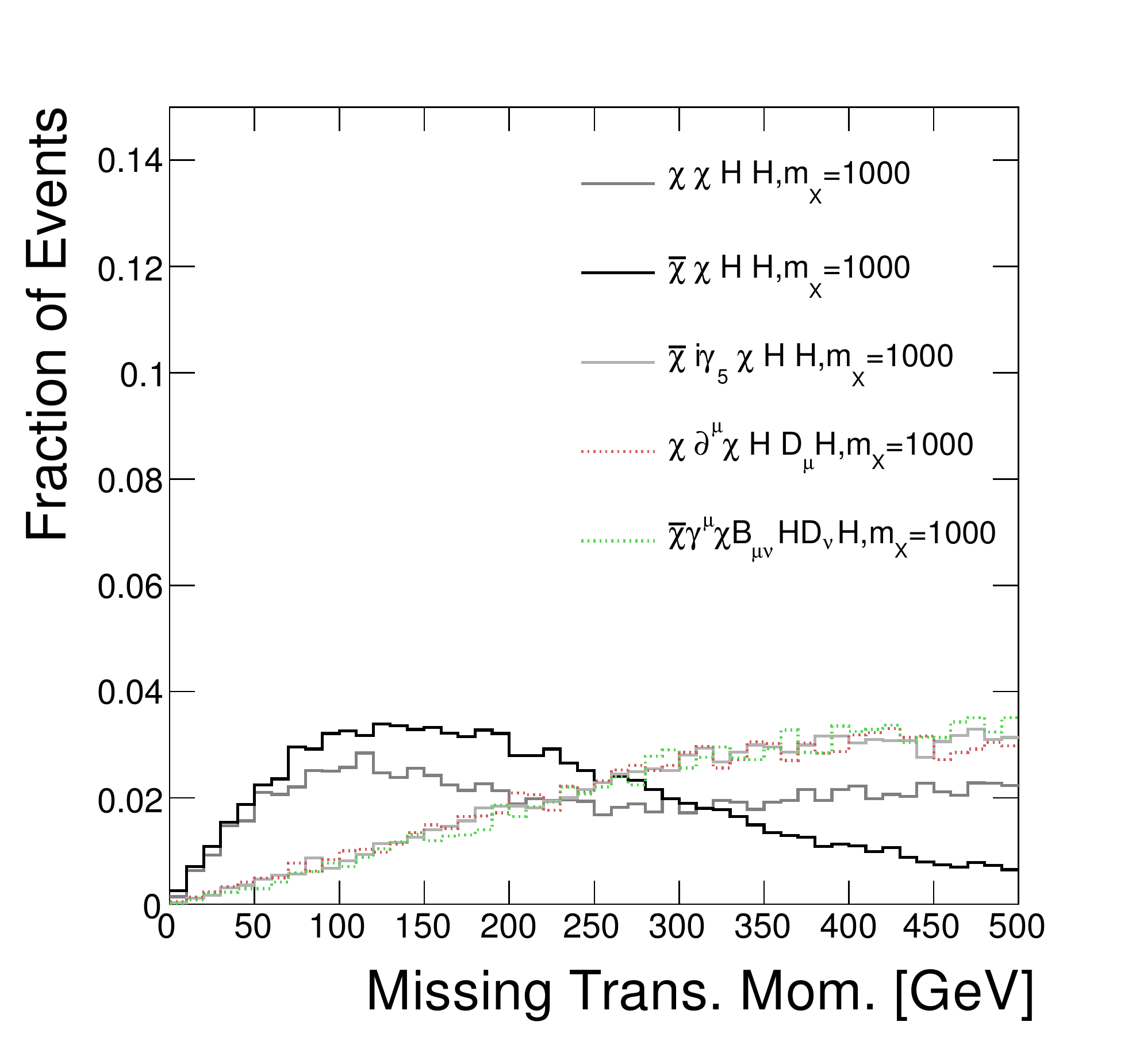}
\includegraphics[width=1.65in]{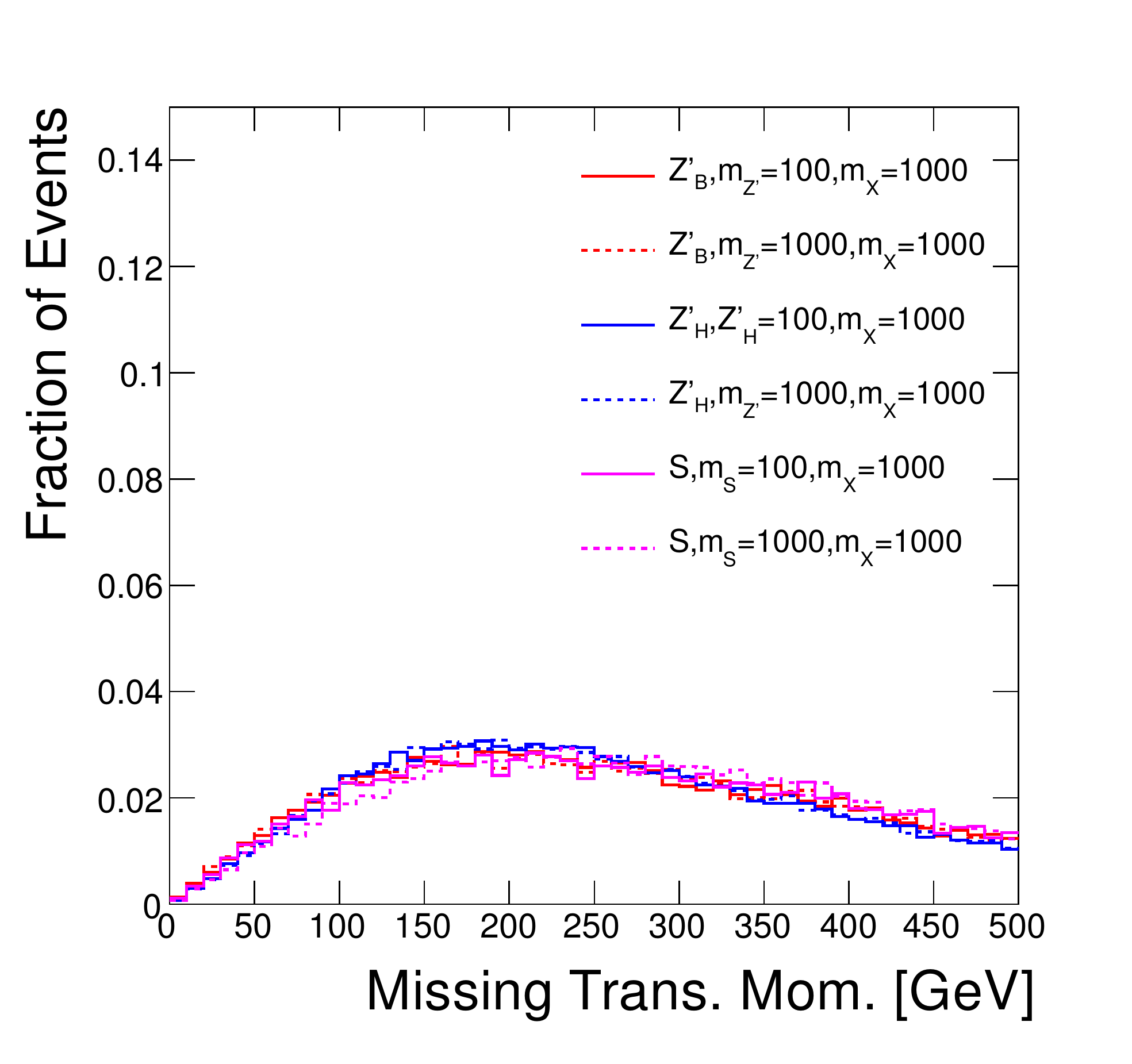}\\
\caption{Distribution of missing transverse momentum for EFT models (top) and simplified models (bottom) for $m_{\chi}=1$ GeV (left) and $m_{\chi}=1000$ GeV (right).}
\label{fig:models}
\end{figure}

\begin{figure}
\includegraphics[width=1.65in]{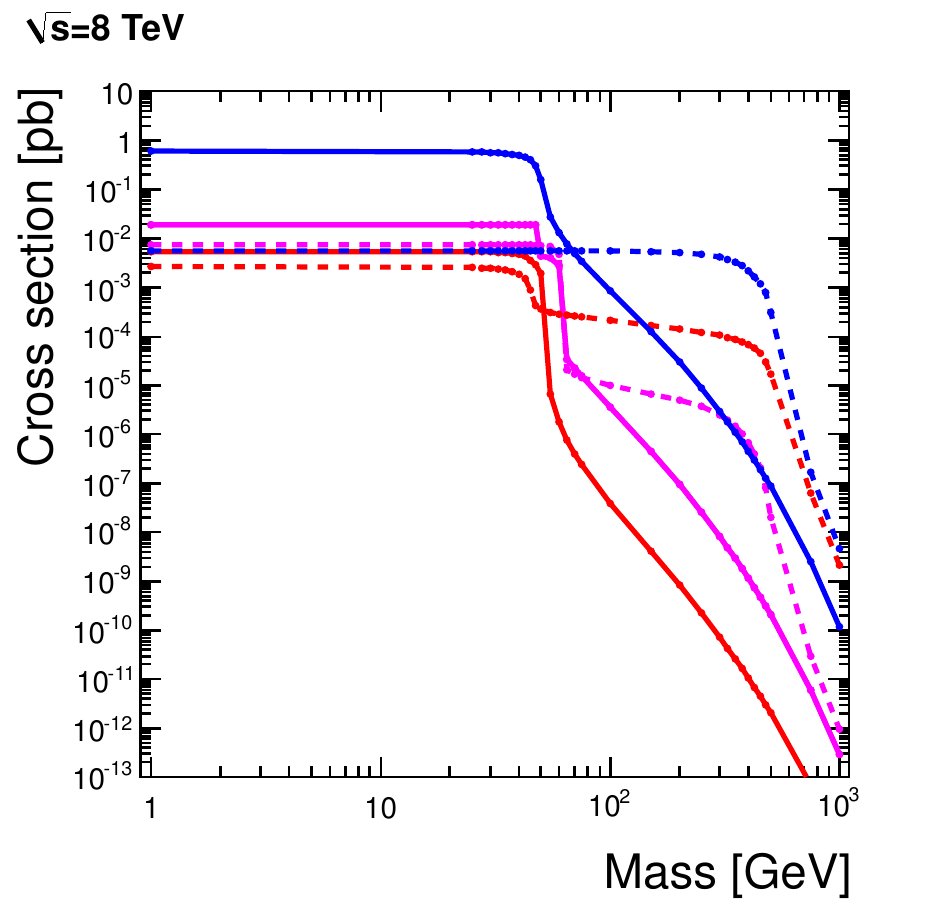}
\includegraphics[width=1.65in]{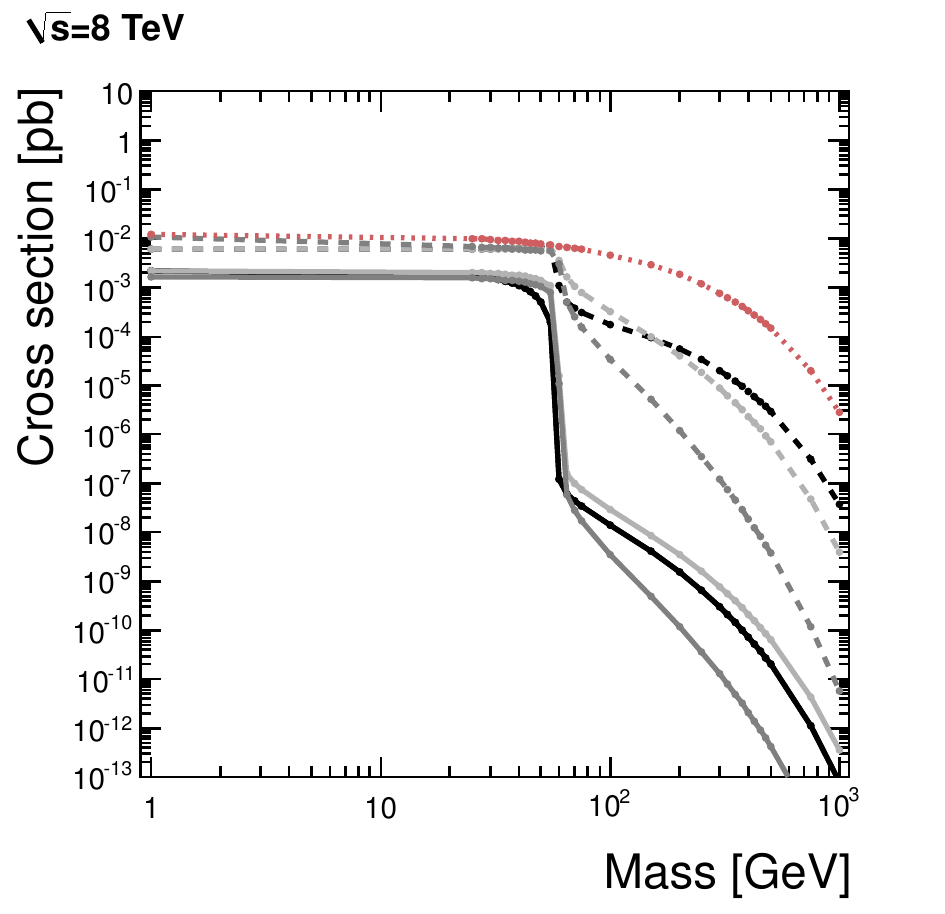}\\
\includegraphics[width=1.65in]{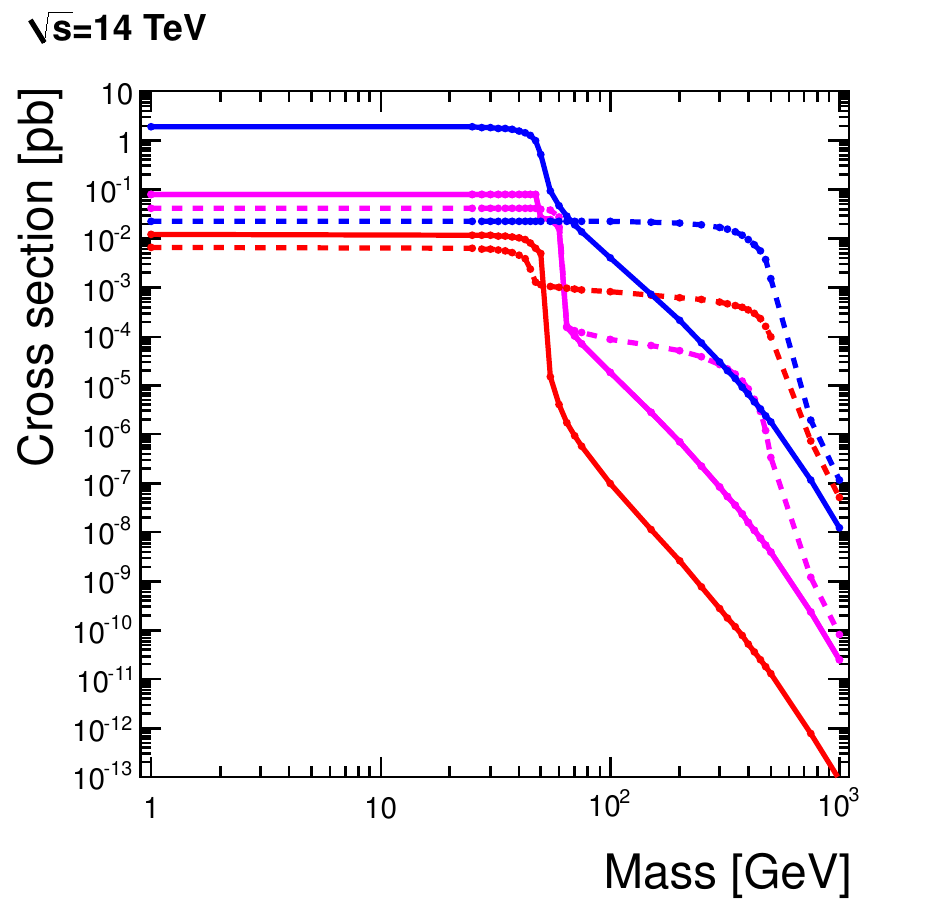}
\includegraphics[width=1.65in]{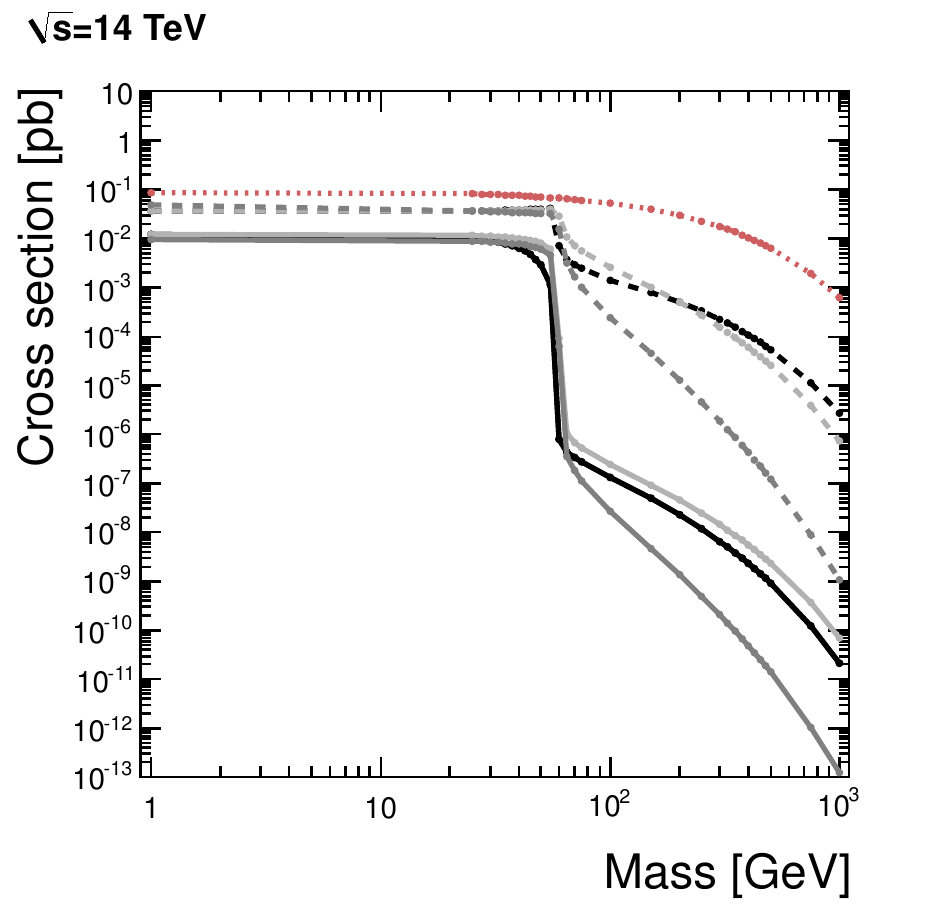}\\
\includegraphics[width=1.65in]{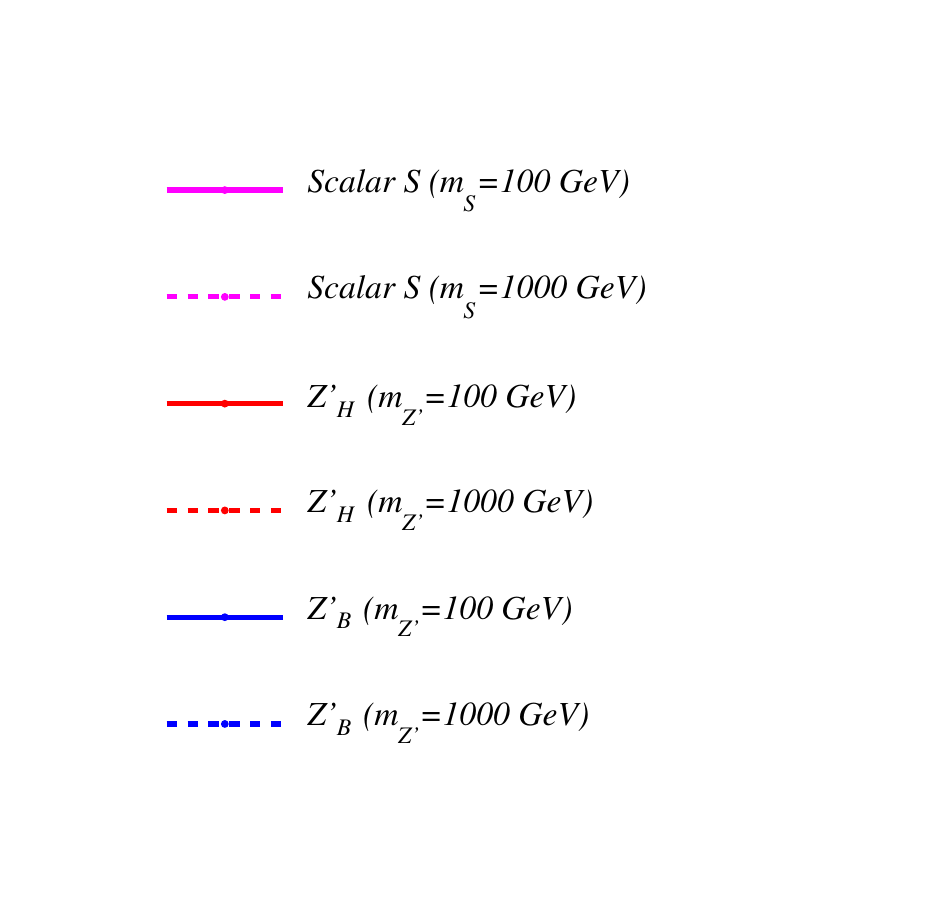}
\includegraphics[width=1.65in]{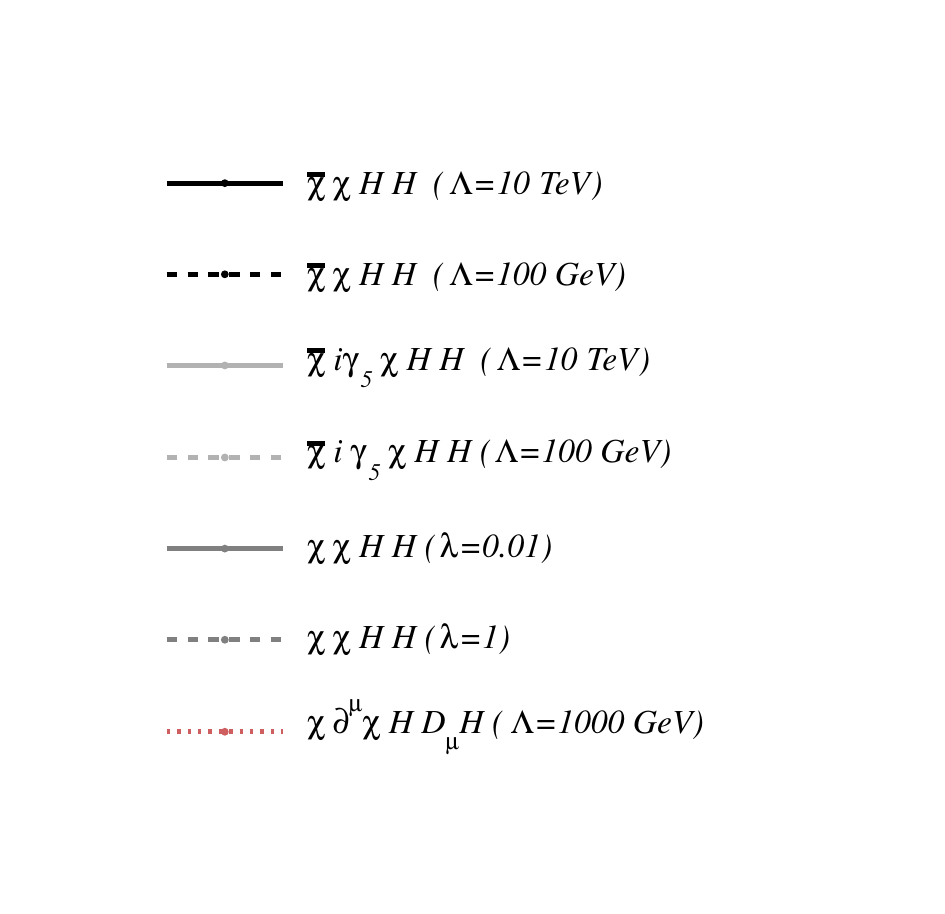}\\
\caption{ Production cross section for $h\chi\chi$ for each model at $\sqrt{s}=8$ TeV (top) and $\sqrt{s}=14$ TeV (bottom) using the benchmark values in Table~\ref{tab:bench}.}
\label{fig:xs}
\end{figure}

The critical experimental quantity is the missing transverse energy; a comparison of the $\missET$ for a few choices of dark matter or mediator masses for the  models under study can be seen in Fig.~\ref{fig:models}. The production cross section for $h\chi\chi$ under the various models are shown in Fig.~\ref{fig:xs}.

In the following sub-sections, we estimate the LHC sensitivity in four Higgs boson decay modes: $\gamma\gamma, 4\ell, b\bar{b},\ell\ell jj$.

\subsection{Two-photon decays}

The $\gamma\gamma$ decay mode has a small branching fraction, $\mathcal{B}(h\rightarrow\gamma\gamma)=2.23 \times 10^{-3}$~\cite{Heinemeyer:2013tqa}, but smaller backgrounds than other final states and well-measured objects, which leads to well-measured $\missET$.

Significant backgrounds to the $\gamma\gamma+\missET$ final state include:

\begin{itemize}
\item $Zh$ production with $Z\rightarrow\nu\bar{\nu}$, an irreducible
  background;
\item $Wh$ production with $W\rightarrow\ell\bar{\nu}$ where the
  lepton is not identified;
\item  $h\rightarrow\gamma\gamma$ or non-resonant $\gamma\gamma$
  production, with $\missET$ from mismeasurement of photons or soft radiation;
\item $Z\gamma\gamma$ with $Z\rightarrow\nu\bar{\nu}$.
\end{itemize}

\begin{figure}
\includegraphics[width=3in]{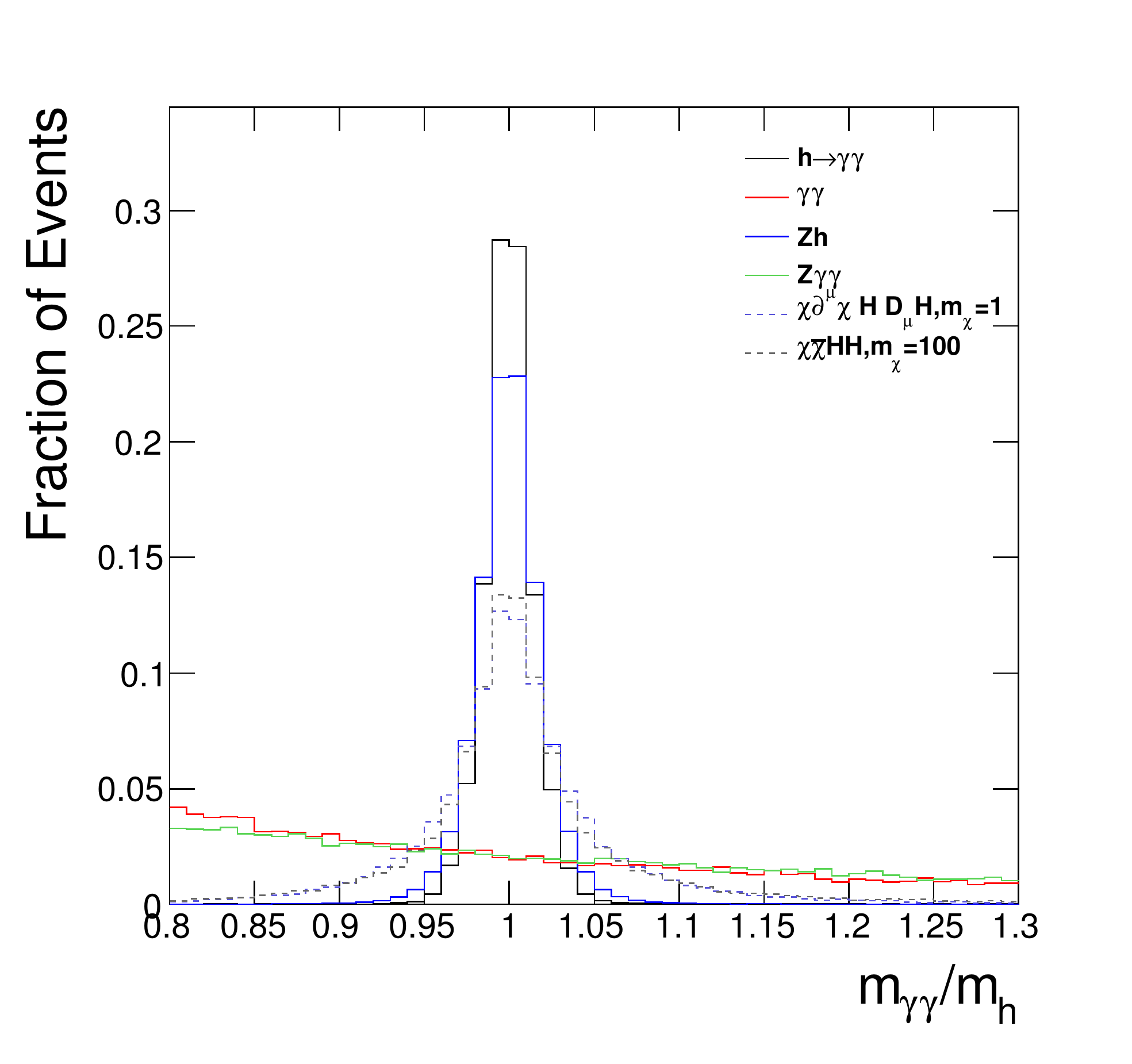}
\includegraphics[width=3in]{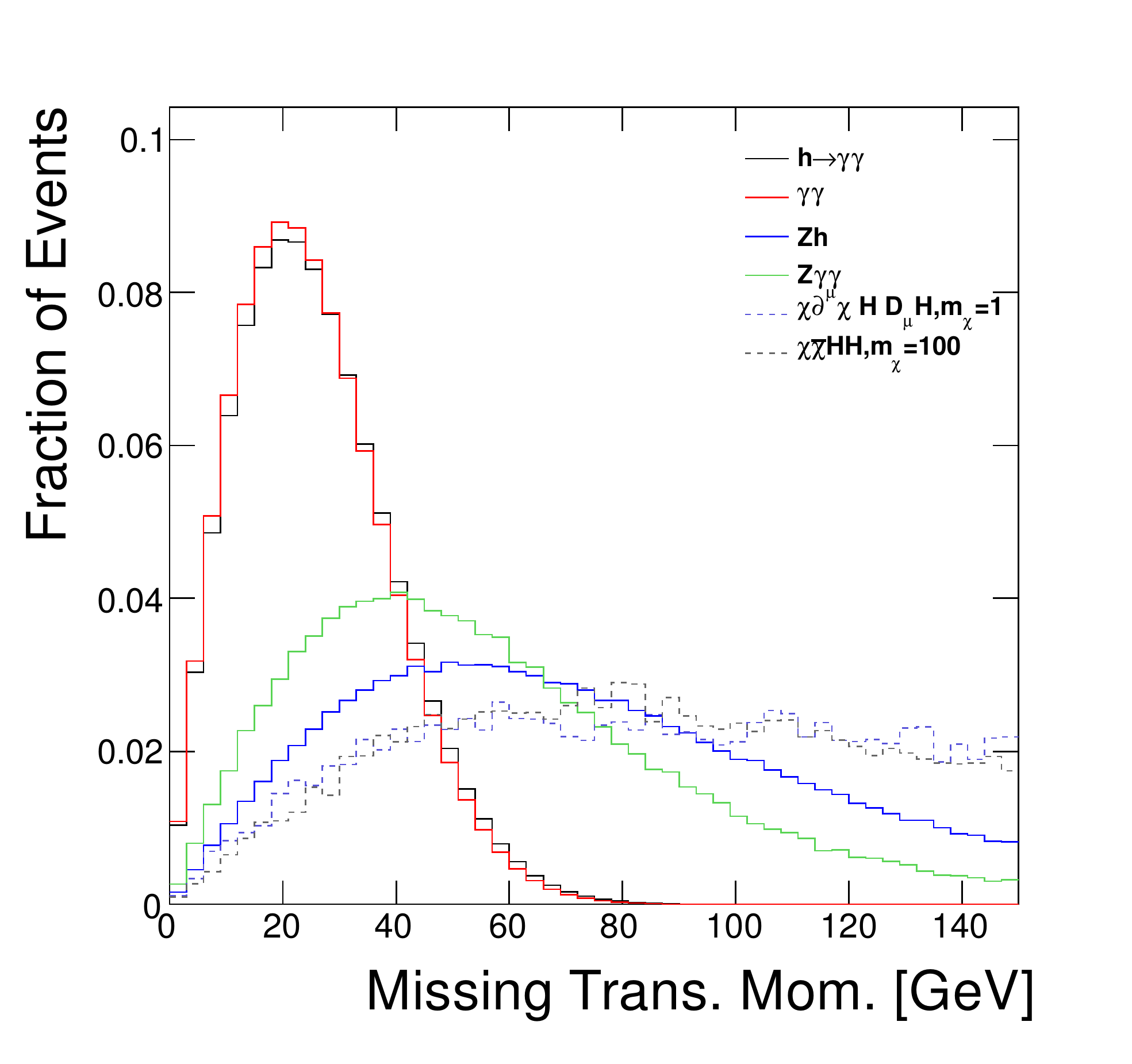}
\caption{ Distributions of diphoton invariant mass, top, and missing
  transverse momentum (bottom) for simulated $\bar{\chi}\chi HH$ signal samples
  with two choices of $m_\chi$, as well as the major background
  processes. All are for $pp$ collisions at $\sqrt{s}=8$ TeV.}
\label{fig:gg_kin}
\end{figure}

Figure~\ref{fig:gg_kin} shows distributions of the diphoton mass
($m_{\gamma\gamma}$) and the missing transverse momentum for two example signal cases and the background processes.

The production cross section for $gg\rightarrow h$ is taken at NNLO+NNLL in QCD plus NLO EW corrections~\cite{Heinemeyer:2013tqa} with 8\% uncertainty due to renormalization and factorization scale dependence and 7\% uncertainty due to parton distribution function (PDF) and $\alpha_s$ uncertanties. For $Wh,Zh$, we use the calculation of Ref.~\cite{Heinemeyer:2013tqa} which employs a zero-width approximation with NNLO QCD + NLO EW in which the dominant uncertainties are 1-3\% due to scales and 4\% due to PDFs and $\alpha_s$.  In each case, we use $\mathcal{B}(h\rightarrow\gamma\gamma)= 2.23\times 10^{-3}$ with a 5\% relative uncertainty~\cite{Heinemeyer:2013tqa}.

 The cross section for $Z\gamma\gamma$ is calculated at LO by {\sc madgraph}5, but normalized to NLO calculations using a $k$-factor of $1.75\pm 0.25$~\cite{Bozzi:2011en}. The cross section for $\gamma\gamma$ production is calculated at leading order by {\sc madgraph}5, corrected using a $k$-factor of $1.6\pm0.7$ extracted by comparing to the measured di-photon cross section~\cite{Aad:2011mh}.

Systematic uncertainties due to photon efficiency and resolution will be small compared to the uncertainty on the backgrounds, and are neglected. A potential significant source of systematic uncertainty is the modeling of the missing transverse momentum spectrum due to mismeasurement, as arises in the $\gamma\gamma$ and $h\rightarrow\gamma\gamma$ backgrounds. For the purposes of this sensitivity study, the thresholds in $\missET$ are designed to suppress these backgrounds to essentially negligible levels. A future experimental analysis must consider these more rigorously.

The event selection is:
\begin{itemize}
\item at least two photons with $p_T>20$ and $|\eta|<2.5$
\item invariant mass $m_{\gamma\gamma} \in [110,130]$ GeV
\item no electons or muons with $p_T>20$ and $|\eta|<2.5$
\item $\missET>100$ or 250 GeV.
\end{itemize}

\begin{figure}
\includegraphics[width=3in]{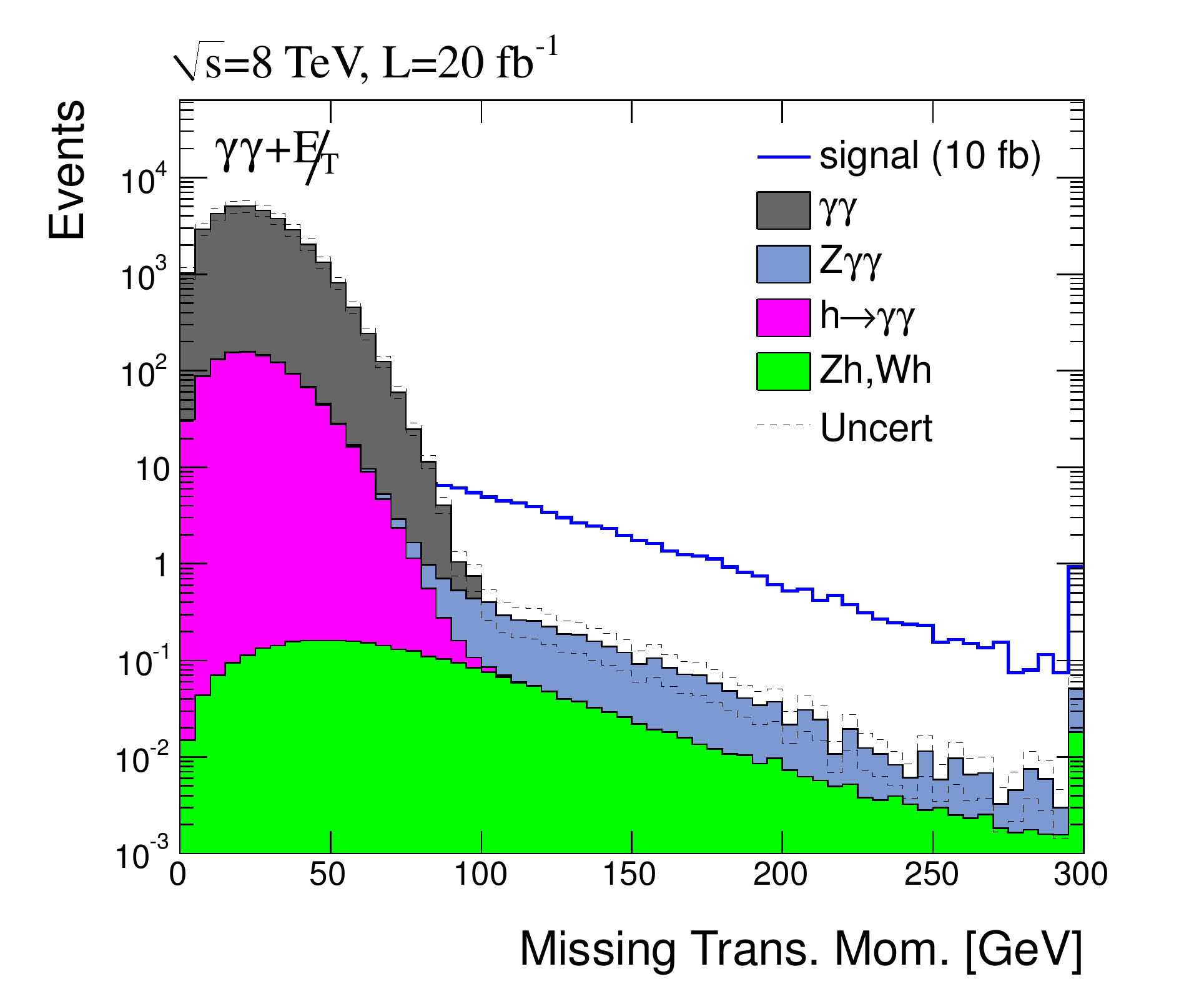}
\includegraphics[width=3in]{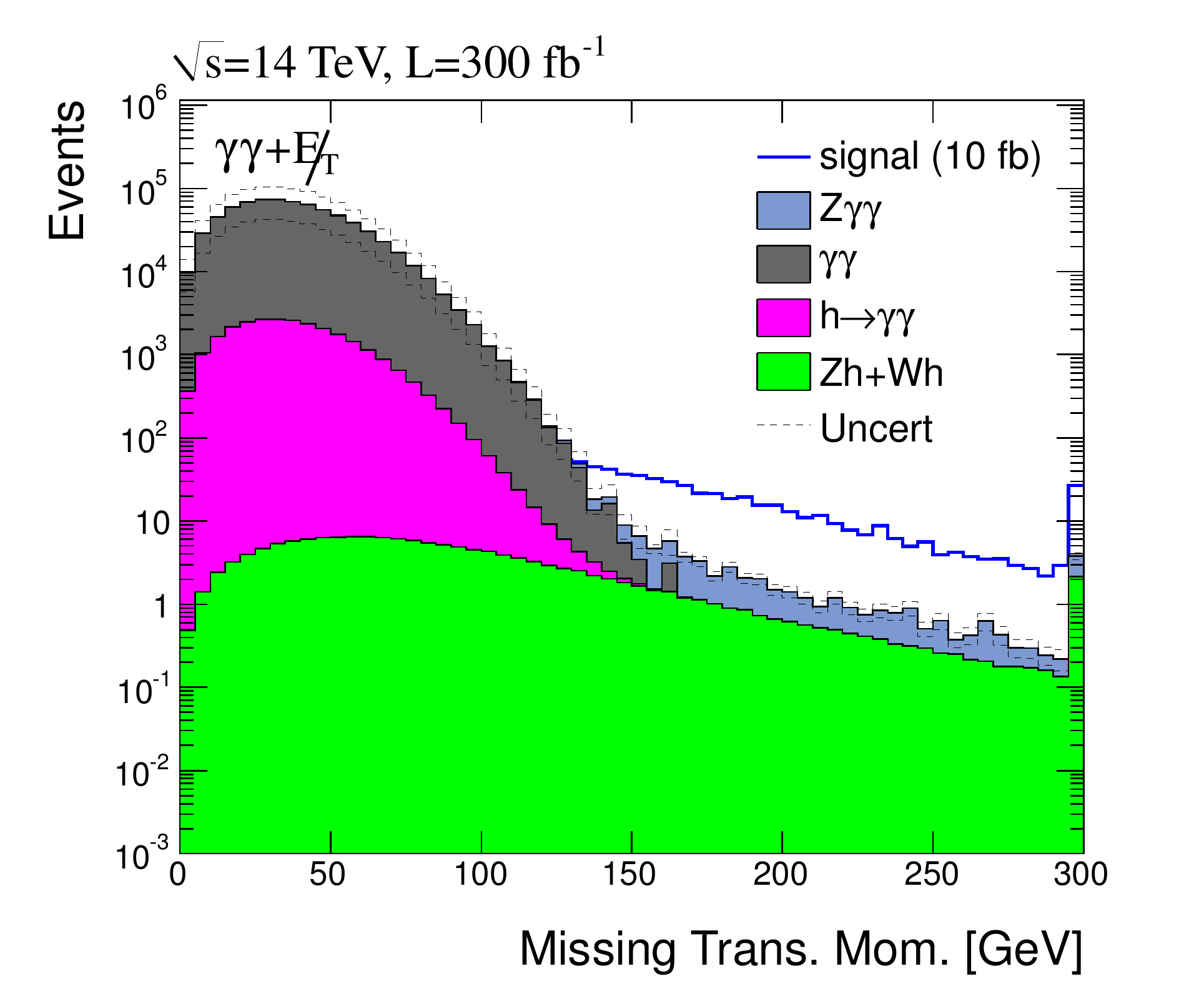}
\caption{ Distributions of  missing
  transverse momentum  in
  the $\gamma\gamma+\missET$ final state for background sources and one example signal process with after requiring  $m_{\gamma\gamma}\in [110,130]$ GeV, normalized to expected luminosity for $\sqrt{s}=8$ TeV (top) and $\sqrt{s}=14$ TeV (bottom).}
\label{fig:gg_met}
\end{figure}

Figure~\ref{fig:gg_met} shows the distribution of expected events at
$\sqrt{s}=8$ and 14 TeV as a function of missing transverse
momentum. We select a minimum $\missET$ threshold by optimizing the
expected cross-section upper limit, finding $\missET>100$~GeV and
$\missET>250$~GeV for the $\sqrt{s}=8$ and 14 TeV cases,
respectively.  Note that the $\missET$ spectrum varies between the models, such that a single global optimal value of $\missET$ is not possible.  We select a single $\missET$ threshold which gives the best aggregrate limits across choices of models and $m_\chi$; further optimization is not warranted given the approximate nature of our background model and systematic uncertainties. Table~\ref{tab:gg} shows the expected event yields for
each of these cases.

\begin{table}
\caption{Expected background and signal yields in the $\gamma\gamma+\missET$ channel for $pp$ collisions at
  $\sqrt{s}=8$ TeV with $\mathcal{L}=20$~fb$^{-1}$, left, or
  $\sqrt{s}=14$ TeV with $\mathcal{L}=300$~fb$^{-1}$, right. The signal
  case corresponds to $\sigma=10$~fb, and $m_\chi=1$ GeV in the $\bar{\chi}\chi HH$ model.}
\label{tab:gg}
\begin{tabular}{lrr}
\hline\hline
& $\sqrt{s}=8$ TeV &\ \ \ \ $\sqrt{s}=14$ TeV \\
&  $\mathcal{L}=20$~fb$^{-1}$ &
$\mathcal{L}=300$~fb$^{-1}$\\
& $\missET>100$ & $\missET>250$ \\
\hline
$Z\gamma\gamma$ & $2.4 \pm 0.3$& $3.4 \pm 0.4$ \\
$\gamma\gamma$ & $0^{+0.5}_{-0.0}$ & $0^{+0.5}_{-0.0}$ \\
$h\rightarrow\gamma\gamma$ & $0^{+0.1}_{-0.0}$ &  $0^{+0.1}_{-0.0}$ \\
$Zh,Wh$ &$0.7 \pm 0.1$ & $3.9 \pm 0.4$\\
\hline
Total Bkg & $3.1 \pm 0.6$ & $7.3 \pm 0.7$ \\
$\bar{\chi}\chi HH$& 50 & 45 \\
\hline\hline
\end{tabular}
\end{table}

\begin{figure}
\includegraphics[width=1.65in]{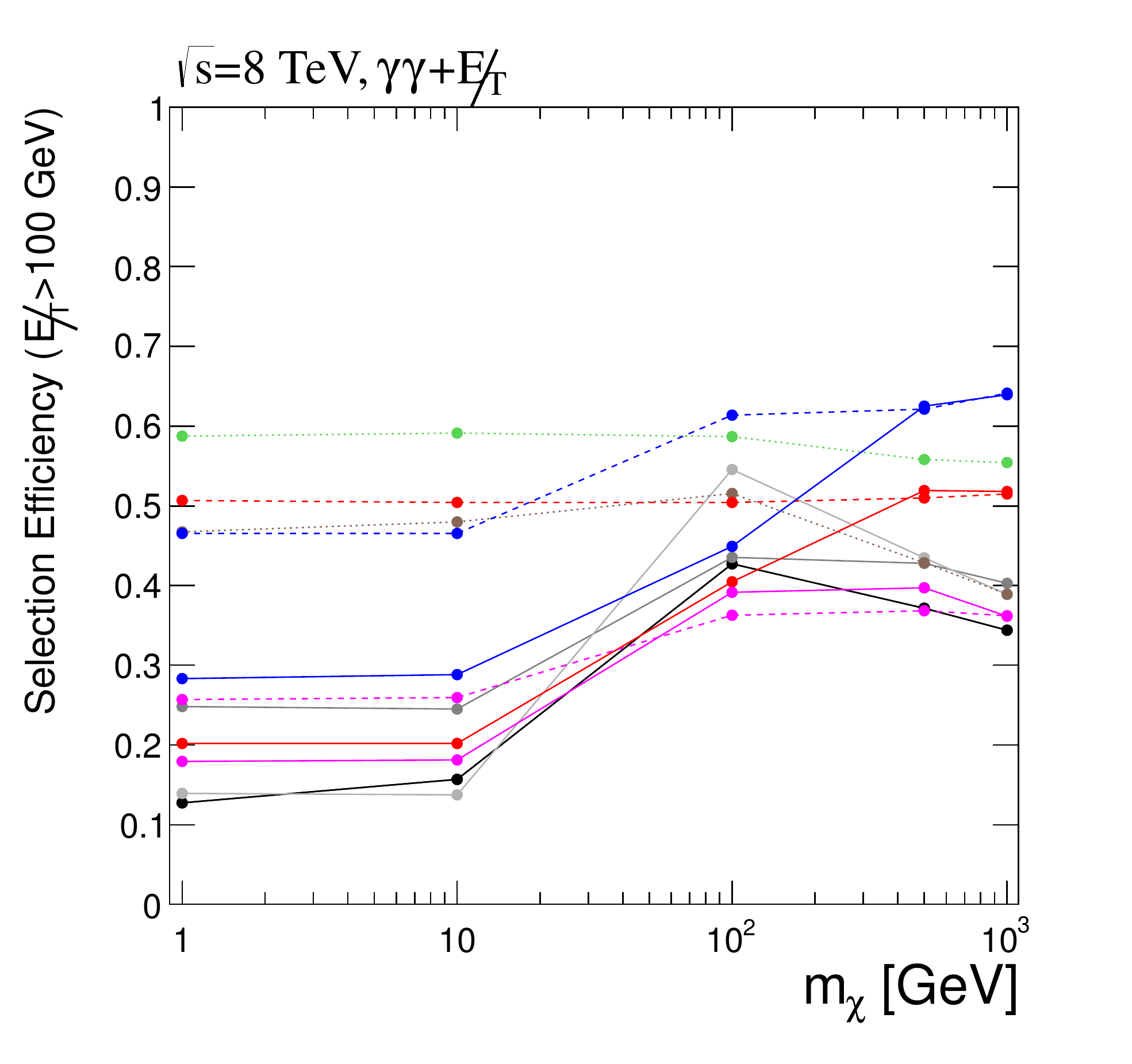}
\includegraphics[width=1.65in]{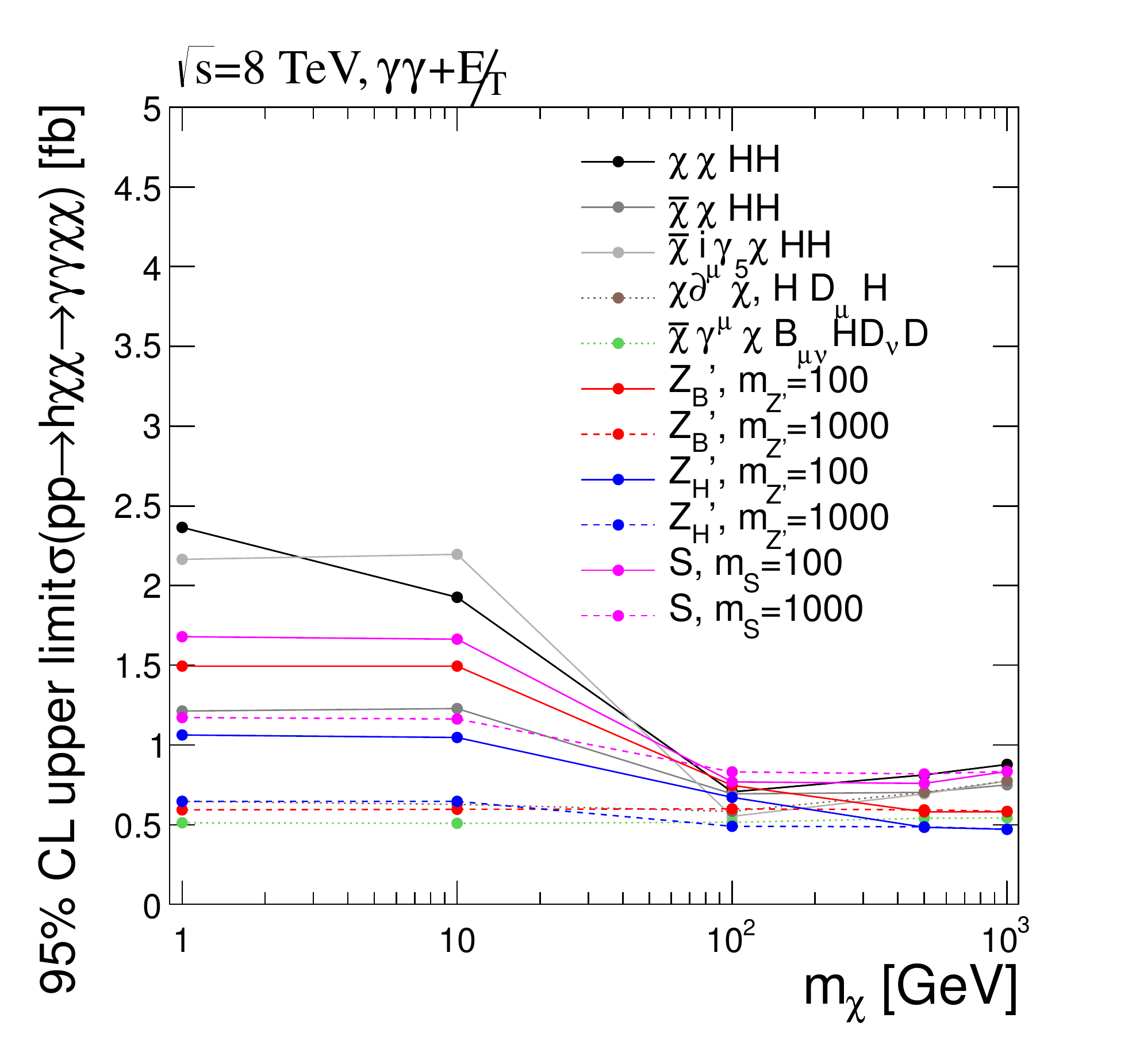}
\includegraphics[width=1.65in]{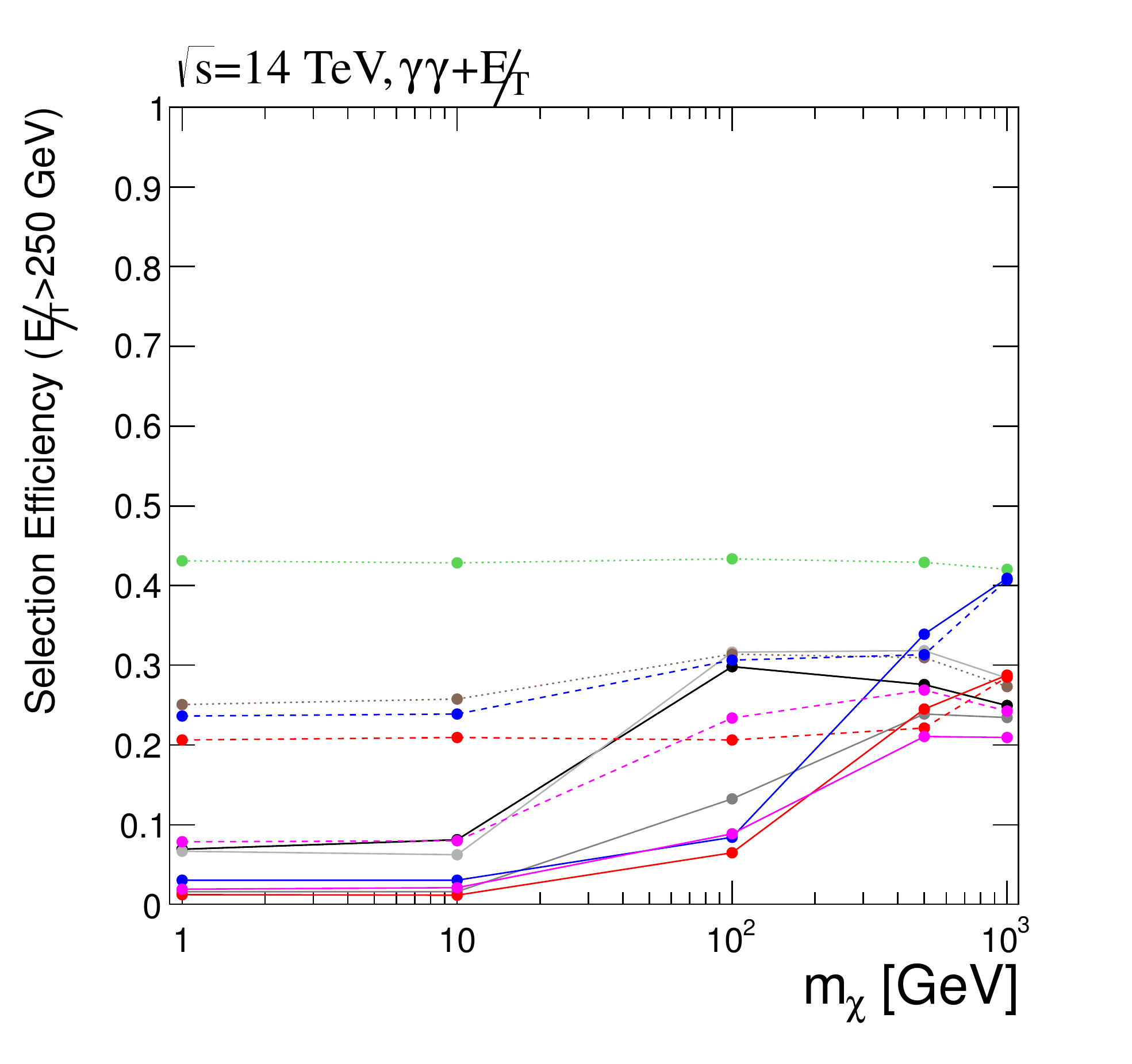}
\includegraphics[width=1.65in]{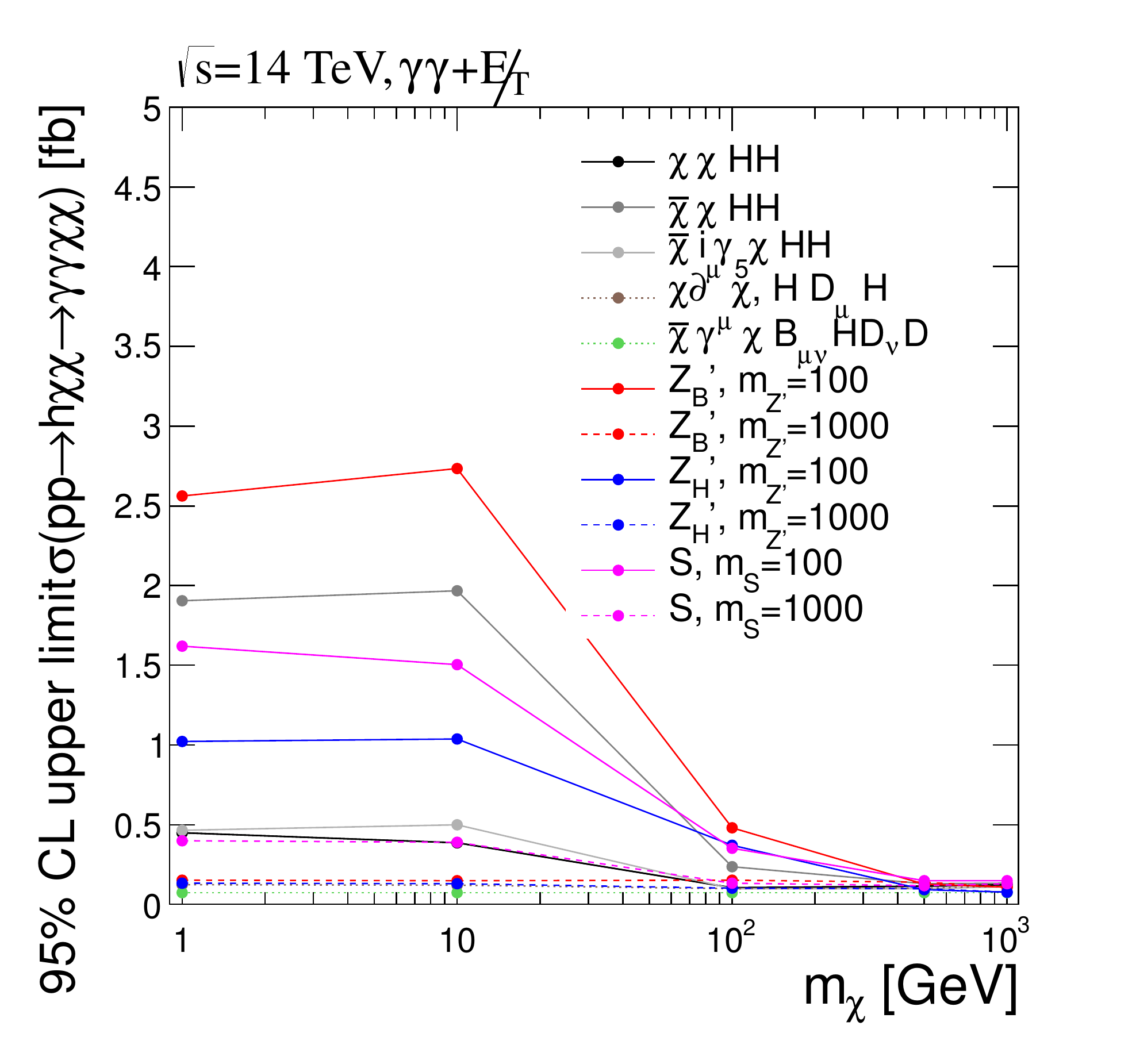}
\caption{ Selection efficiency in the $\gamma\gamma+\missET$ channel (left) and upper limits (right) on
  $\sigma(pp\rightarrow h\chi\bar{\chi}\rightarrow
  \gamma\gamma\chi\bar{\chi})$ for $\sqrt{s}=8$ TeV (top) and
  $\sqrt{s}=14$ TeV (bottom).  }
\label{fig:gg_lim}
\end{figure}

Limits are calculated using the CLs method with the asymptotic
approximation~\cite{cranmer}. Selection efficiency and upper limits on
 $\sigma(pp\rightarrow h\chi\bar{\chi}\rightarrow
  \gamma\gamma\chi\bar{\chi})$ are shown in Fig.~\ref{fig:gg_lim}.

\subsection{Four-lepton decays}

The four-lepton decay mode, via $h\rightarrow ZZ^*\rightarrow 4\ell$, has the smallest branching ratio of the modes considered here, but also offers the smallest backgrounds.

Backgrounds to the $4\ell+\missET$ final state include:

\begin{itemize}
\item $Zh$ production with $Z\rightarrow\nu\bar{\nu}$, an irreducible
  background;
\item $Zh$ production with $Z\rightarrow\ell\ell$ and $h\rightarrow\ell\ell\nu\nu$;
\item $Wh$ production with $W\rightarrow\ell\bar{\nu}$ where the
  lepton from the $W$ decay is not identified;
\item  $h\rightarrow ZZ^*\rightarrow 4\ell$ or the continuum $(Z^{(*)}/\gamma^*)(Z^{(*)}/\gamma^*)\rightarrow4\ell$
  production, with $\missET$ from mismeasurement of leptons or soft radiation.
\end{itemize}

As in the case for two-photon decays, the cross sections and uncertainties for $gg\rightarrow h$, $Wh$, and $Zh$ production, and the $h$ branching fractions are taken from Ref.~\cite{Heinemeyer:2013tqa}. We take branching fractions $\mathcal{B}(h\rightarrow 4\ell)=1.26\times10^{-4}$, and $\mathcal{B}(h\rightarrow 2\ell2\nu)=1.06\times10^{-2}$. The considerably larger branching fraction involving neutrinos results in a significant contribution from the non-resonant $Zh\ (Z\rightarrow \ell\ell)$ background.

Simulated samples of $(Z^{(*)}/\gamma^*)(Z^{(*)}/\gamma^*)$ events, hereafter referred to simply as $ZZ^*$, are generated by \textsc{madgraph5} at LO. The yield is compared against NLO values calculated with {\sc powheg} and gg2ZZ in \cite{Aad:2013hdb}, and the difference is assigned as a systematic.

To improve the accuracy of the modeling of lepton reconstruction efficiency by {{\sc delphes}, we scale the per-lepton efficiencies to match those reported by ATLAS~\cite{Aad:2013hdb} in the $4e, 4\mu, 2e2\mu$ final states and apply these efficiences to all simulated samples.

\begin{figure}
\includegraphics[width=1.65in]{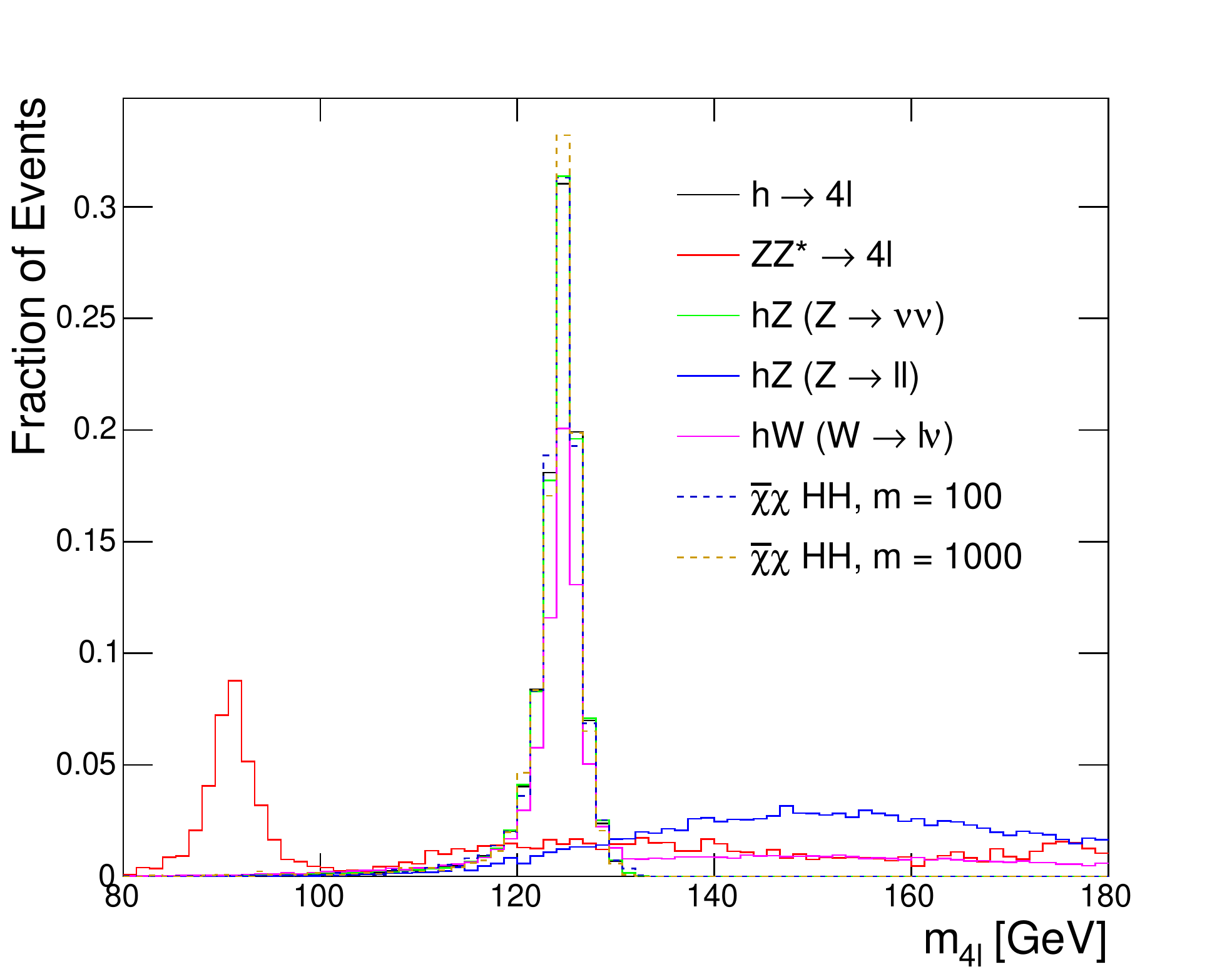}
\includegraphics[width=1.65in]{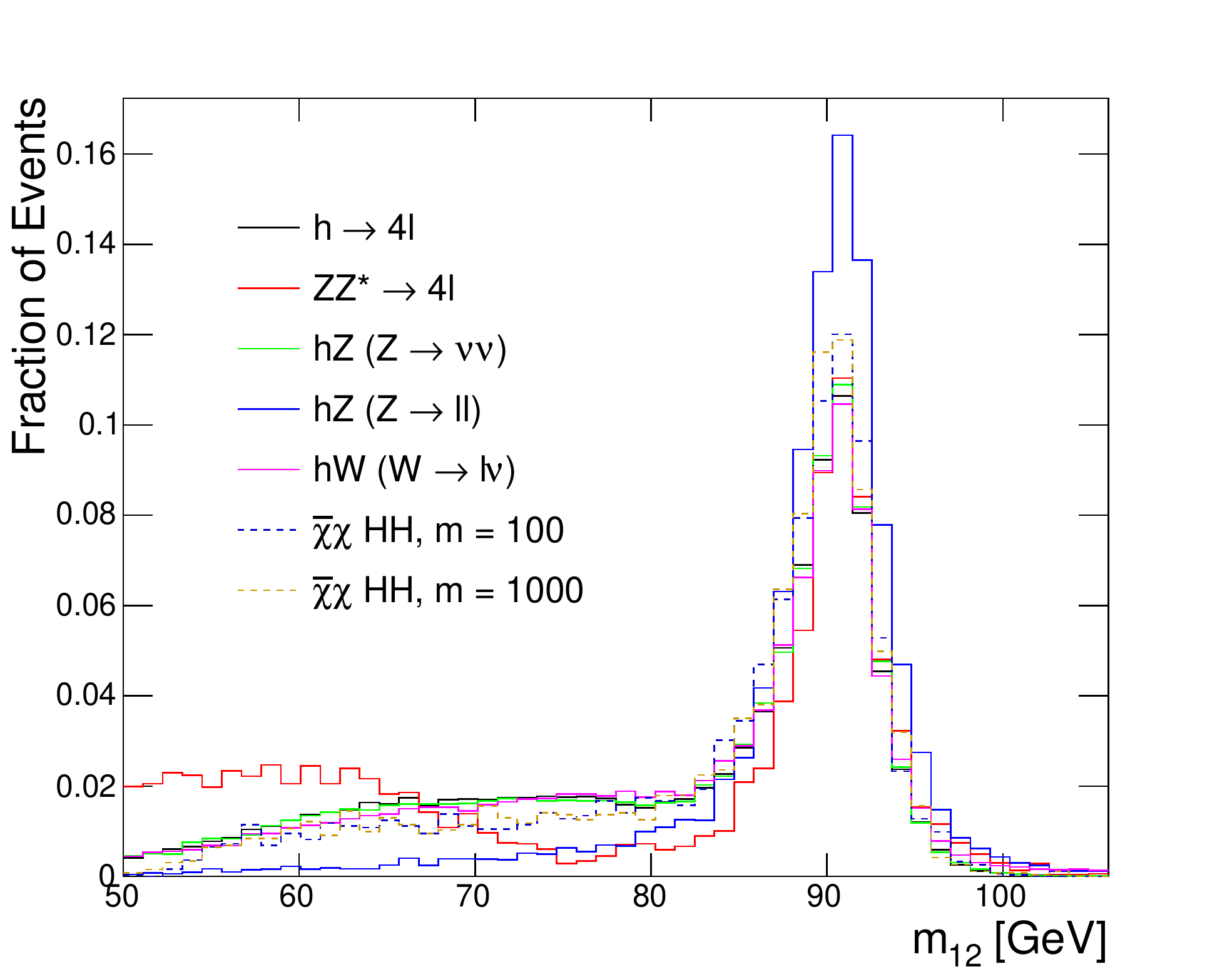}
\includegraphics[width=1.65in]{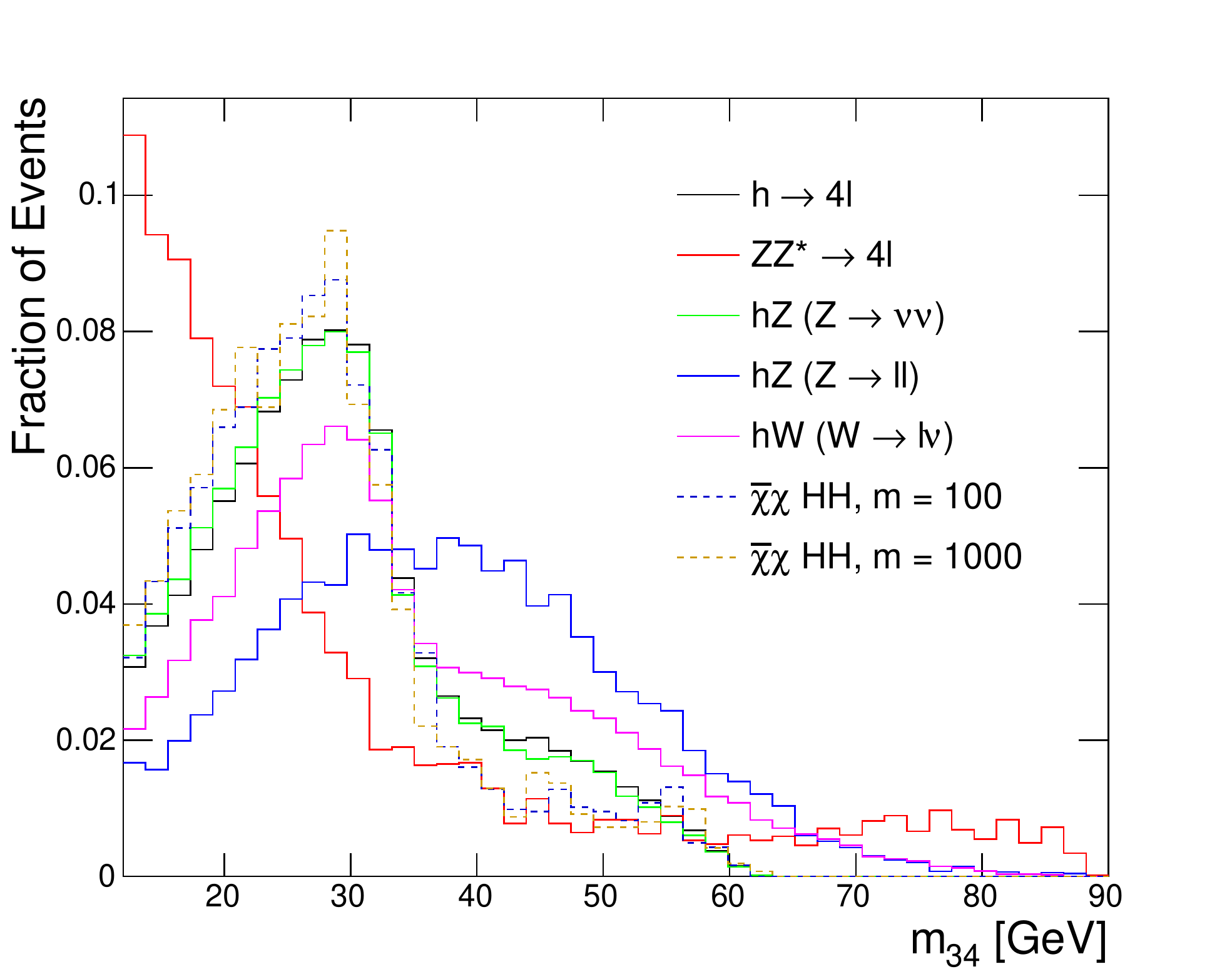}
\includegraphics[width=1.65in]{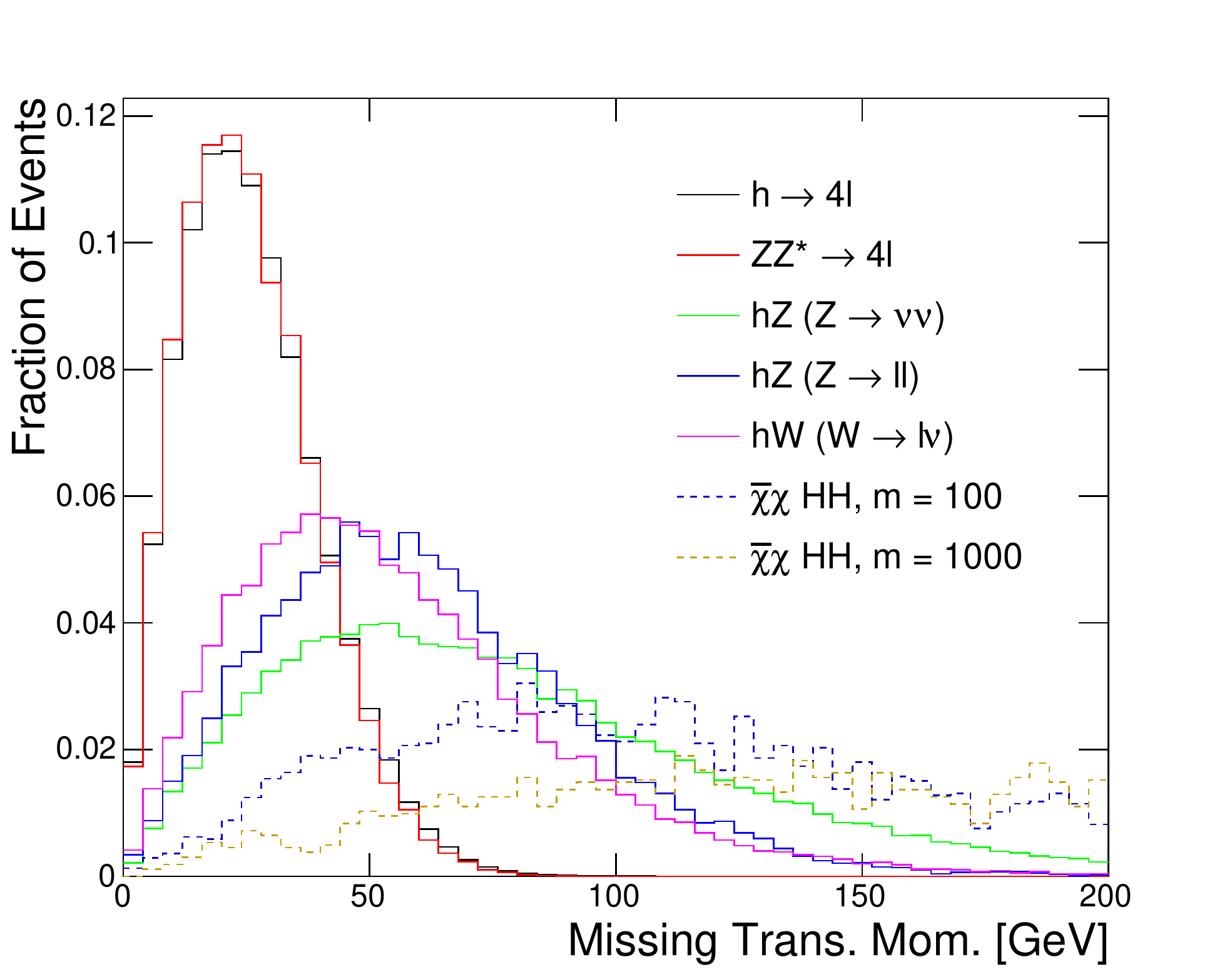}
\caption{ Distributions of four-lepton invariant mass $m_{4\ell}$, leading  ($m_{12}$) and subleading ($m_{34}$) dilepton mass, and missing
  transverse momentum for simulated $h\chi\bar{\chi}$ signal samples
  with two choices of $m_\chi$, as well as the major background
  processes. All are for $pp$ collisions at $\sqrt{s}=8$ TeV.}
\label{fig:4l_kin}
\end{figure}

We define the leading lepton pair to be the same-flavor, opposite-sign pair with invariant mass $m_{12}$ closest to the $Z$-boson mass. The sub-leading pair's invariant mass, $m_{34}$, is the next closest to the $Z$-boson mass. Figure~\ref{fig:4l_kin} shows distributions of the lepton pair masses, $m_{12}$ and $m_{34}$, the four-lepton invariant mass, $m_{4\ell}$, and the missing transverse momentum. We also define $m_{\textrm{min}}$ as a function which is a constant $12\ \mathrm{GeV}$ for $m_{4\ell}<140\ \mathrm{GeV}$, then rises linearly to $50\ \mathrm{GeV}$ for $m_{4\ell}<190\ \mathrm{GeV}$ and remains constant. Then each event must satisfy:
\begin{itemize}
\item at least four leptons with each electron (muon) satisfying:
  \begin{itemize}
     \item $p_T>7\ \mathrm{GeV}$ ($p_T>6\ \mathrm{GeV}$)
     \item $|\eta|<2.47$ ($|\eta|<2.7$)
  \end{itemize}
\item highest $p_T$ lepton is an electron (muon) with $p_T > 20\ \mathrm{GeV}$, and the second (third) lepton satisfies $p_T > 15\ \mathrm{GeV}$ ($p_T > 10\ \mathrm{GeV}$)
\item $50\ \mathrm{GeV} <m_{12}<106\ \mathrm{GeV}$
\item $m_{\textrm{min}}<m_{34}<115\ \mathrm{GeV}$
\item $105\ \mathrm{GeV} < m_{4\ell} < 145\ \mathrm{GeV}$
\item $\missET>75$ or 150 GeV.
\end{itemize}

Figure~\ref{fig:4l_met} shows the distribution of expected events at
$\sqrt{s}=8$ and 14 TeV as a function of missing transverse
momentum. We select a minimum $\missET$ threshold by optimizing the
expected cross-section upper limit, finding $\missET>75$~GeV and
$\missET>150$~GeV for the $\sqrt{s}=8$ and 14 TeV cases,
respectively. Table~\ref{tab:4l} shows the expected event yields for
each of these cases.

\begin{figure}
\includegraphics[width=3.3in]{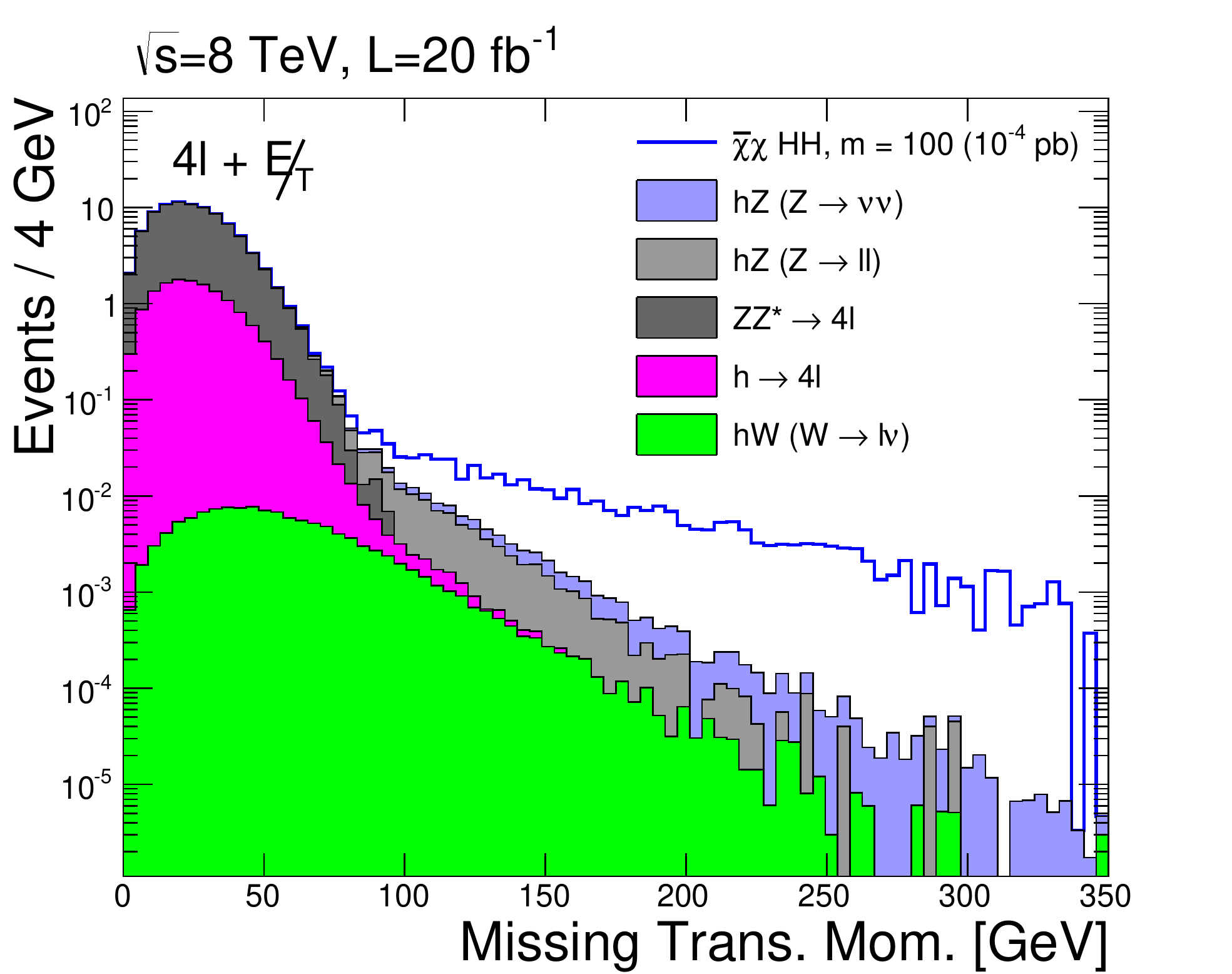}
\includegraphics[width=3.3in]{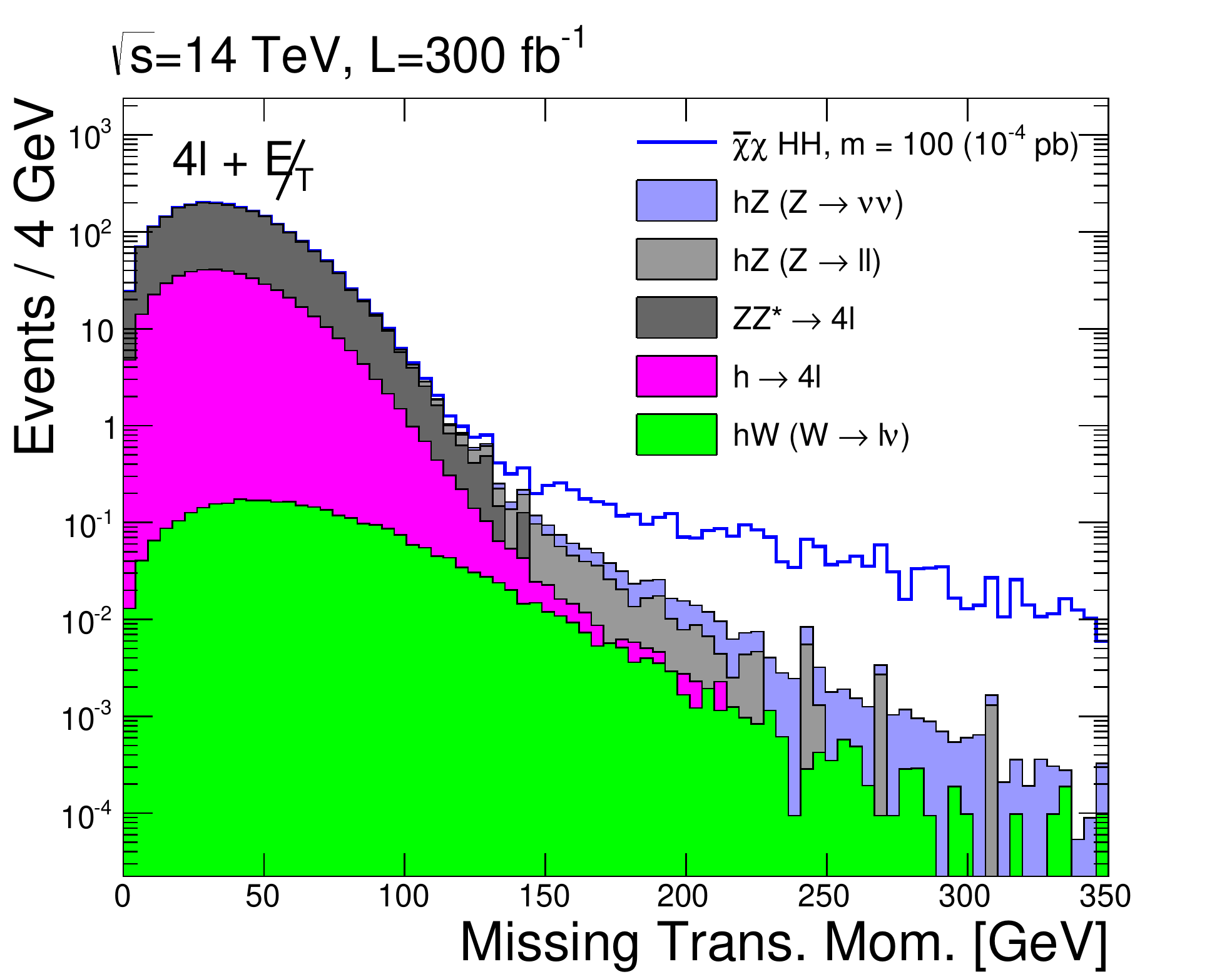}
\caption{ Distributions of  missing
  transverse momentum  in
  the $4\ell+\missET$ final state for simulated signal and background samples with normalized to expected luminosity.}
\label{fig:4l_met}
\end{figure}

\begin{table}
\caption{Expected background and signal yields in the $4\ell+\missET$ channel for $pp$ collisions at
  $\sqrt{s}=8$ TeV with $\mathcal{L}=20$~fb$^{-1}$, left, or
  $\sqrt{s}=14$ TeV with $\mathcal{L}=300$~fb$^{-1}$, right. The signal
  case corresponds to $\sigma=10^{-4}$~pb, and $m_\chi=1$ GeV.}
\label{tab:4l}
\begin{tabular}{lrr}
\hline\hline
& $\sqrt{s}=8$ TeV &\ \ \ \ $\sqrt{s}=14$ TeV \\
&  $\mathcal{L}=20$~fb$^{-1}$ &
$\mathcal{L}=300$~fb$^{-1}$\\
& $\missET>75$ & $\missET>150$ \\
\hline
$ZZ^{*}$ 				& $(5.21 \pm 0.05) \times 10^{-1}$		& $0^{+0.45}_{-0.00} \times 10^{-1}$ \\
$hZ (Z\rightarrow \ell\ell)$ & $(2.09 \pm 0.09) \times 10^{-1}$	& $(6.30 \pm 0.49) \times 10^{-1}$\\
$h\rightarrow 4\ell$		& $(1.83 \pm 0.20) \times 10^{-1}$	& $(2.33 \pm 0.57) \times 10^{-1}$\\
$hZ (Z\rightarrow \nu\nu)$ & $(3.11 \pm 0.13) \times 10^{-2}$	& $(1.89 \pm 0.13) \times 10^{-1}$\\
$hW$ 				& $(3.29 \pm 0.09) \times10^{-2}$	& $(9.73 \pm 0.52) \times 10^{-2}$\\
\hline
Total Bkg & $0.977 \pm 0.023$ & $1.15 \pm 0.09$ \\
$\chi\bar{\chi}HH$  & 0.279 & 0.866 \\
\hline\hline
\end{tabular}
\end{table}

Selection efficiency and upper limits on
 $\sigma(pp\rightarrow h\chi\bar{\chi}\rightarrow
  4\ell\chi\bar{\chi})$ are shown in Fig.~\ref{fig:4l_lim}.

\begin{figure}
\includegraphics[width=1.65in]{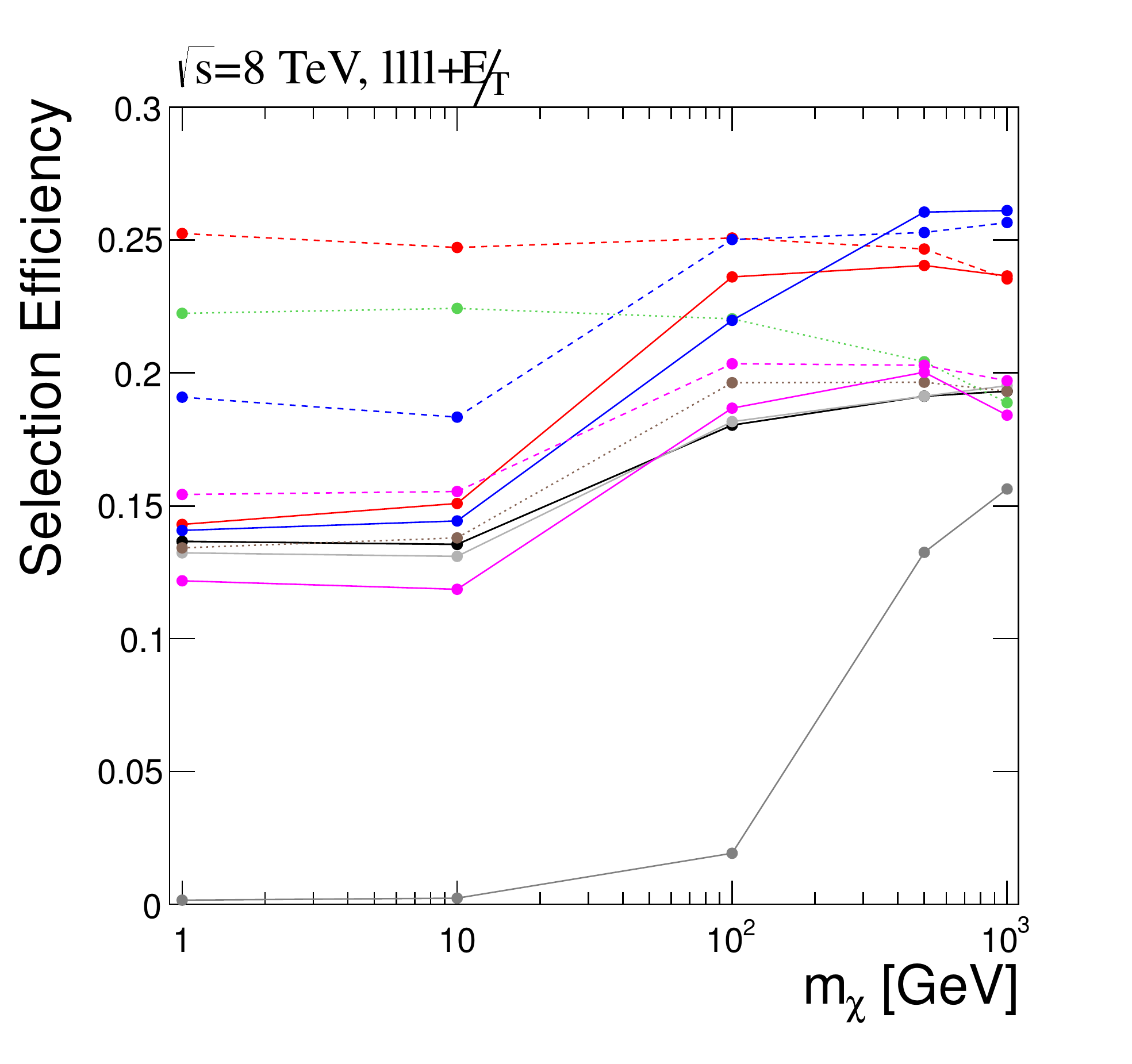}
\includegraphics[width=1.65in]{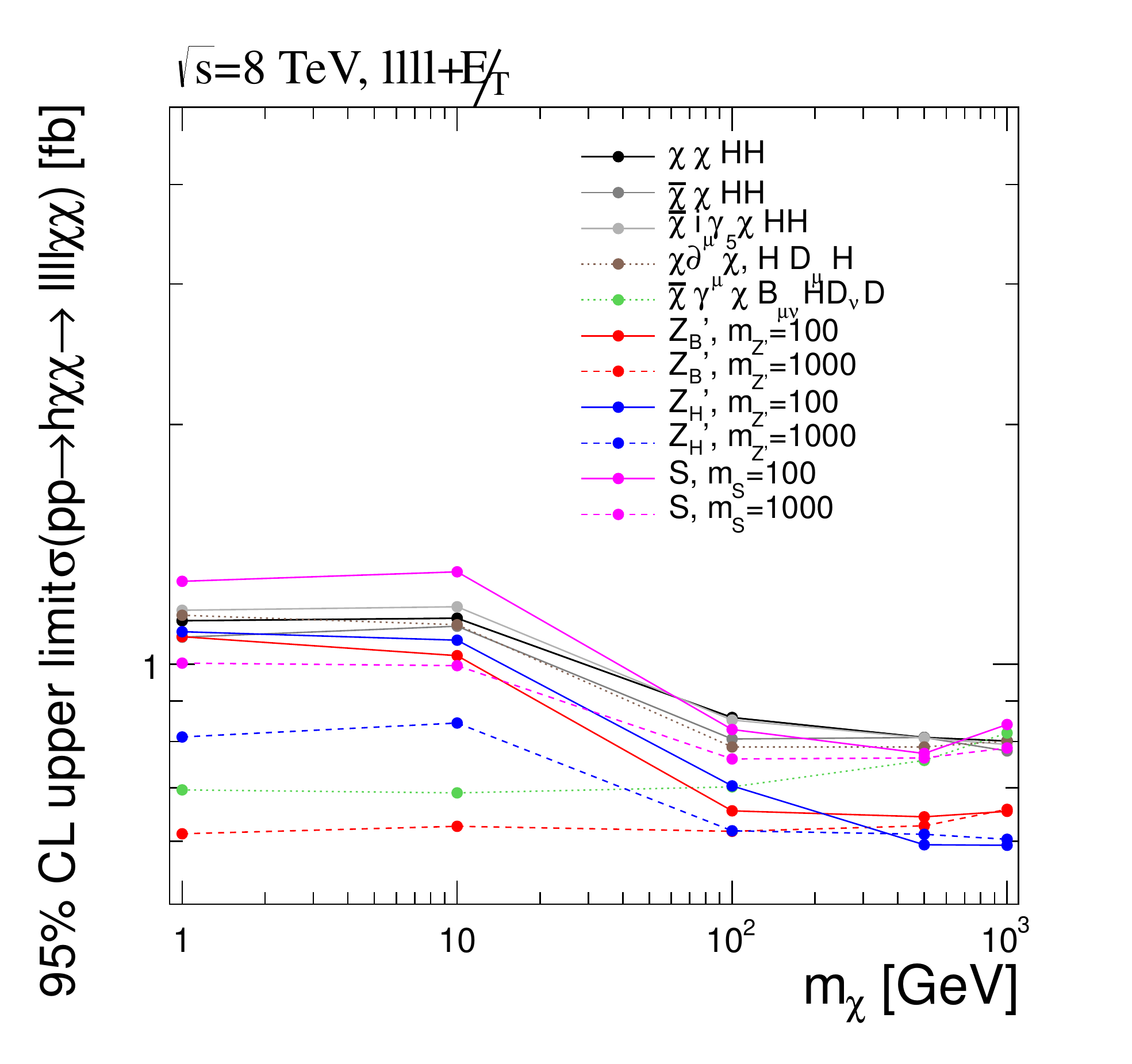}
\includegraphics[width=1.65in]{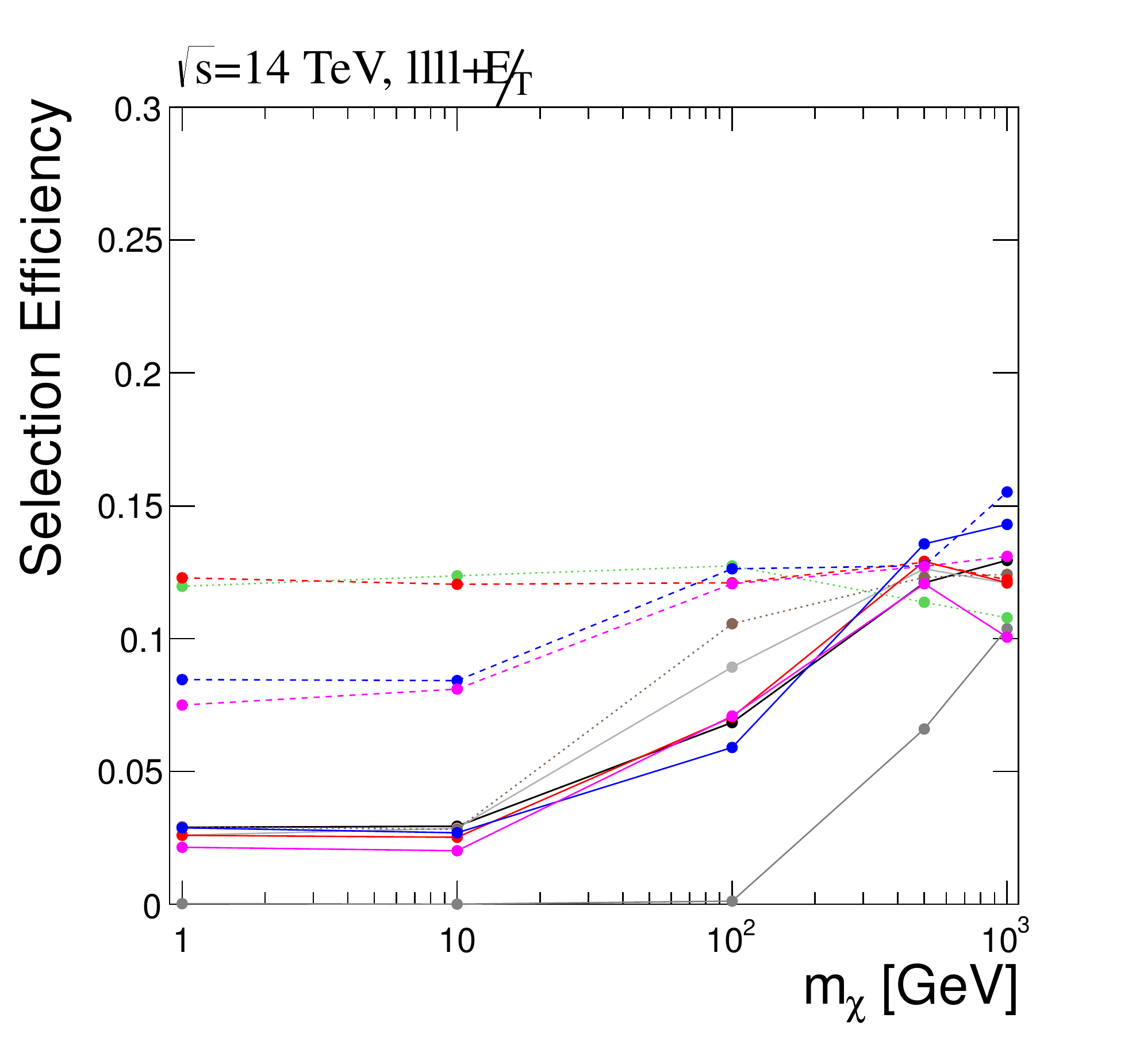}
\includegraphics[width=1.65in]{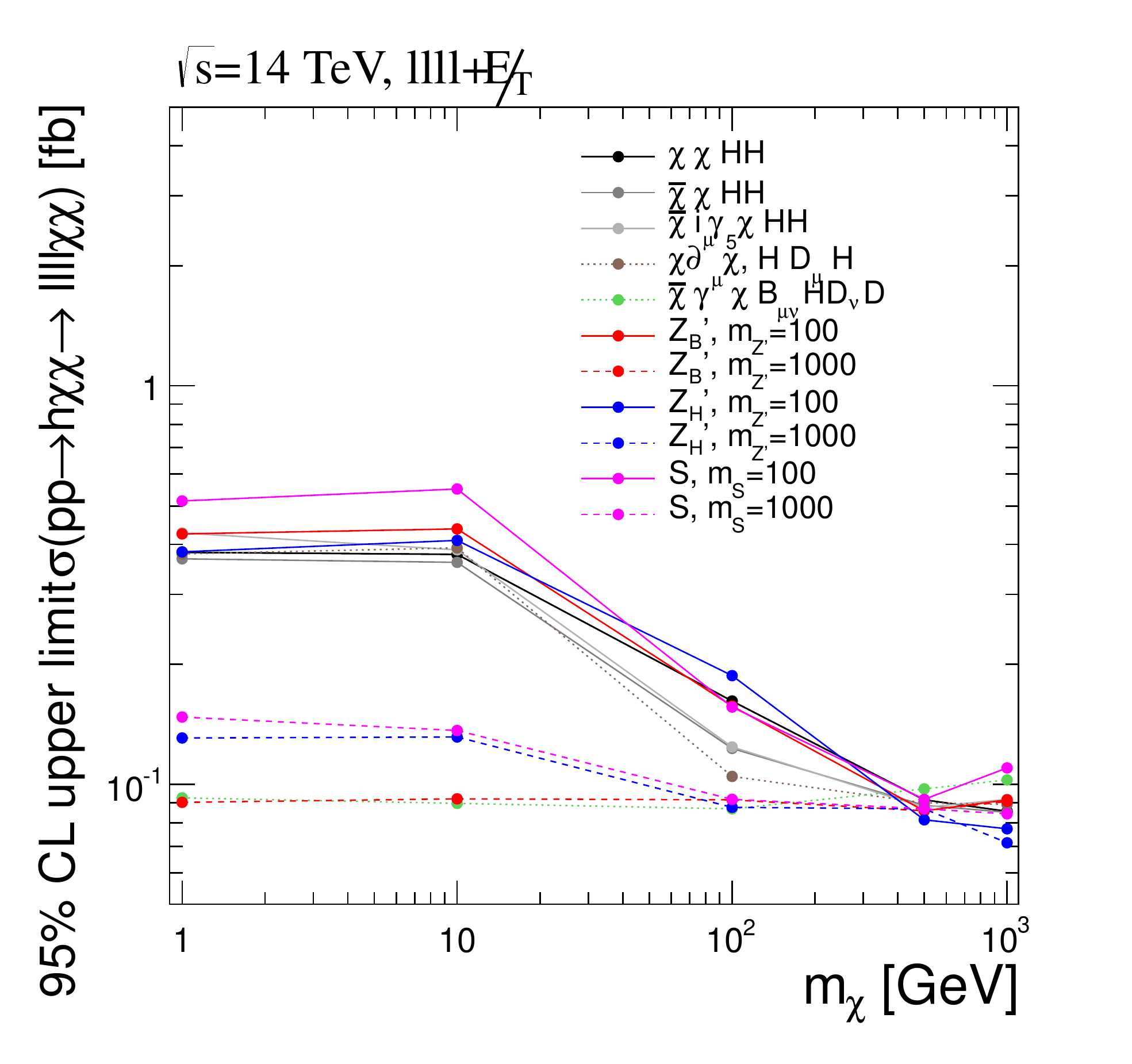}
\caption{ Selection efficiency in the $4\ell+\missET$ channel (left) and upper limits on
  $\sigma(pp\rightarrow h\chi\bar{\chi}\rightarrow
  4\ell\chi\bar{\chi})$ for $\sqrt{s}=8$ TeV (top) and
  $\sqrt{s}=14$ TeV (bottom).}
\label{fig:4l_lim}
\end{figure}

\subsection{Two-$b$-quark decays}

The two-$b$-quark mode is the dominant Higgs boson decay mode, but suffers from a very large background due to strong production of dijets as well as the poorest $\missET$ resolution.

Backgrounds to the $b\bar{b}+\missET$ final state include:

\begin{itemize}
\item $Zh$ production with $Z\rightarrow\nu\bar{\nu}$, an irreducible
  background;
\item $Wh$ production with $W\rightarrow\ell\bar{\nu}$ where the
  lepton from the $W$ decay is not identified;
\item $Zb\bar{b}$ and $Wb\bar{b}$ production;
\item  $h\rightarrow b\bar{b}$ or non-resonant $b\bar{b}$
  production, with $\missET$ from mismeasurement of leptons or soft radiation;
\item top-quark pair production $t\bar{t}$
\end{itemize}

\begin{figure}
\includegraphics[width=1.65in]{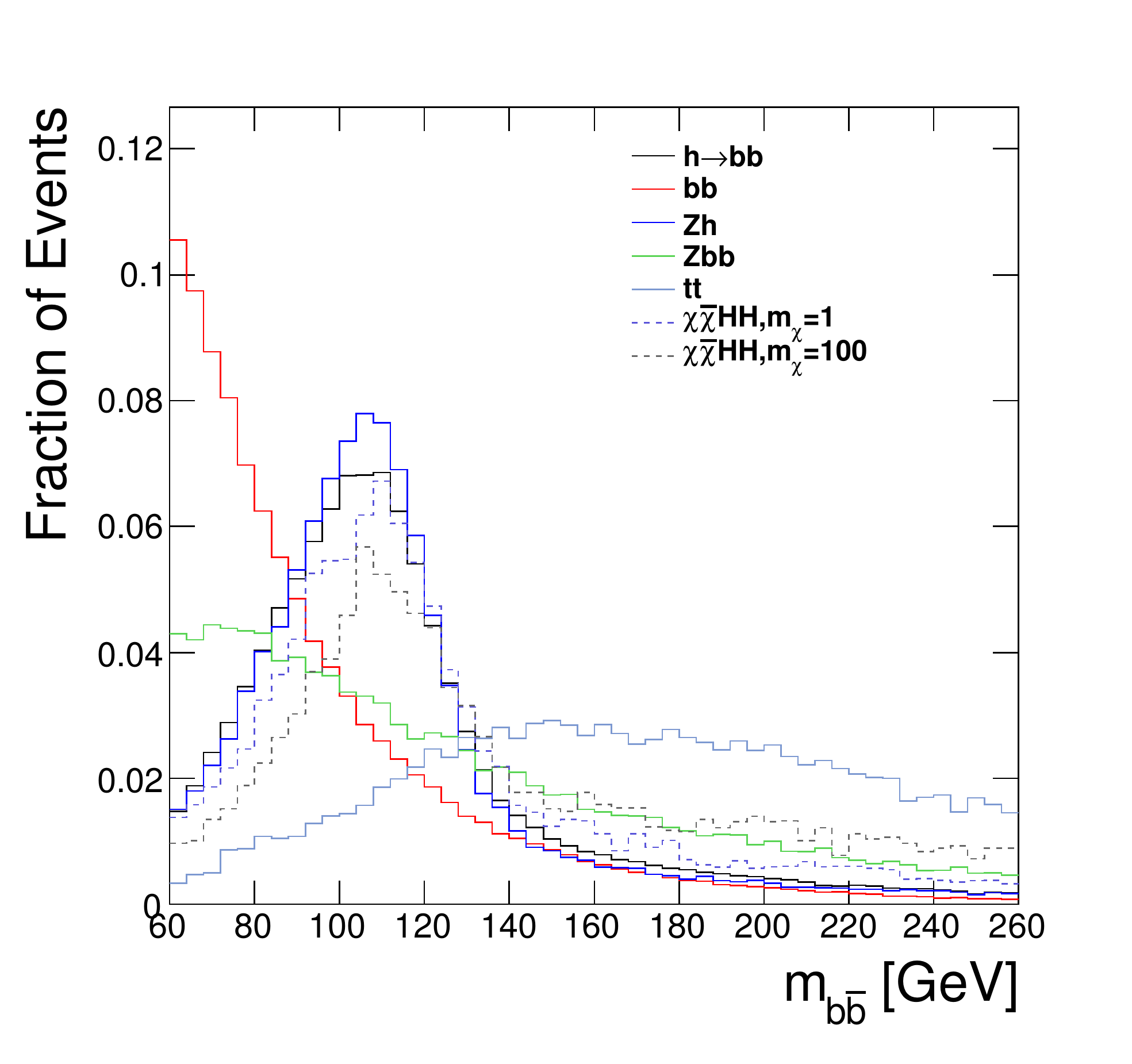}
\includegraphics[width=1.65in]{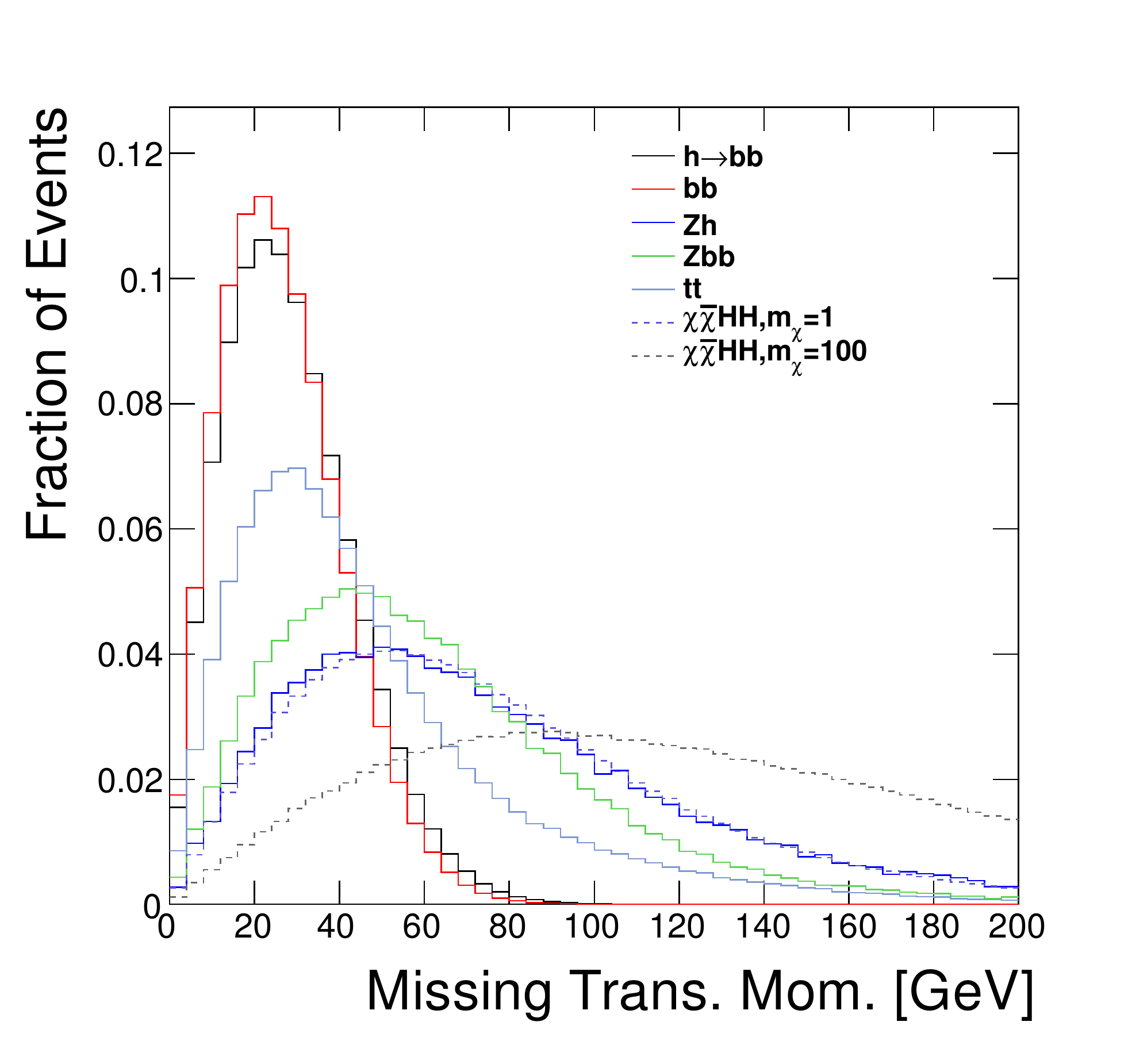}
\includegraphics[width=1.65in]{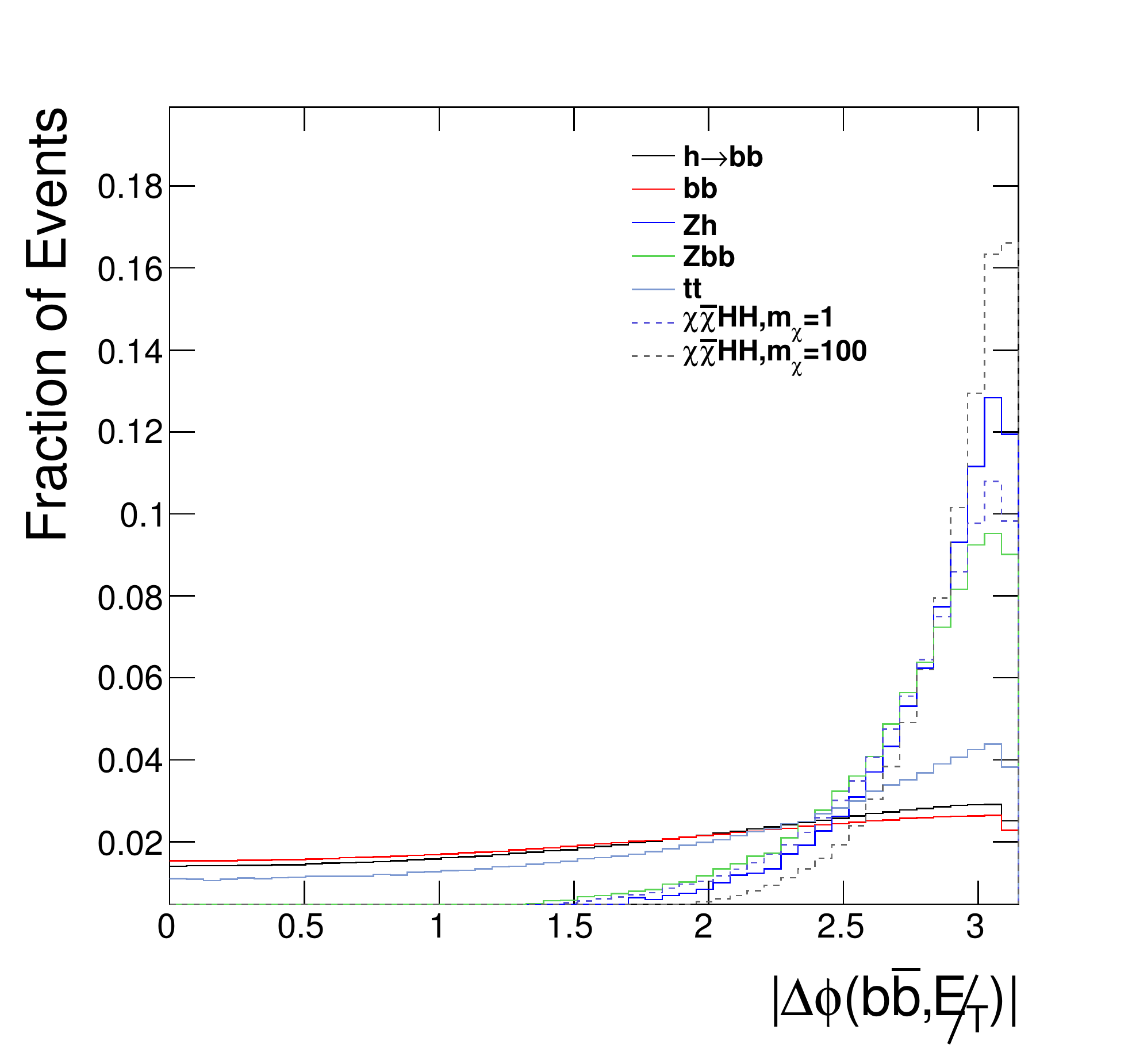}
\includegraphics[width=1.65in]{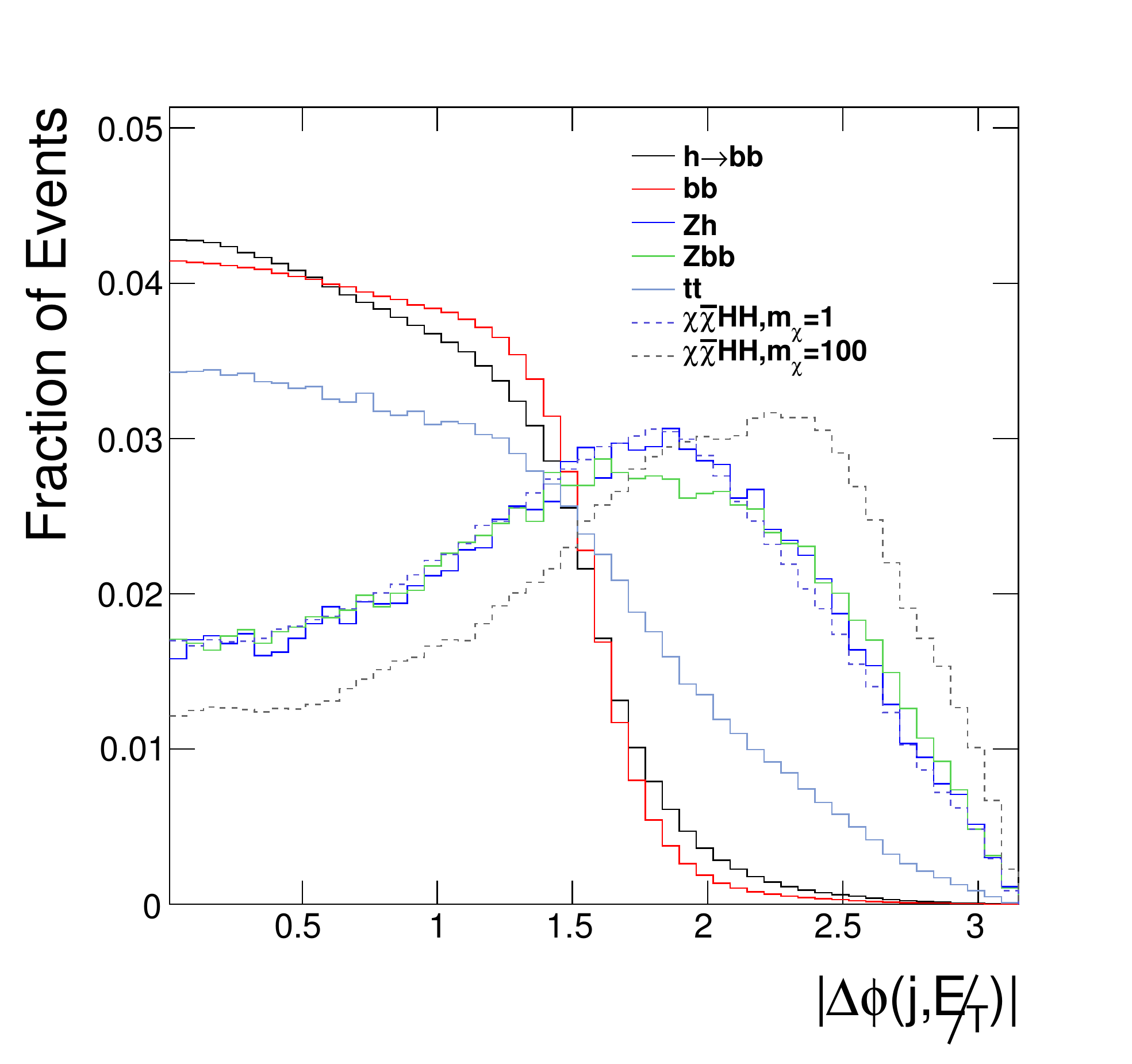}
\caption{ Distributions of $bb$ invariant mass, missing
  transverse momentum and the angle between the $\missET$ and the $b\bar{b}$ system and the nearest jet,
for simulated $h\chi\bar{\chi}$ signal samples
  with two choices of $m_\chi$, as well as the major background
  processes. All are for $pp$ collisions at $\sqrt{s}=8$ TeV.}
\label{fig:bb_kin}
\end{figure}

The event selection is:
\begin{itemize}
\item two $b$-tagged jets with $p_T>50, 20$ GeV and $|\eta|<2.5$
\item invariant mass $m_{bb} \in [50,130]$ GeV
\item no electons or muons with $p_T>20$ and $|\eta|<2.5$
\item no more than one additional jet with $p_T>20$ GeV and $|\eta|<2.5$
\item $\Delta\phi(b\bar{b},\missET)>2.5$ and $\Delta\phi(j,\missET)>1$ to
  suppress false $\missET$
\item $\missET>250$~GeV ($\sqrt{s}=8$ TeV) or $\missET>300$~GeV \mbox{($\sqrt{s}=14$ TeV)}.
\end{itemize}

Figure~\ref{fig:gg_met} shows the distribution of expected events at
$\sqrt{s}=8$ and 14 TeV as a function of missing transverse
momentum.

The production cross section and uncertaintaties for $gg\rightarrow h$, $Zh$ and $Wh$ are calculated as above, with branching fraction $\mathcal{B}(h\rightarrow b\bar{b}) = 0.57$ with a 3\% relative uncertainty~\cite{Heinemeyer:2013tqa}.  The cross section for $t\bar{t}$ production is calculated at NNLO~\cite{Czakon:2013goa}. The $Z/W+b\bar{b}$ cross sections are calculated at LO with {\sc madgraph}5 and scaled using the inclusive $Z$ and $W$ boson production cross section $k$-factors~\cite{FebresCordero:2008ci,Cordero:2009kv}. The $b\bar{b}$ cross section is calculated at leading order with {\sc madgraph}5 scaled to NLO using a $k$-factor~\cite{Banfi:2007gu}.

We select a minimum $\missET$ threshold by optimizing the
expected cross-section upper limit, finding $\missET>85$~GeV and
$\missET>250$~GeV for the $\sqrt{s}=8$ and 14 TeV cases,
respectively. Table~\ref{tab:gg} shows the expected event yields for
each of these cases.

\begin{figure}
\includegraphics[width=3.3in]{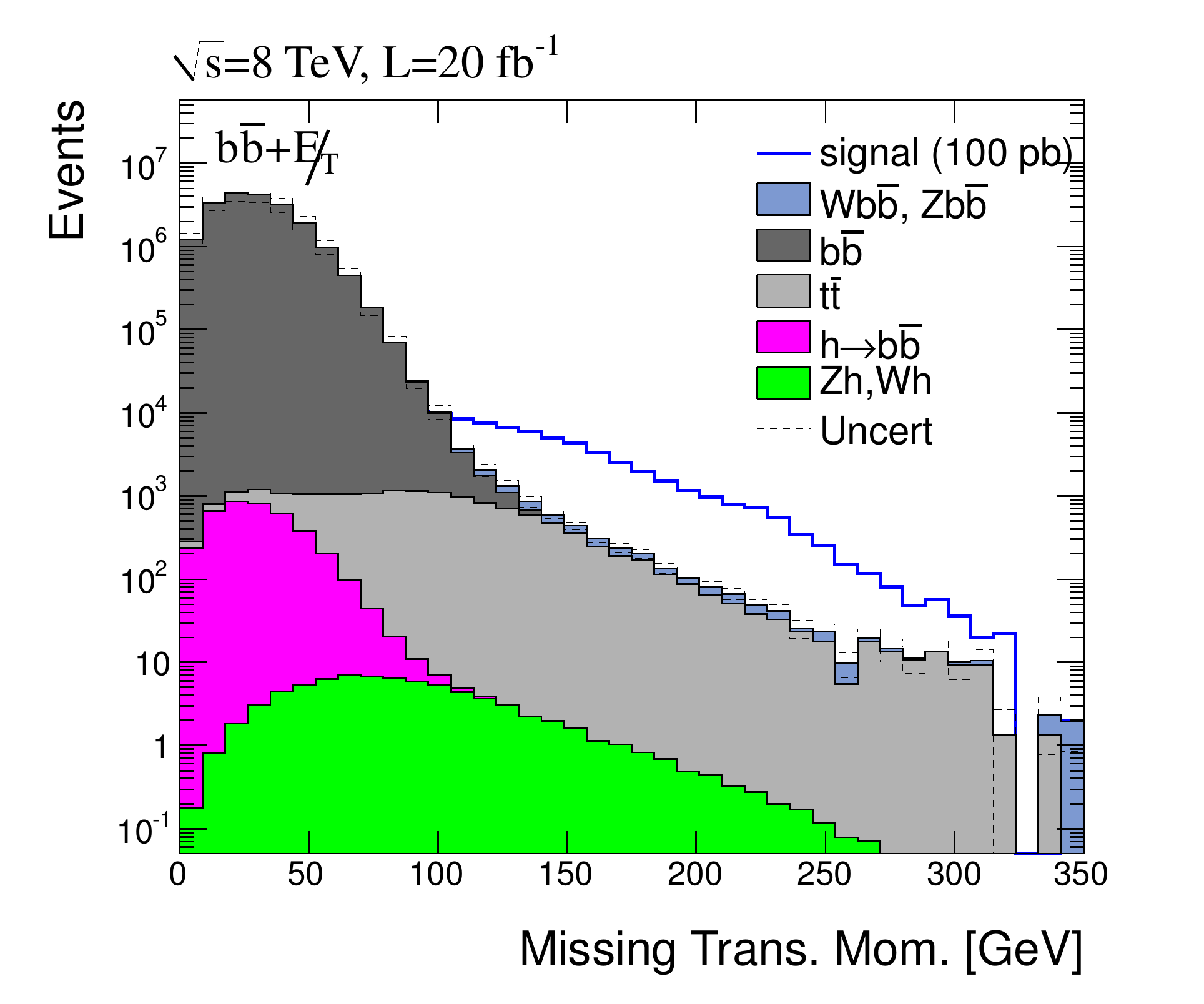}
\includegraphics[width=3.3in]{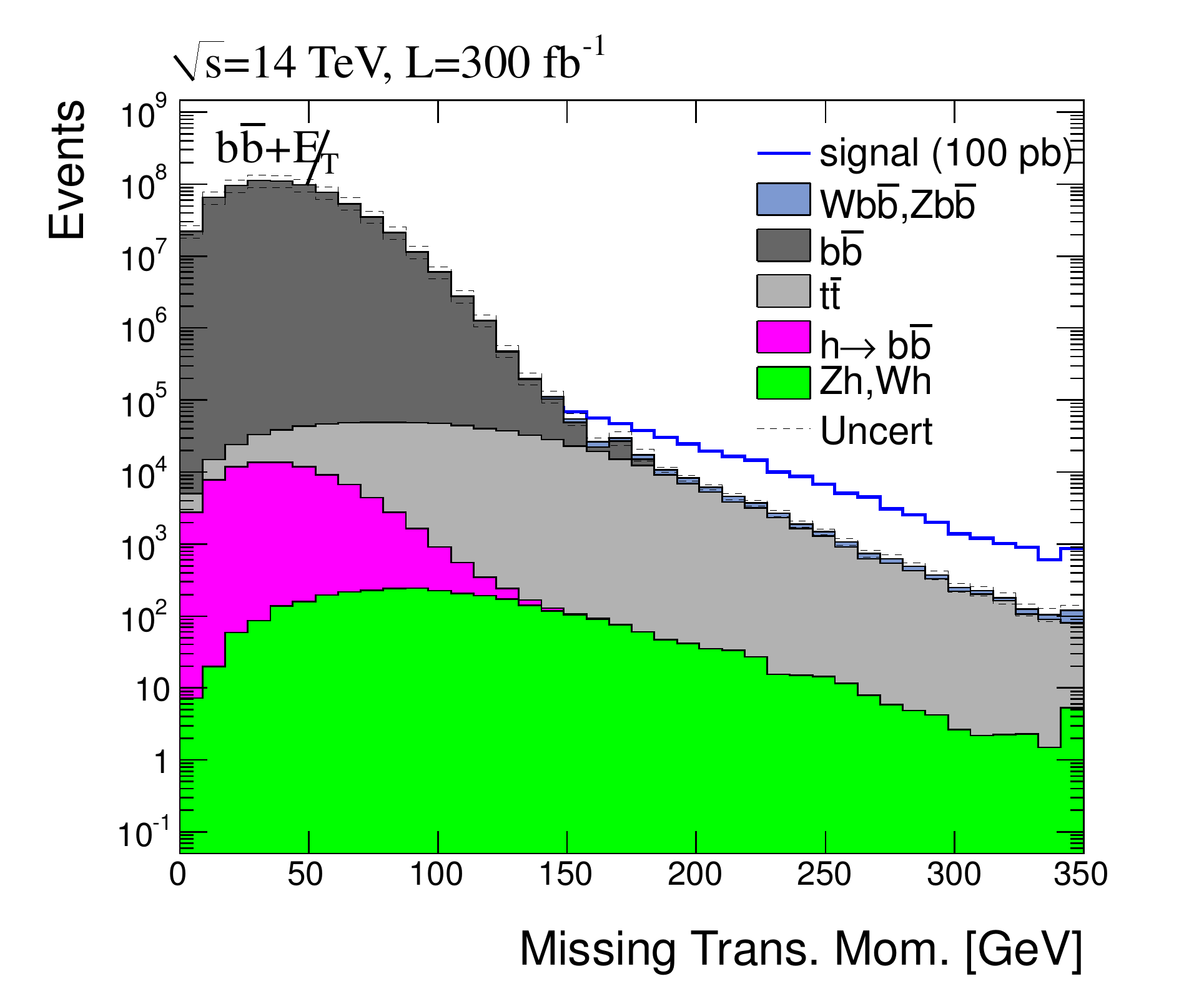}
\caption{ Distributions of  missing
  transverse momentum for simulated signal and background samples in
  the $bb+\missET$ final state with
  with all selection other than the $\missET$ threshold, normalized to expected luminosity.}
\label{fig:bb_met}
\end{figure}

\begin{table}
\caption{Expected background and signal yields in the $b\bar{b}+\missET$ channel for $pp$ collisions at
  $\sqrt{s}=8$ TeV with $\mathcal{L}=8$~fb$^{-1}$, left, or
  $\sqrt{s}=14$ TeV with $\mathcal{L}=14$~fb$^{-1}$, right. The signal
  case corresponds to $\sigma=1$~pb, and $m_\chi=1$ GeV.}
\label{tab:bb}
\begin{tabular}{lrr}
\hline\hline
& $\sqrt{s}=8$ TeV &\ \ \ \ $\sqrt{s}=14$ TeV \\
&  $\mathcal{L}=20$~fb$^{-1}$ &
$\mathcal{L}=300$~fb$^{-1}$\\
& $\missET>250$ & $\missET>300$ \\
\hline
$Zb\bar{b}+Wb\bar{b}$ & $15 \pm 3$& $ 130 \pm 15$ \\
$b\bar{b}$ & $0^{+5}_{-0}$ &  $0^{+5}_{-0}$ \\
$t\bar{t}$ & $90 \pm 10$ & $750\pm 75 $ \\
$h\rightarrow b\bar{b}$ & $0^{+5}_{-0}$ & $0^{+5}_{-0}$   \\
$Zh,Wh$ &$1 \pm 0.5$ & $15 \pm 5$\\
\hline
Total Bkg & $105 \pm 11$ & $ 900 \pm 140$ \\
$\chi\bar{\chi}HH (\sigma=10$ pb$)$  & 63 &  60\\
\hline\hline
\end{tabular}
\end{table}

\begin{figure}
\includegraphics[width=1.65in]{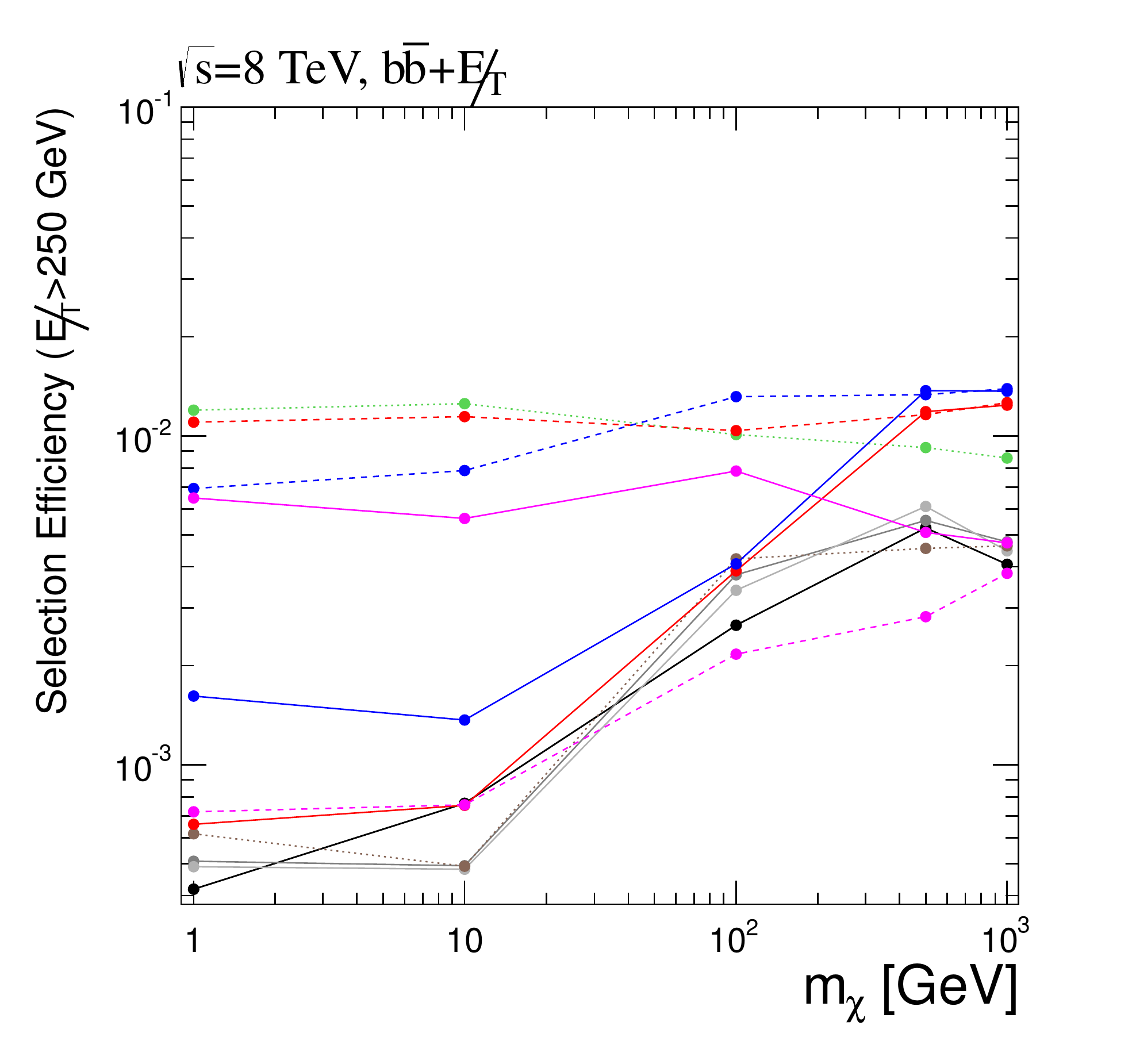}
\includegraphics[width=1.65in]{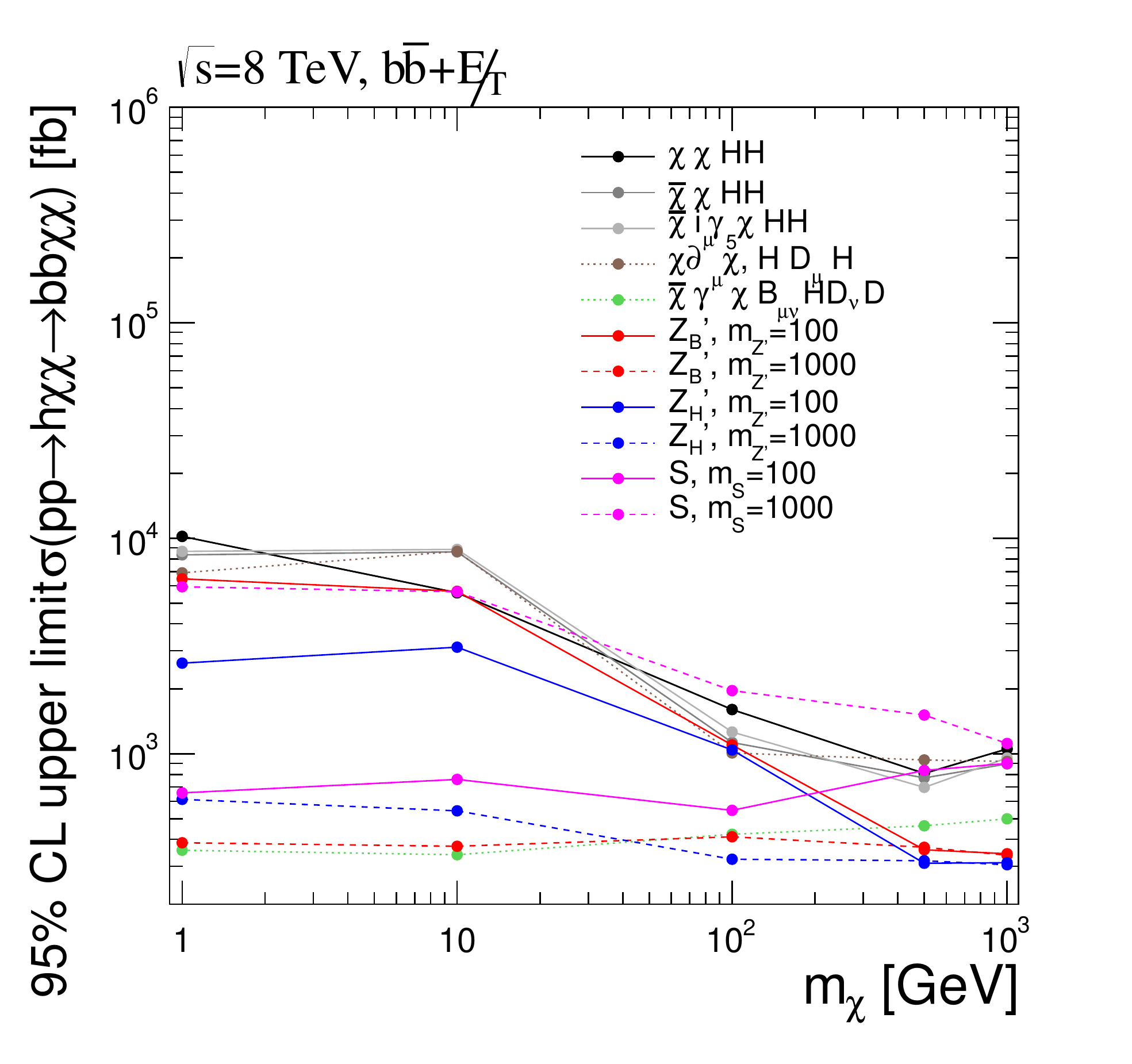}
\includegraphics[width=1.65in]{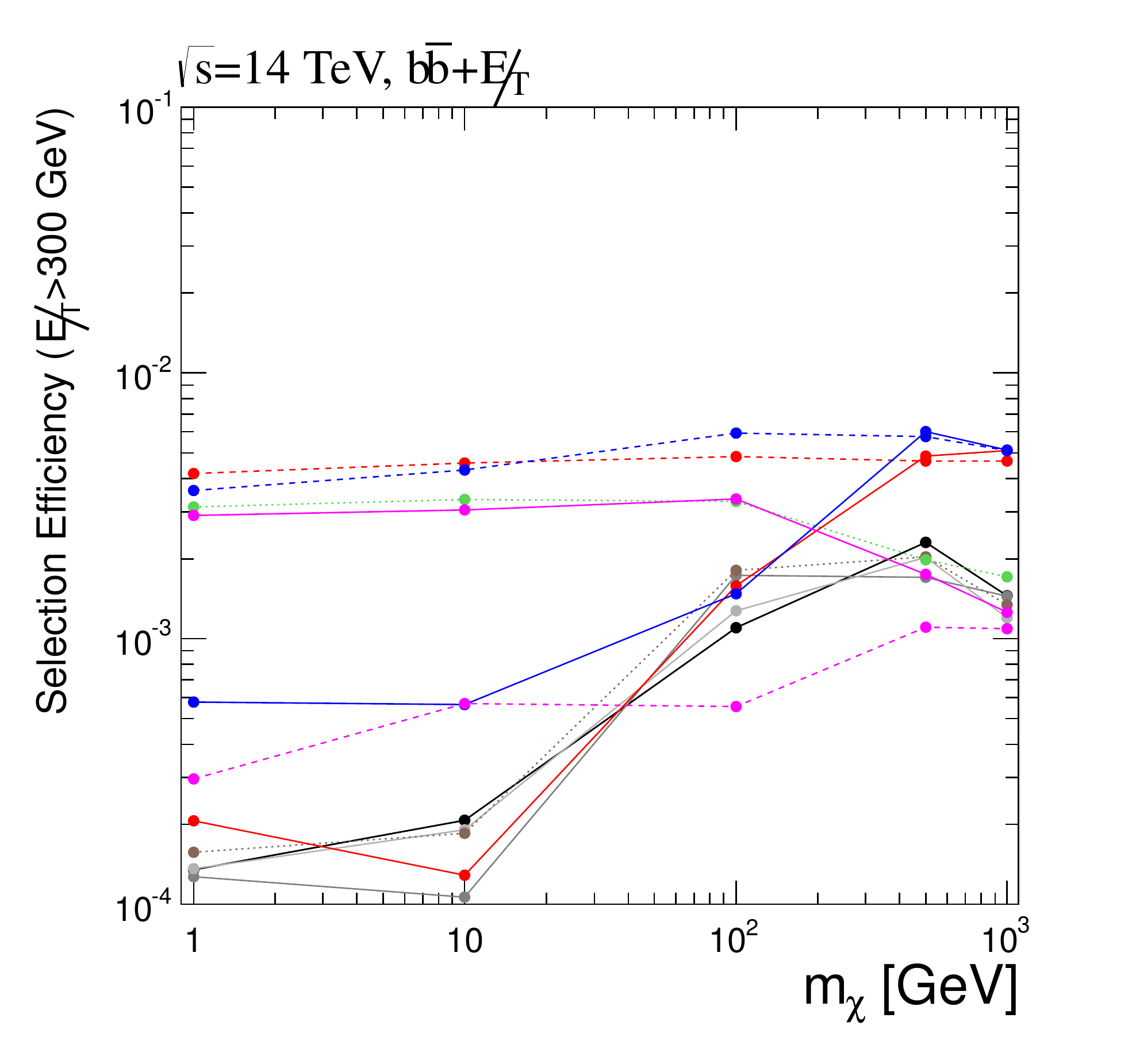}
\includegraphics[width=1.65in]{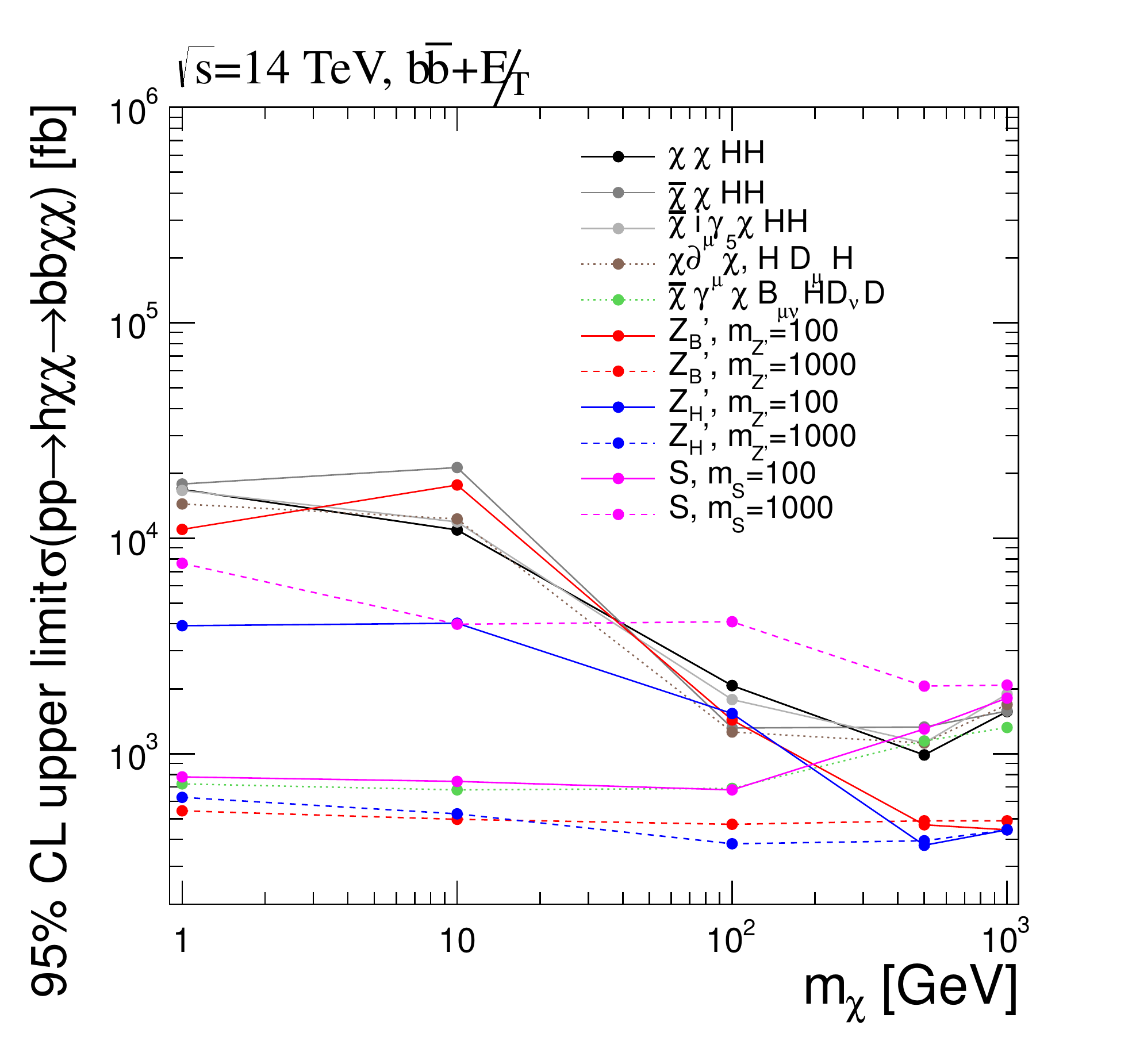}
\caption{ Selection efficiency in the $b\bar{b}+\missET$ channel (left) and upper limits on
  $\sigma(pp\rightarrow h\chi\bar{\chi}\rightarrow
  b\bar{b}\chi\bar{\chi})$ for $\sqrt{s}=8$ TeV (top) and
  $\sqrt{s}=14$ TeV (bottom).}
\label{fig:bb_lim}
\end{figure}

Selection efficiency and upper limits on
 $\sigma(pp\rightarrow h\chi\bar{\chi}\rightarrow b\bar{b}\chi\bar{\chi})$ are shown in Fig.~\ref{fig:bb_lim}. Note that the small signal efficiency is largely due to the need for a high minimum threshold on $\missET$ to supress the backgrounds. Similar missing energy thresholds and efficiencies are seen in mono-jet analyses.

\subsection{Two-lepton and two-jet decays}

The branching fraction of $ZZ^*$ to four leptons is quite small due to the small charged-lepton decay fraction relative to hadronic decay modes. To balance that, we consider the $h\rightarrow ZZ^*\rightarrow \ell\ell jj$ mode.

The backgrounds to the $\ell \ell j j + \missET$ final state include:

\begin{itemize}

\item $Zh$ production with $Z\rightarrow\nu\bar{\nu}$ and $h \to \ell \ell j j$,
  an irreducible background;

\item Additional decay modes of $Zh$ and $Wh$ production, all with final state
  $j j \ell \ell \nu \nu$;

\item Higgs boson production with $h \to ZZ^* \to \ell \ell jj$;

\item Diboson production:  $ZZ \to \ell \ell jj$ and $ZW \to \ell \ell jj$ ;

\item Production of $WW$ plus additional jets, with $WW \to \ell \nu \ell \nu$;

\item $Z$ boson plus jets production, with $Z \to \ell \ell$;

\item $W$ boson plus jets production, with $W \to \ell \nu$ and one jet misreconstructed as an isolated lepton;

\item $t \bar{t}$ production with $t \to \ell^+ \nu b$ and $\bar{t} \to \ell^- \bar{\nu} \bar{b}$.

\end{itemize}

\begin{figure}
\includegraphics[width=1.65in]{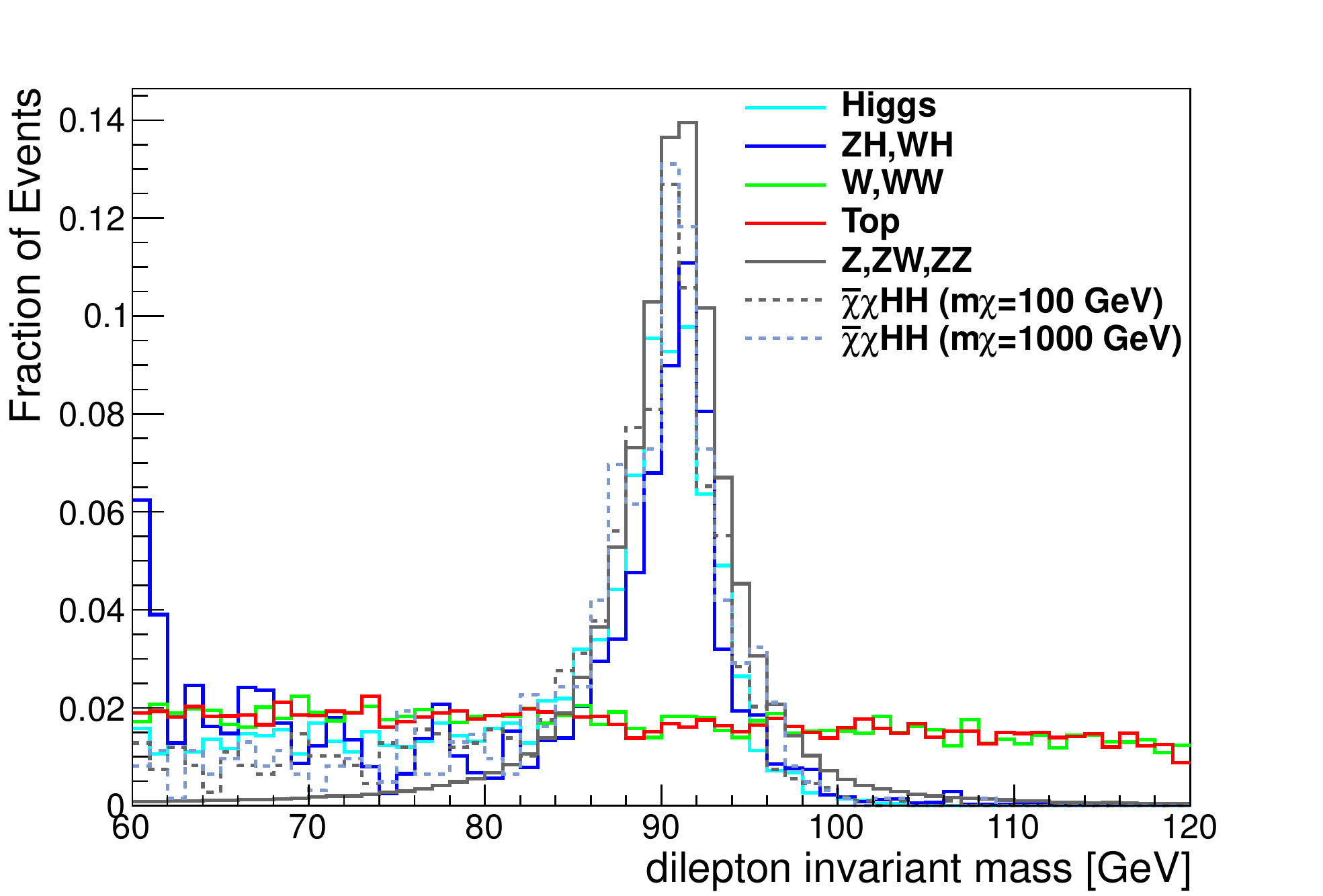}
\includegraphics[width=1.65in]{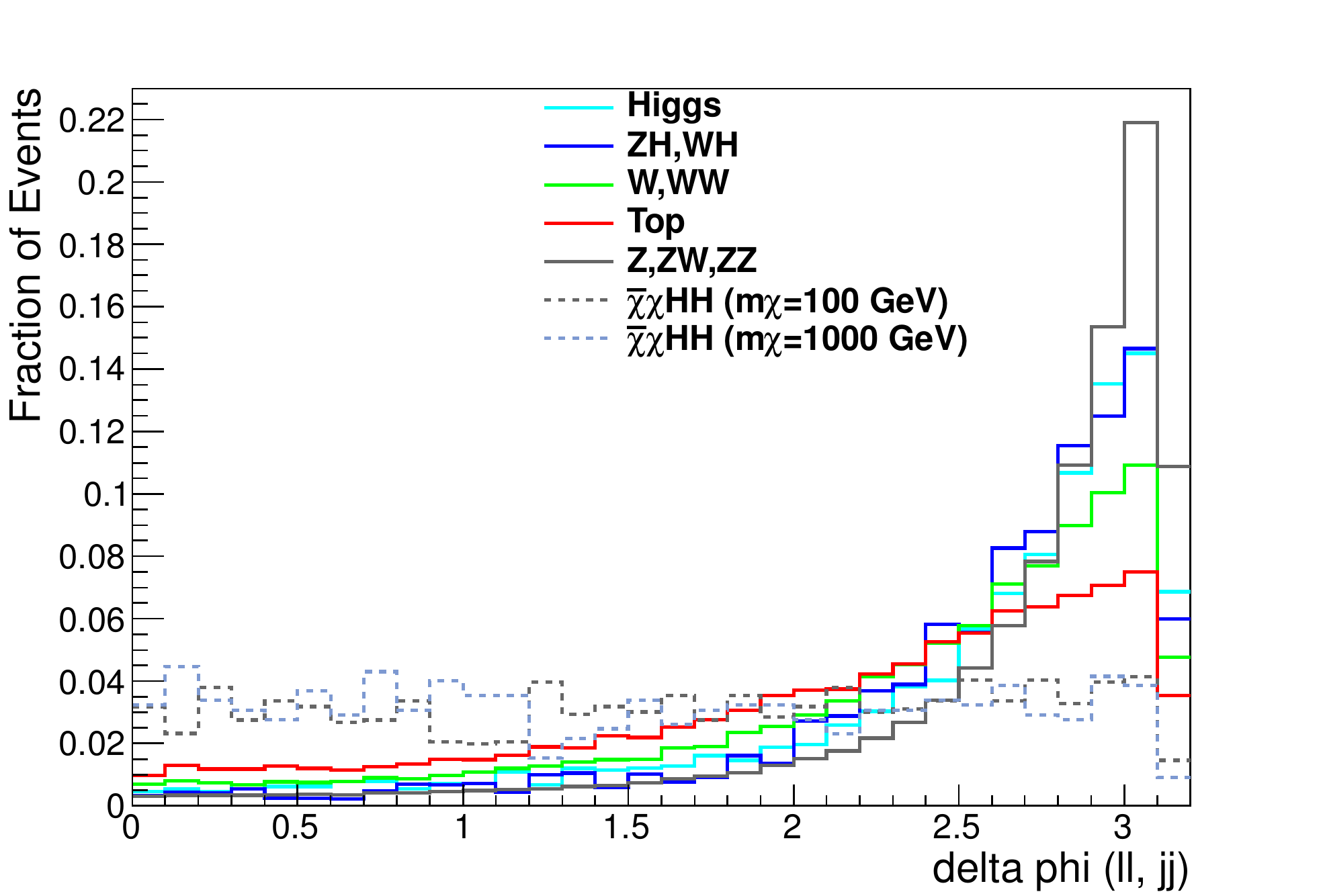}
\includegraphics[width=1.65in]{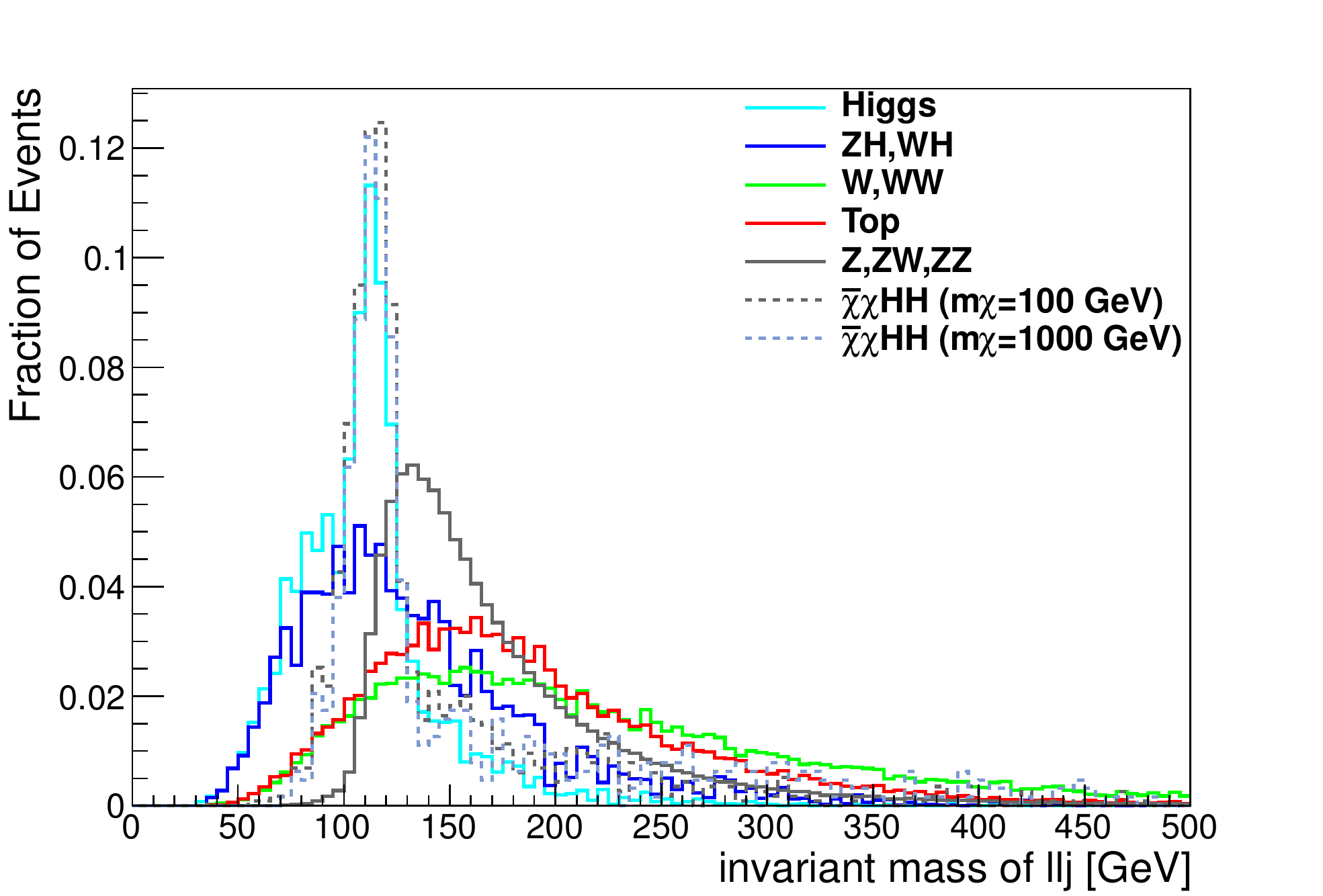}
\includegraphics[width=1.65in]{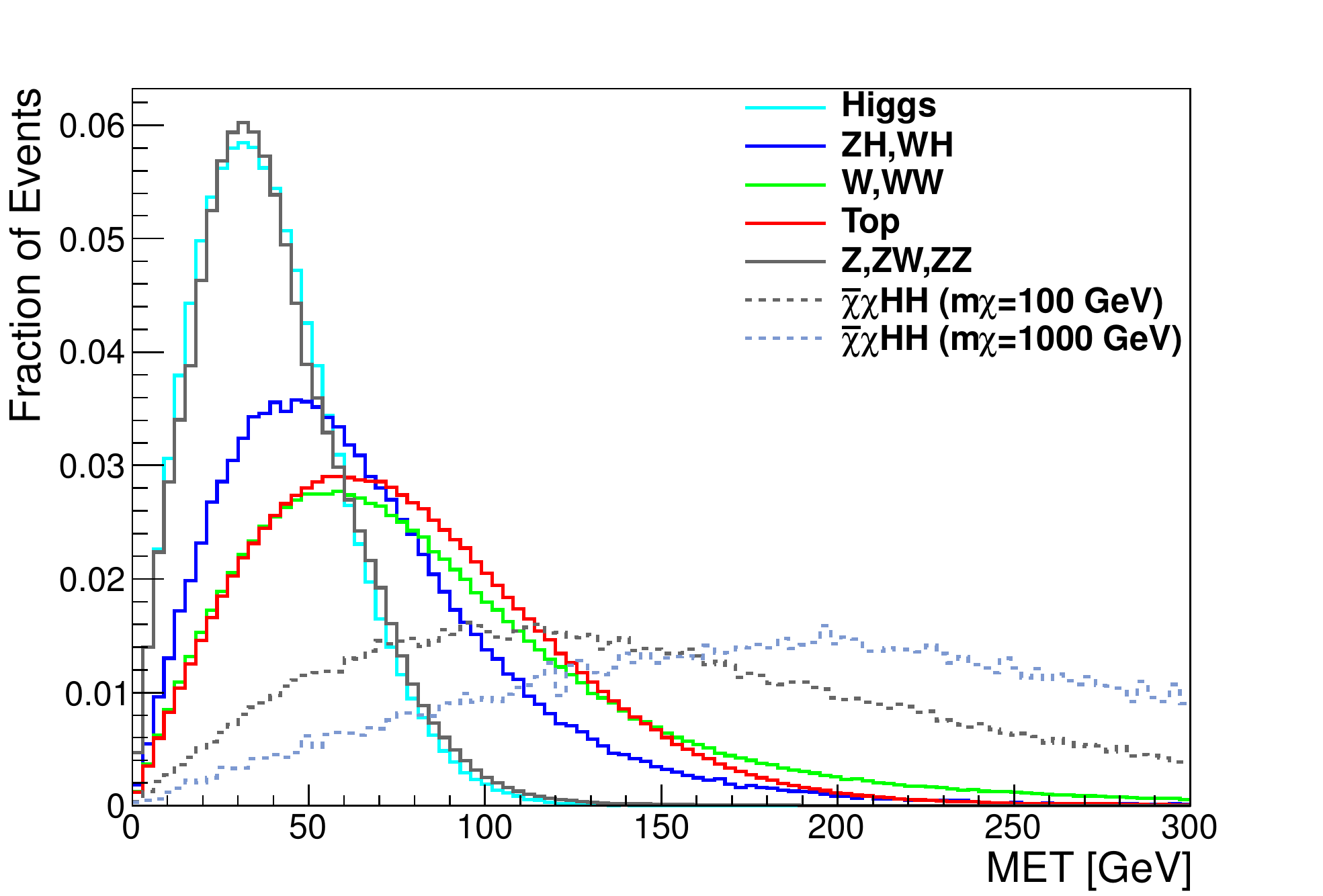}
\caption{ Distributions of dilepton invariant mass, $\Delta \phi$ between dilepton and the
dijet system formed by the two highest-$p_T$ jets, invariant mass of the two
leptons plus the jet nearest the direction of the dilepton system, and missing
transverse momentum, for simulated $h\chi\bar{\chi}$ signal samples with two
choices of $m_\chi$, as well as the major background processes. All are for $pp$
collisions at $\sqrt{s}=8$ TeV.}
\label{fig:2l2j_kin}
\end{figure}

\begin{figure}
\includegraphics[width=3.3in]{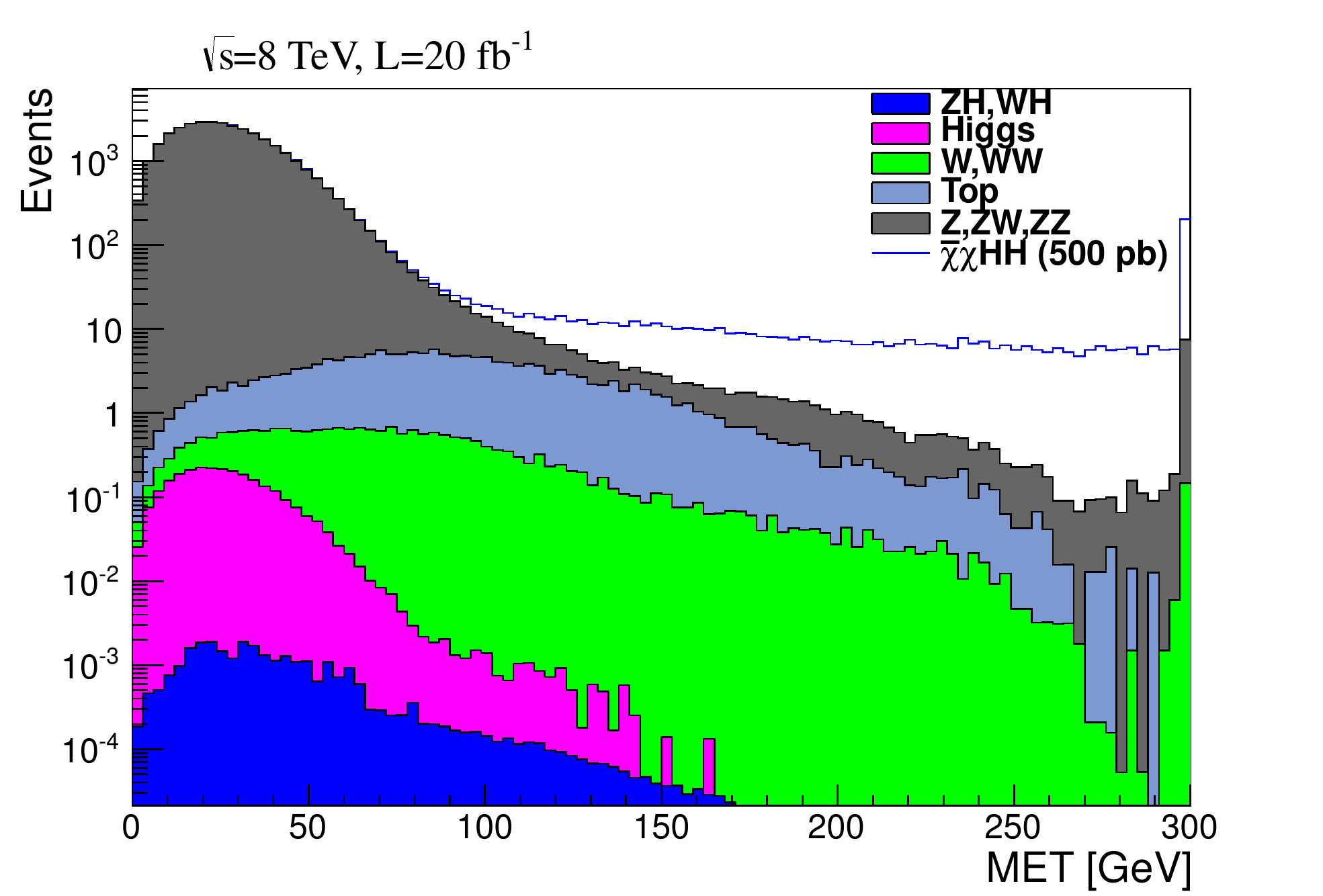}
\includegraphics[width=3.3in]{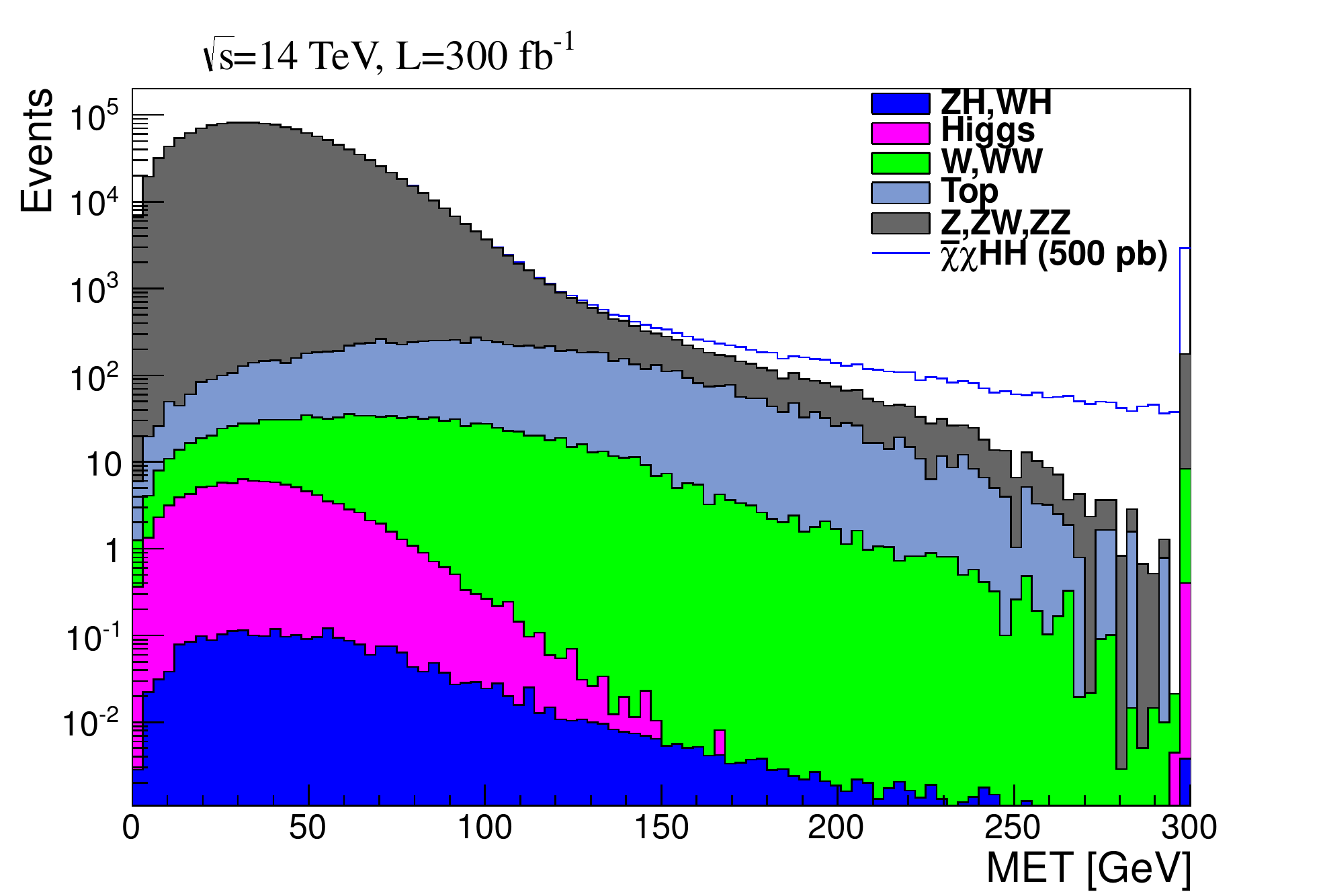}
\caption{ Distributions of  missing
  transverse momentum for simulated signal and background samples in
  the $lljj+\missET$ final state with
  with all selection other than the $\missET$ threshold, normalized to expected luminosity.}
\label{fig:lljj_met}
\end{figure}

The event selection is
\begin{itemize}
\item Two opposite-sign leptons of the same flavor with leading lepton $p_T > 20$ GeV, second leading lepton
$p_T > 15$, and $|\eta|<2.5$.
\item No additional leptons with $p_T > 10$ GeV.
\item Two or more jets with $p_T > 15$ and $|\eta|<2.5$.
\item Dilepton invariant mass between $82$ and $98$ GeV.
\item $\Delta \phi$ between the dilepton system and the dijet system (formed by the two
  highest-$p_T$ jets) less than 2.25 radians.
\item Invariant mass of both leptons plus the jet in the direction with smallest
  $\Delta R$ from the dilepton system less than 124~GeV.
\end{itemize}

Figure~\ref{fig:2l2j_kin} shows distributions of kinematic variables used in the
event selection, and missing transverse momentum.  Figure~\ref{fig:lljj_met} shows
distributions of missing transverse momentum, after event selection, for both
$\sqrt{s}=8$ and $14~$TeV.

To increase the number of simulated events used to model the $W$-boson+jets background, where one jet is misreconstructed as a lepton, the $W$+jet events were weighted by a fake rate for a randomly choosen jet, to match the event yield determined from {\sc delphes}.  As this represents a small contribution to the final
selection, a large uncertainty here has only a small effect on the calculated limits.

\begin{figure}
\includegraphics[width=1.65in]{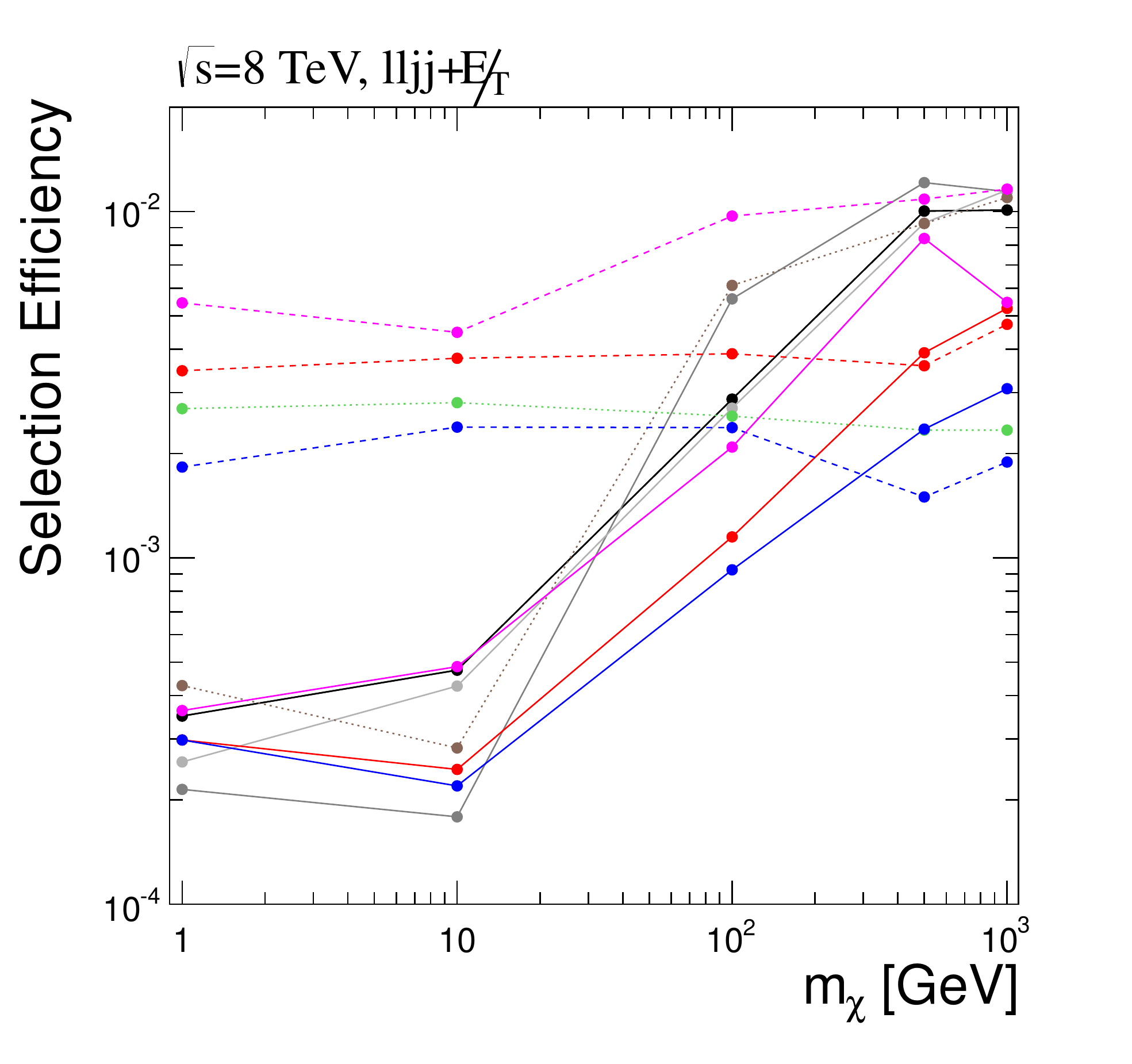}
\includegraphics[width=1.65in]{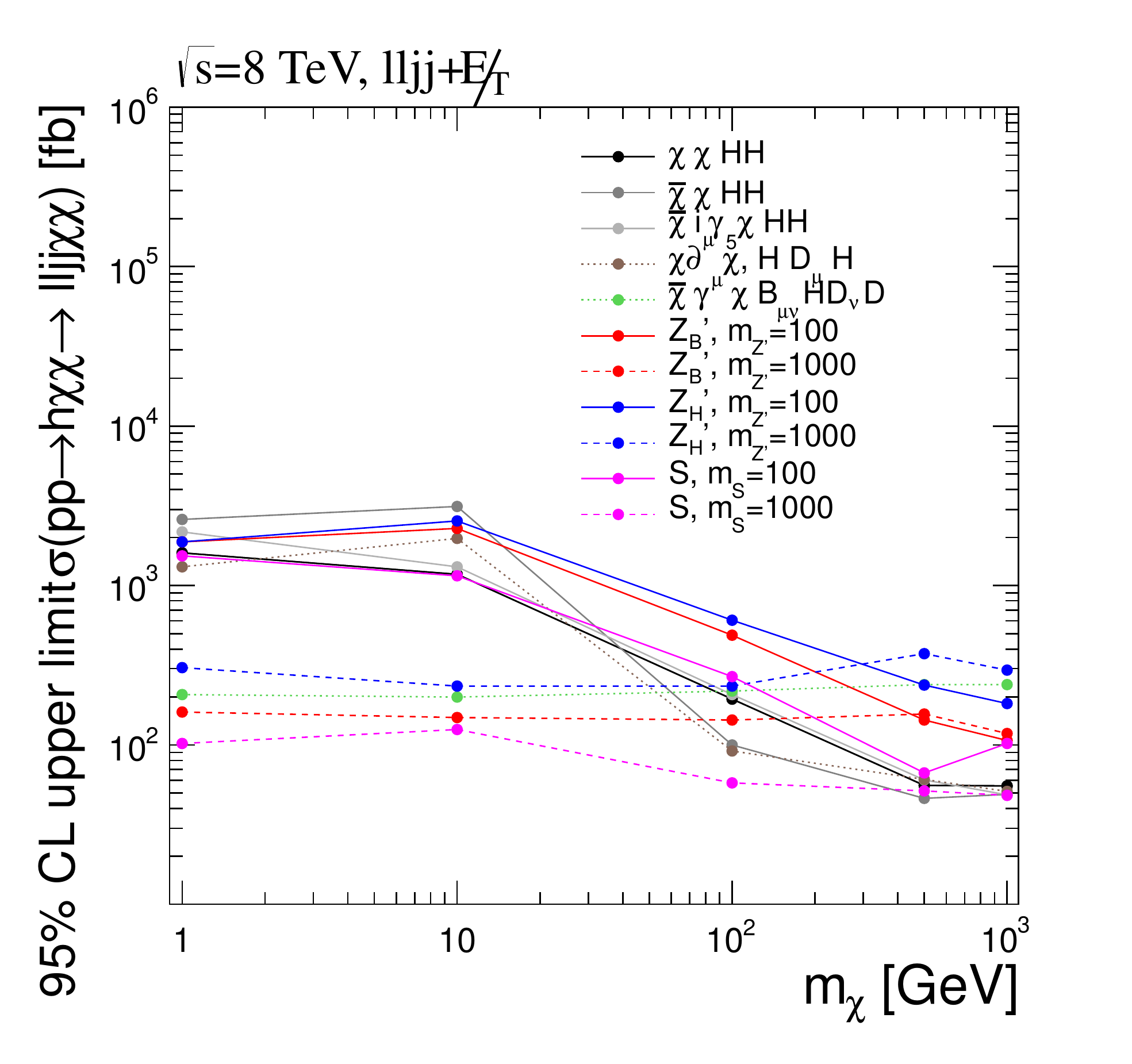}
\includegraphics[width=1.65in]{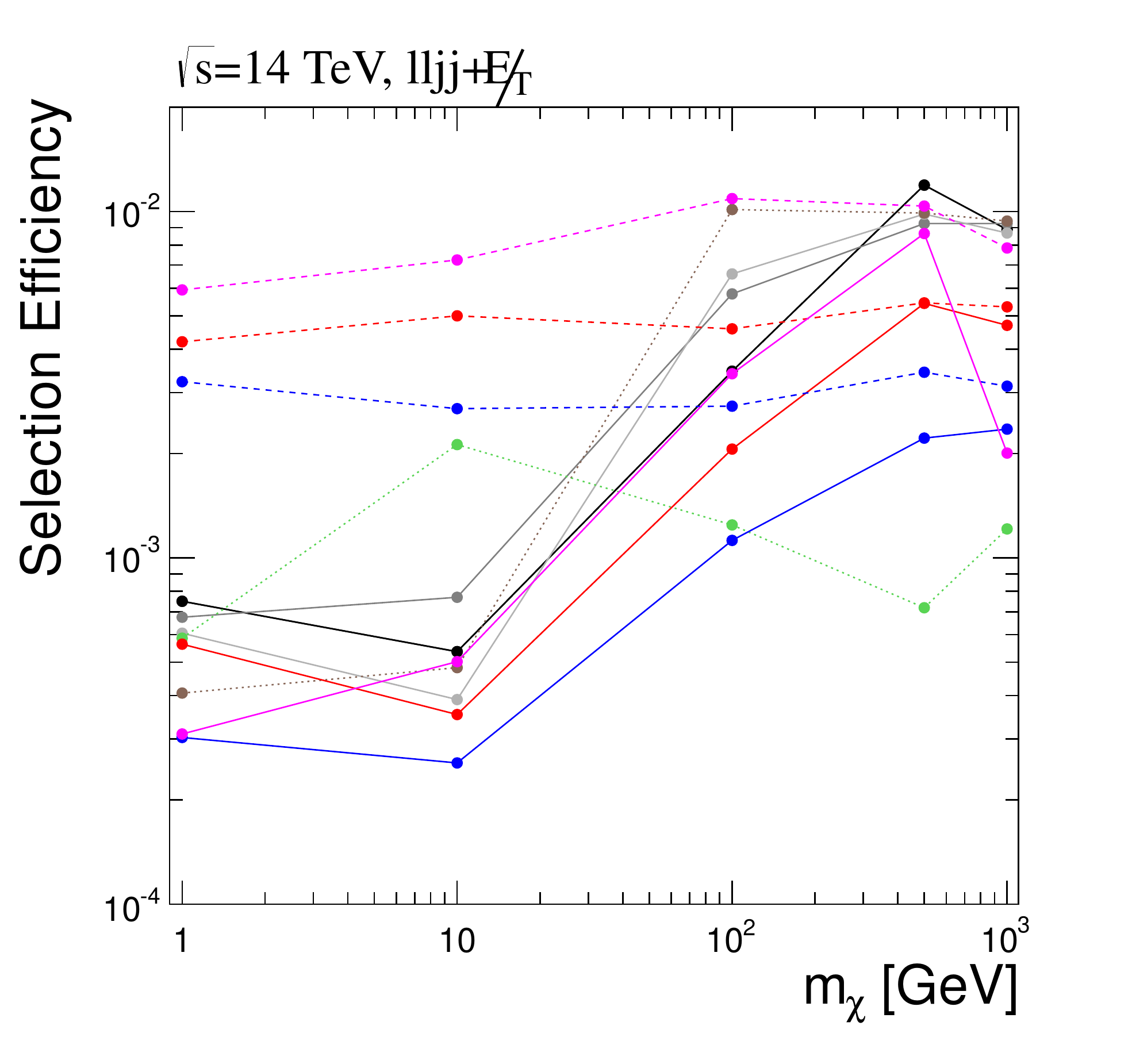}
\includegraphics[width=1.65in]{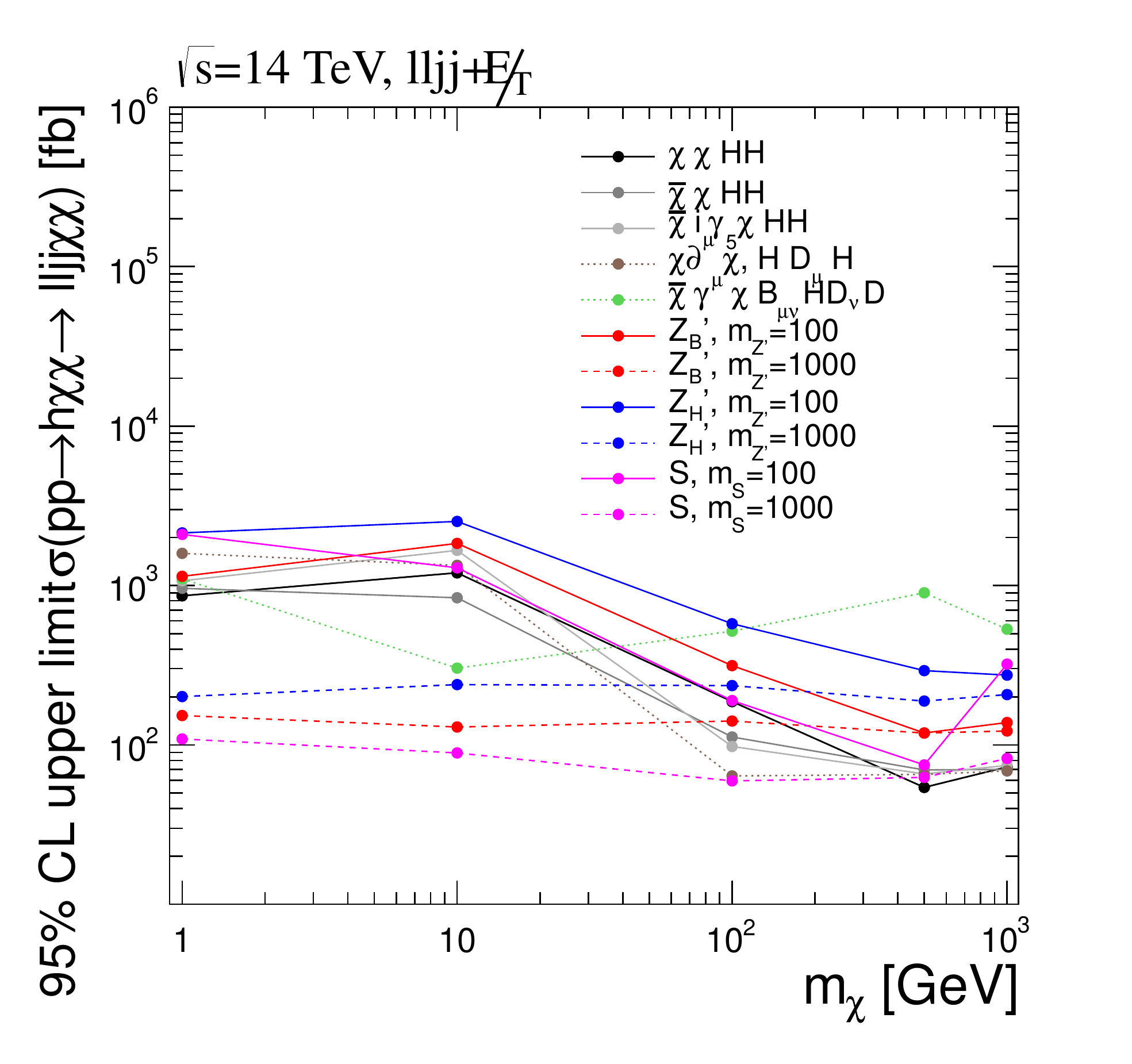}
\caption{ Selection efficiency in the $\ell\ell jj+\missET$ channel (left) and upper limits on
  $\sigma(pp\rightarrow h\chi\bar{\chi}\rightarrow
  \ell\ell jj \chi\bar{\chi})$ for $\sqrt{s}=8$ TeV (top) and
  $\sqrt{s}=14$ TeV (bottom). }
\label{fig:2l2j_lim}
\end{figure}

\begin{table}
\caption{Expected background and signal yields in the $\ell\ell jj+\missET$ channel for $pp$ collisions at
  $\sqrt{s}=8$ TeV with $\mathcal{L}=20$~fb$^{-1}$, left, or
  $\sqrt{s}=14$ TeV with $\mathcal{L}=300$~fb$^{-1}$, right. The signal
  case corresponds to $\sigma=500$~pb, and $m_\chi=500$ GeV.}
\label{tab:2l2j}
\begin{tabular}{lrr}
\hline\hline
& $\sqrt{s}=8$ TeV &\ \ \ \ $\sqrt{s}=14$ TeV \\
&  $\mathcal{L}=20$~fb$^{-1}$ &
$\mathcal{L}=300$~fb$^{-1}$\\
& $\missET>250$ & $\missET>250$ \\
\hline
$Z$,$ZW$,$ZZ$      &  $9.2   \pm 0.2$  & $211   \pm 6$     \\
Higgs              &  $0.17  \pm 0.01$ & $0.39  \pm 0.04$  \\
$WW$,$W$+jets      &  $0.26  \pm 0.06$ & $9.5   \pm 0.9$   \\
$t\bar{t}$         &  $0.26  \pm 0.5$  & $21    \pm 4$     \\
$WH$,$ZH$          &  $-$              & $0.013 \pm 0.001$ \\ 
\hline                                   
Total Bkg          & $9.7 \pm 1.0$            & $242 \pm 8$ \\          
$\chi\bar{\chi}HH$ & $56 \pm 0.5$      & $684 \pm 7$ \\       
\hline\hline
\end{tabular}
\end{table}

%INFO:  settings for sqrt(s) = 14 TeV
%Cutflow:  Stage 4 (after MET cut)
%   Z,ZZ,ZW    events:  211.063 +/-5.91559
%     Higgs    events:  0.3996 +/-0.03996
%      W,WW    events:  9.55403 +/-0.856121
%       Top    events:  20.9361 +/-4.02915
%        VH    events:  0.0130112 +/-0.00075577
% signal064    events:  684.85 +/-7.1525
%Total Background:  241.966
%
%INFO:  settings for sqrt(s) = 8 TeV
%   Z,ZZ,ZW    events:  9.23256 +/-0.195117
%      W,WW    events:  0.177048 +/-0.0159099
%       Top    events:  0.264516 +/-0.0577221
%        VH    events:  5.83656e-05 +/-2.41106e-06
% signal064    events:  56.4035 +/-0.52999
%Total Background:  9.67418

We select a minimum $\missET$ threshold by optimizing the
expected cross-section upper limit, finding
$\missET>250$~GeV for both the $\sqrt{s}=8$ and 14 TeV cases. Table~\ref{tab:2l2j} shows the expected event yields for
each of these cases.

Selection efficiency and upper limits on
 $\sigma(pp\rightarrow h\chi\bar{\chi}\rightarrow \ell\ell jj\chi\bar{\chi})$ are shown in Fig.~\ref{fig:2l2j_lim}.

\subsection{Comparison}

A comparison of sensitivities between final states is shown in
Figs.~\ref{fig:all_lim8} and ~\ref{fig:all_lim14}.  The di-photon final state has the strongest sensitivity across all models and masses.  The two-$b$-quark final state also has significant power, which may be improved by more aggressive rejection of the $t\bar{t}$ background and use of jet-substructure techniques to capture events with large Higgs boson $p_T$.

Note that these comparisons assume the SM Higgs boson branching fractions, which may be diluted in cases where $\mathcal{B}(h\rightarrow\chi\chi)$ is large, but the relative BFs will be unaltered, allowing a comparison of the relative power of each channel.

The systematic uncertainty on the background estimate typically controls the sensitivity of each channel. In this study, we have used simulated samples to describe the background contributions. In future experimental analyses, many of these backgrounds can be estimated by extrapolating from signal-depleted control regions, which significantly reduces the systematic error due to modeling of the $\missET$ tail.  For example, in the $b\bar{b}+\missET$ final state, one can measure the rate of $Wh\rightarrow \ell\nu b\bar{b}$ and $Zh\rightarrow \ell\ell b\bar{b}$ in final states with one or two leptons, respectively.

\begin{figure}
\includegraphics[width=1.65in]{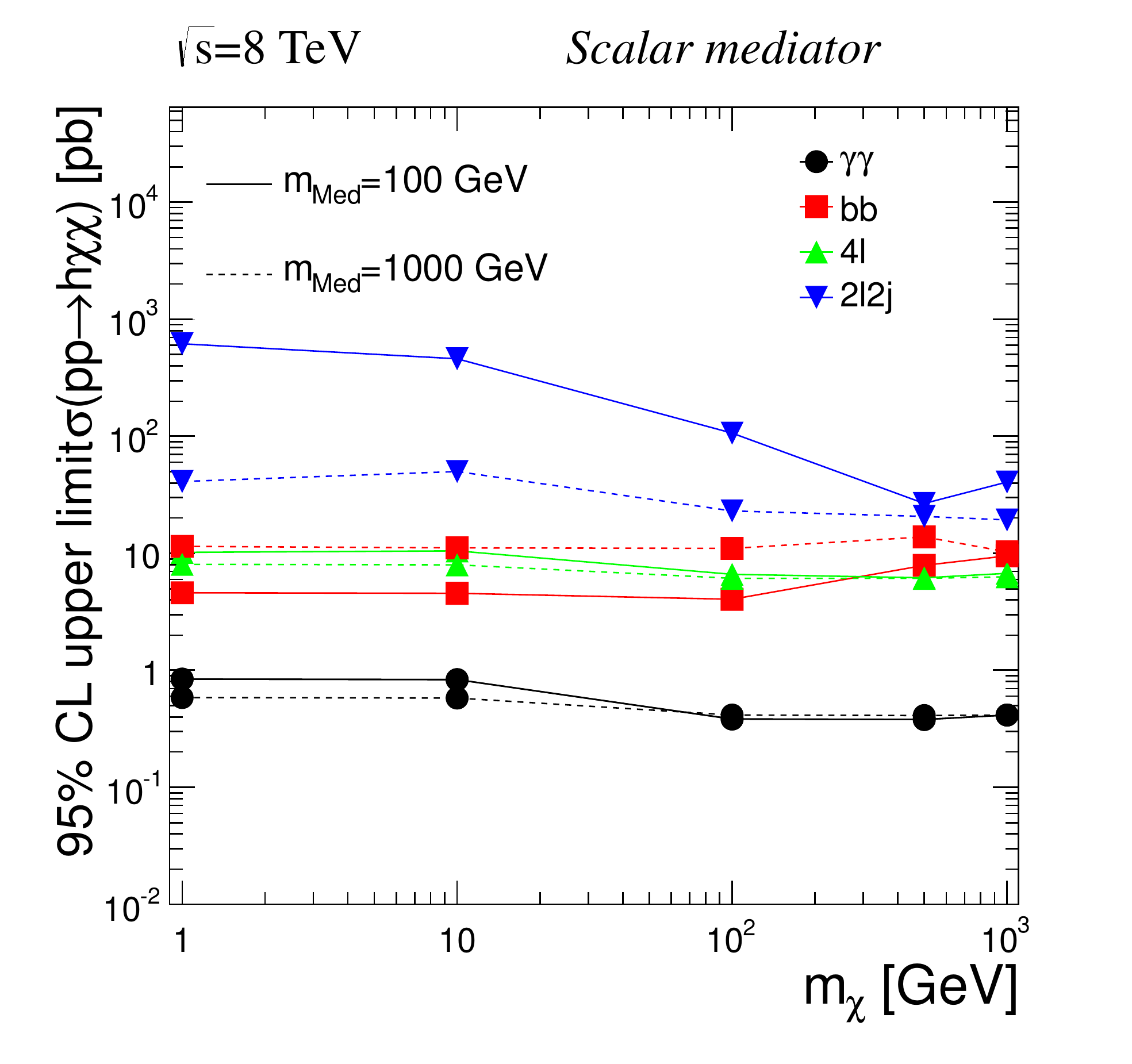}
\includegraphics[width=1.65in]{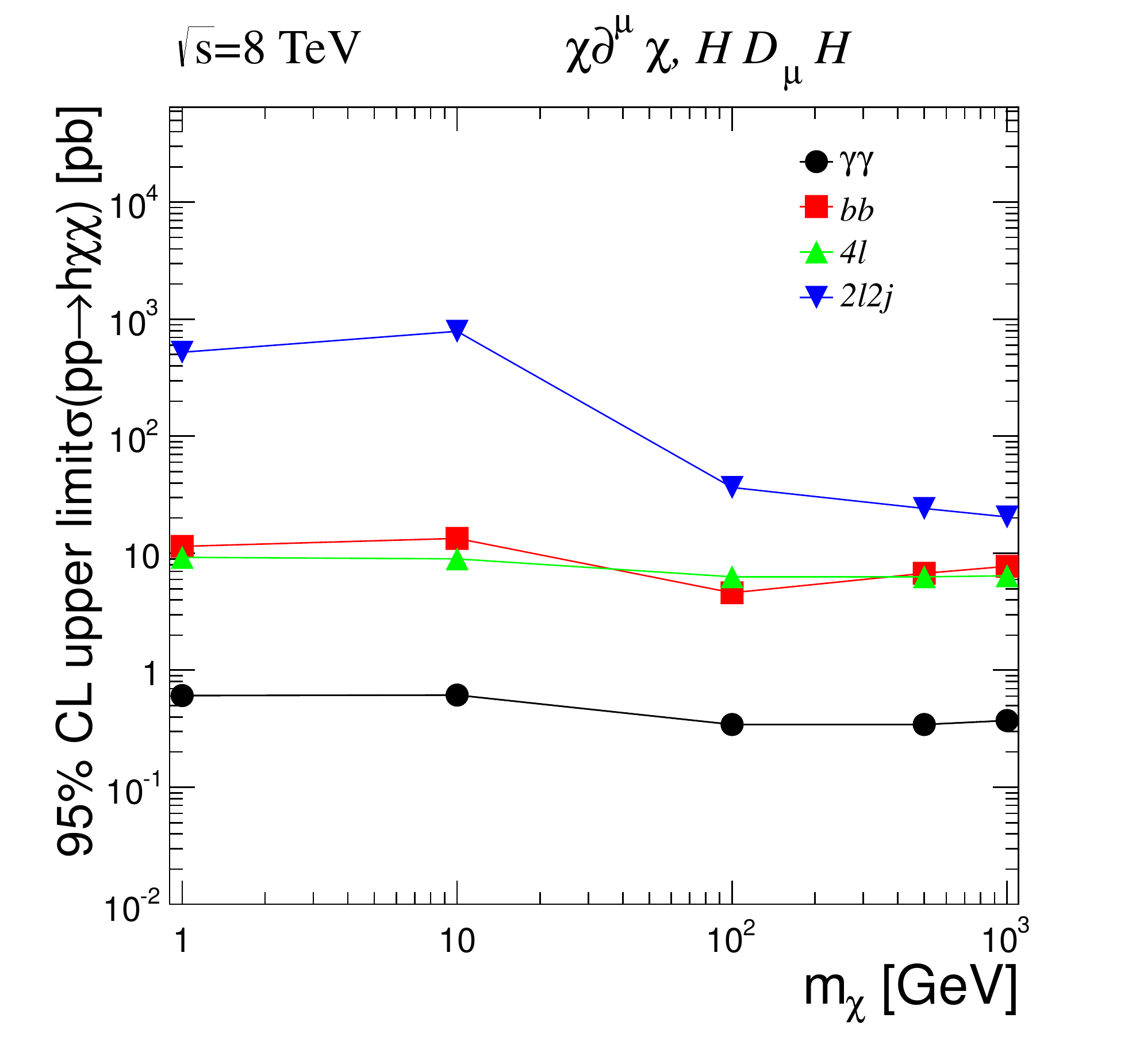}
\includegraphics[width=1.65in]{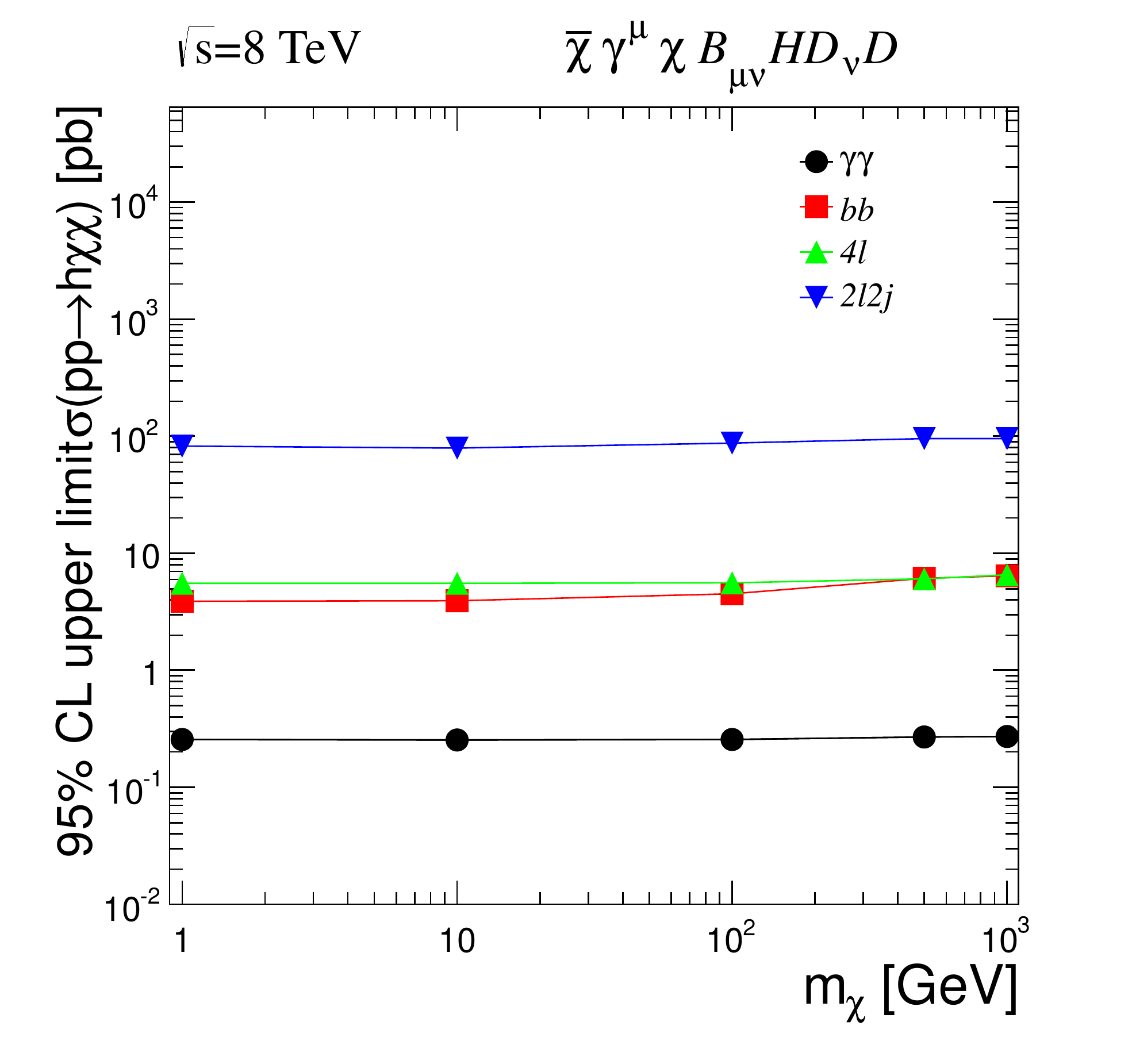}
\includegraphics[width=1.65in]{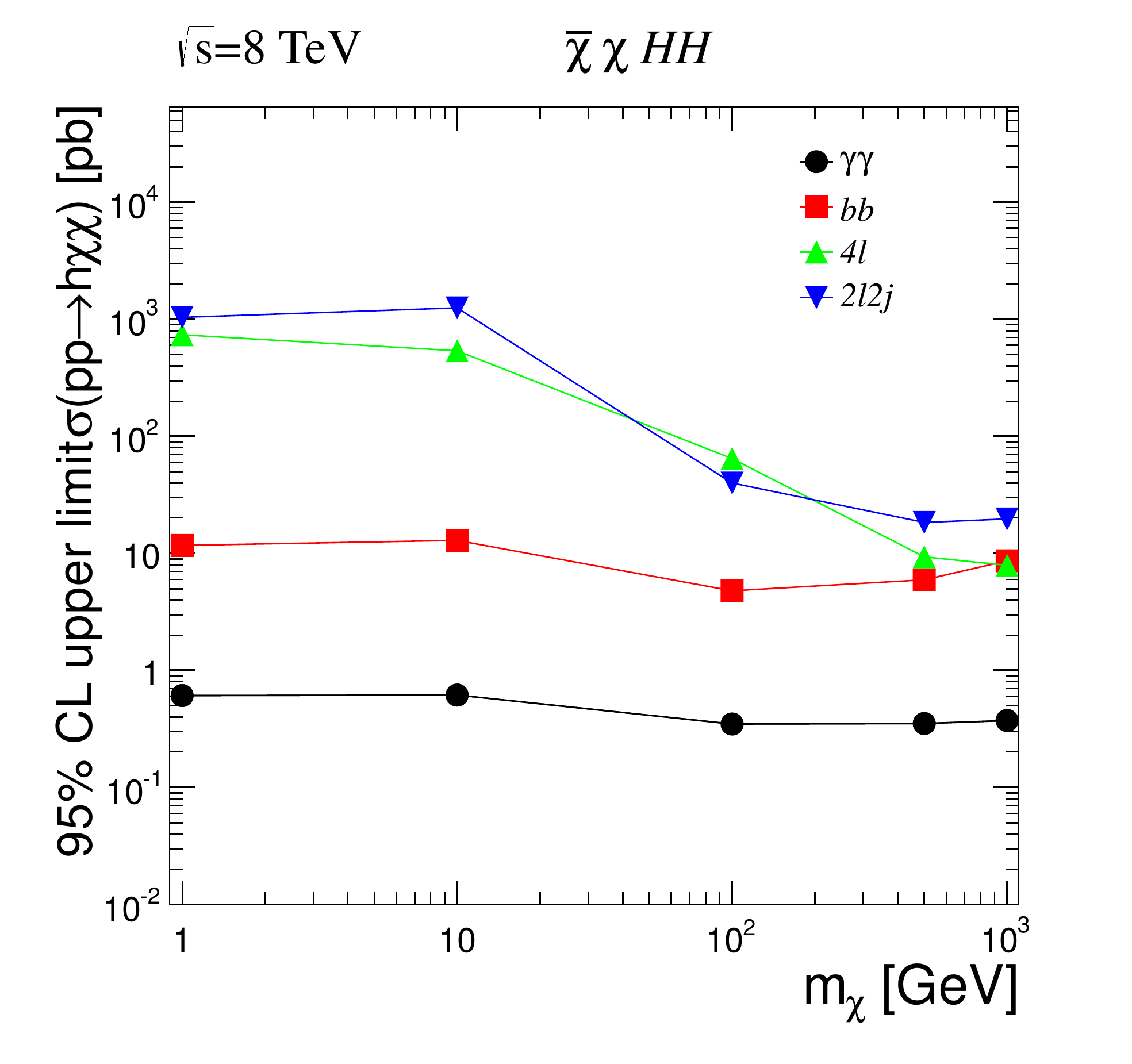}
\includegraphics[width=1.65in]{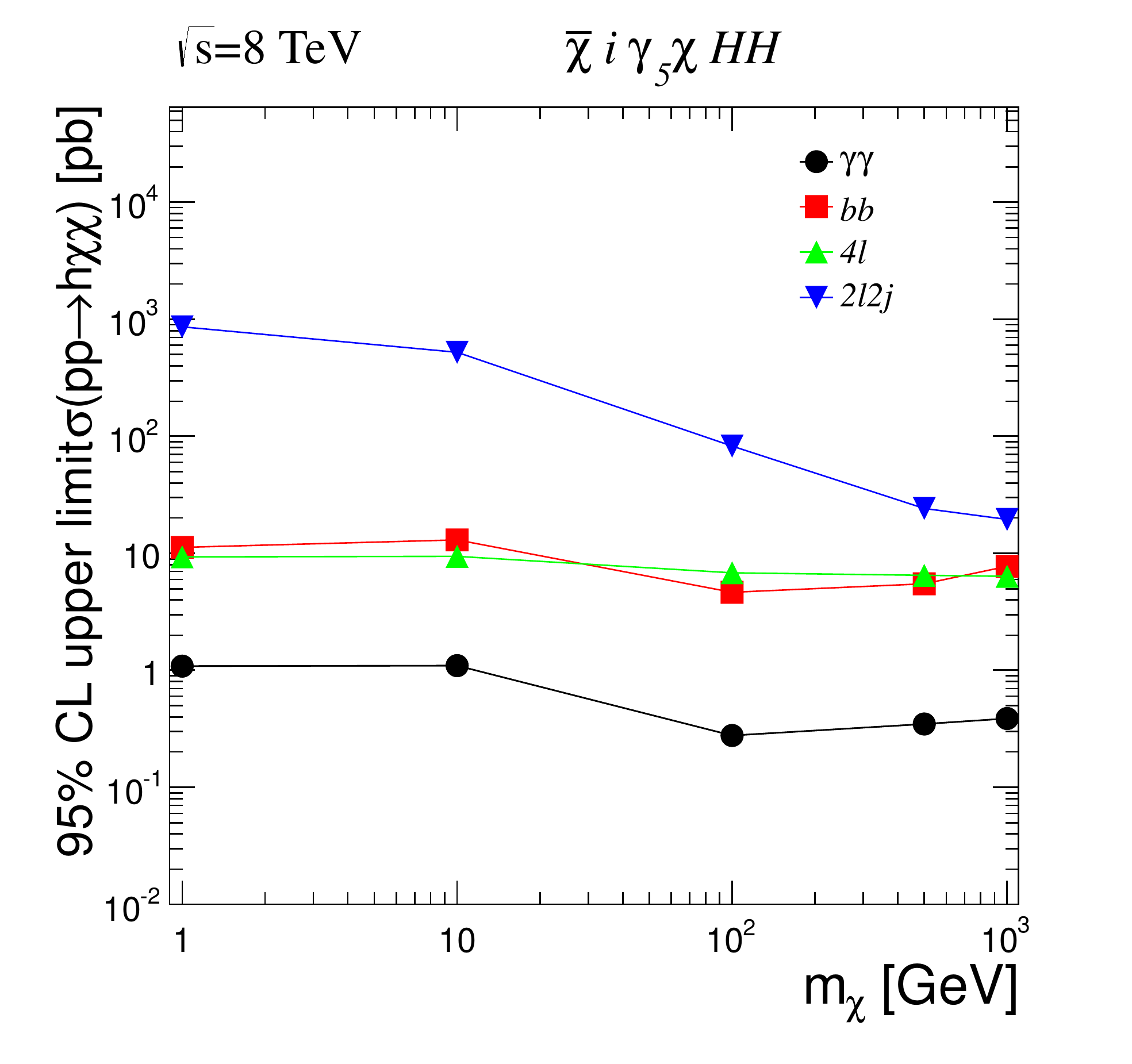}
\includegraphics[width=1.65in]{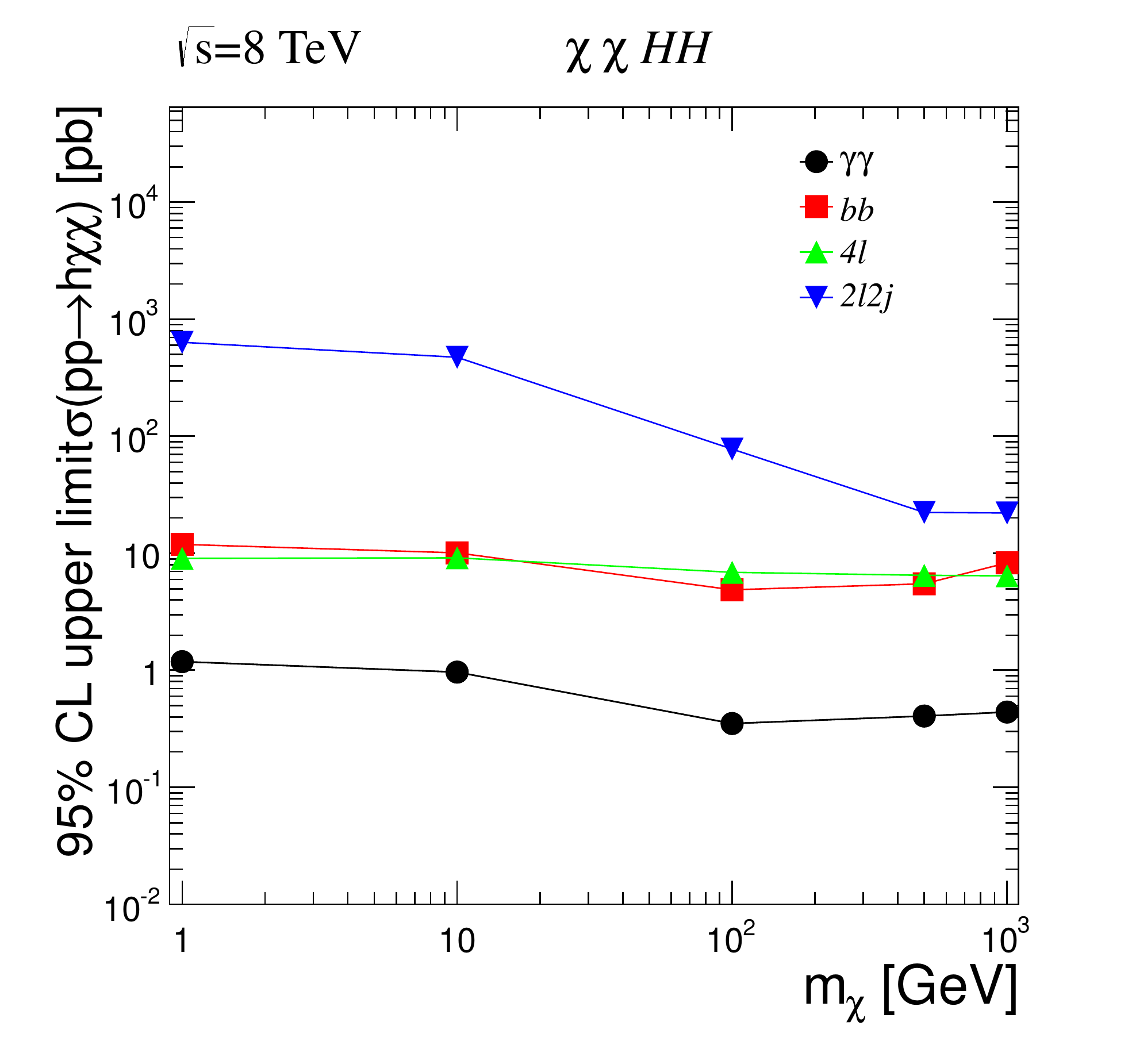}
\includegraphics[width=1.65in]{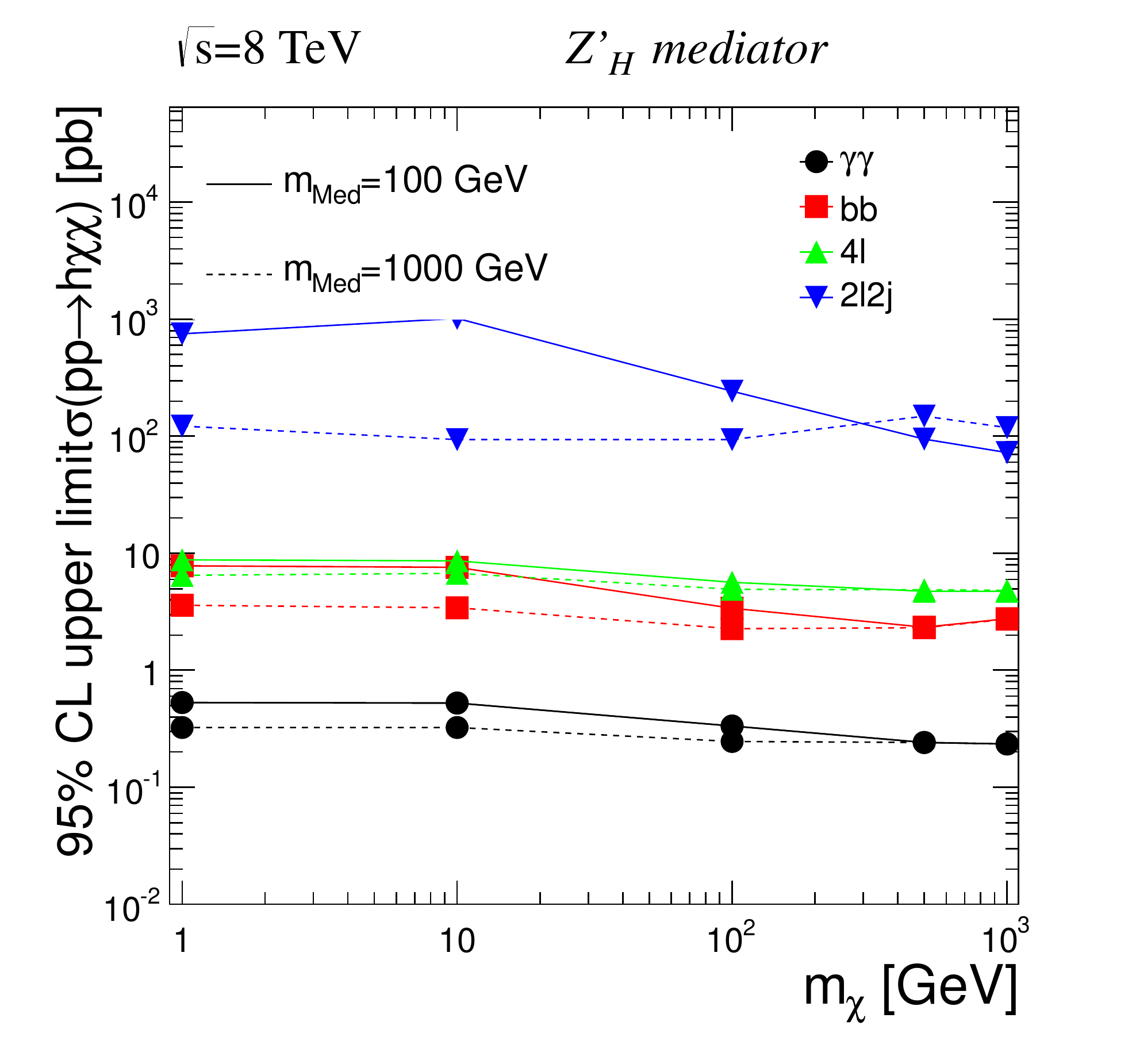}
\includegraphics[width=1.65in]{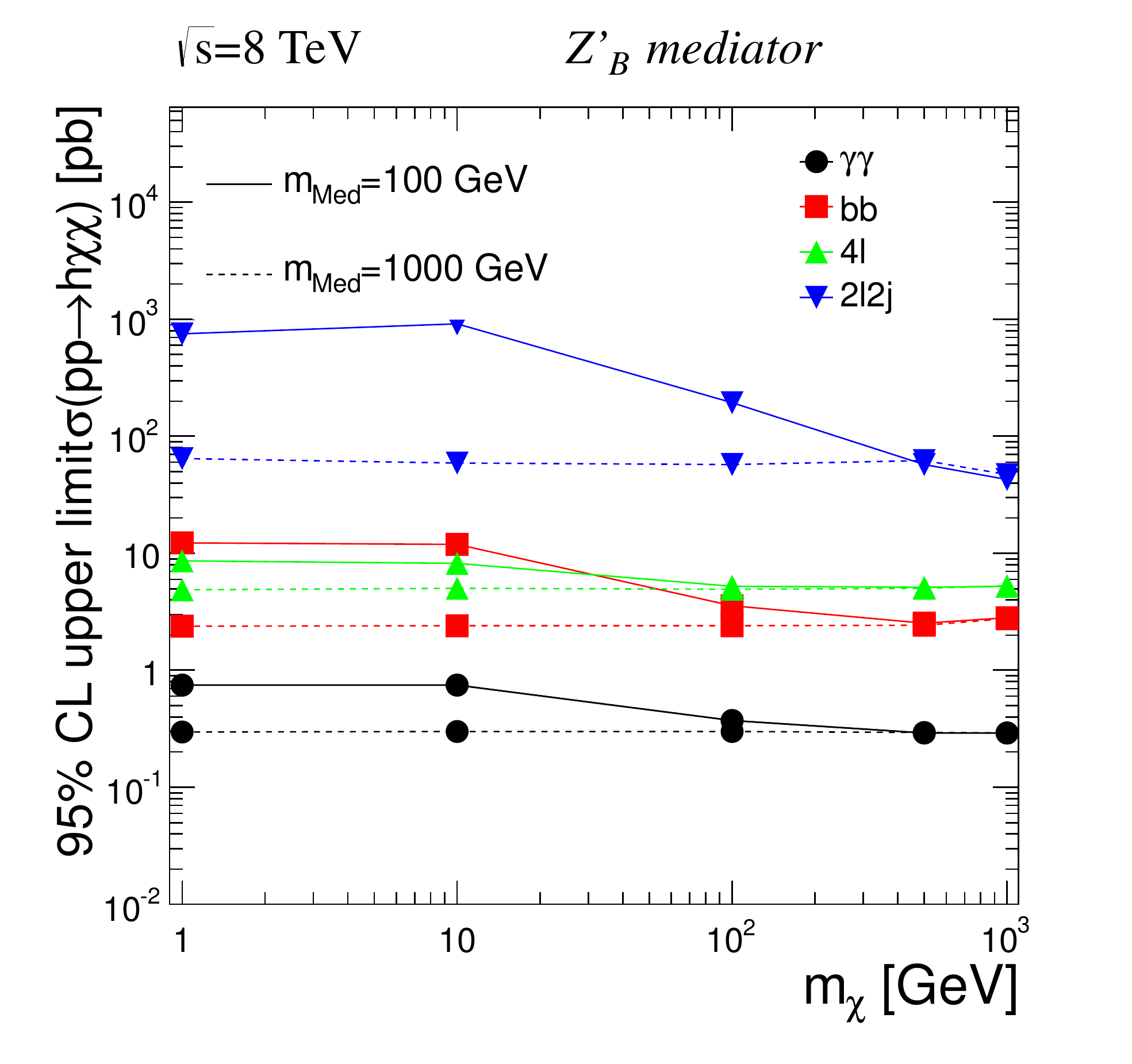}
\caption{ Upper limits on
  $\sigma(pp\rightarrow h\chi\bar{\chi})$ for $\sqrt{s}=8$ TeV in
  different decay modes and different models.  For simplified models with explicit mediators, solid lines are for 100 GeV mediator, and dashed for 1000 GeV.}
\label{fig:all_lim8}
\end{figure}

\begin{figure}
\includegraphics[width=1.65in]{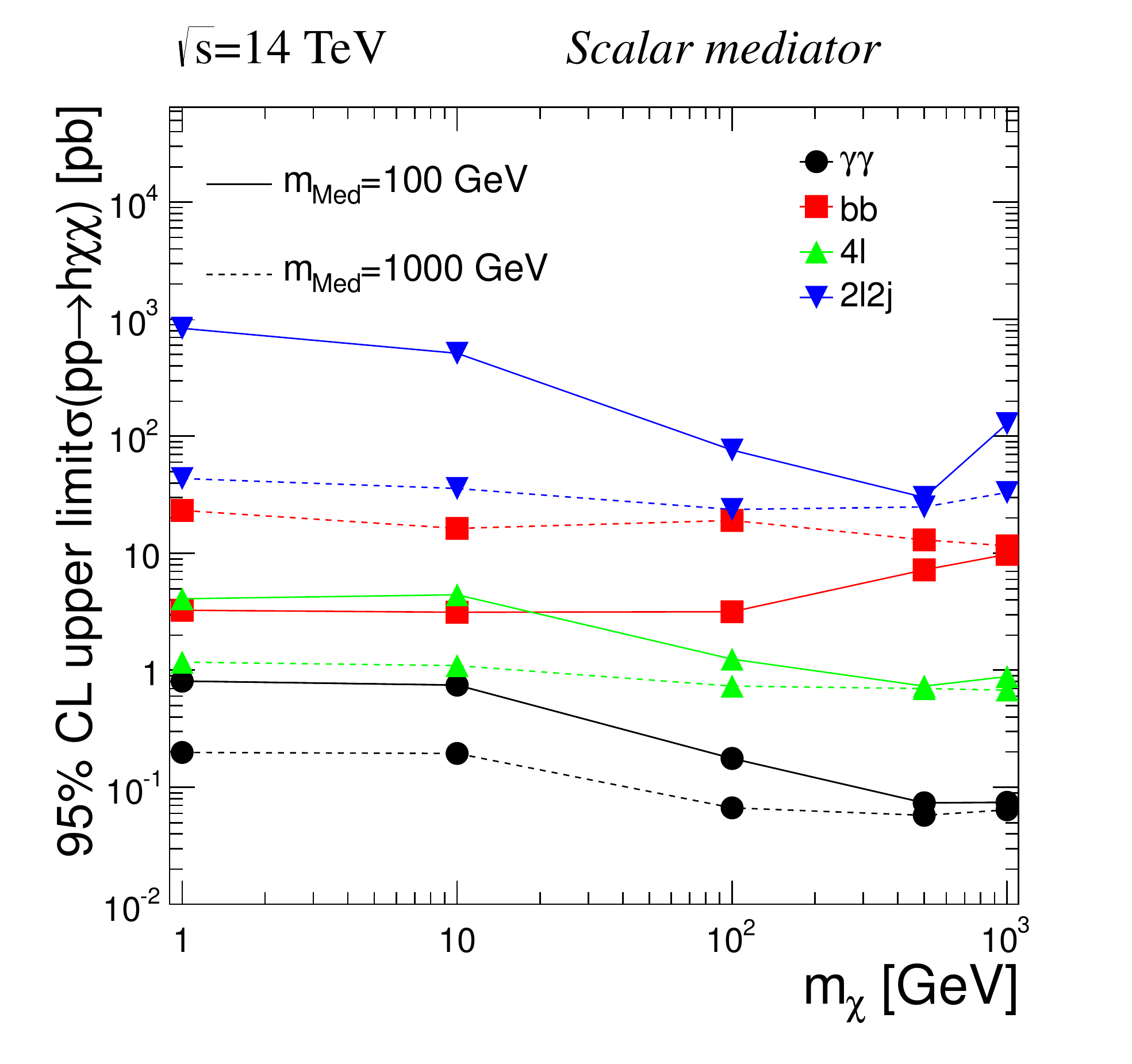}
\includegraphics[width=1.65in]{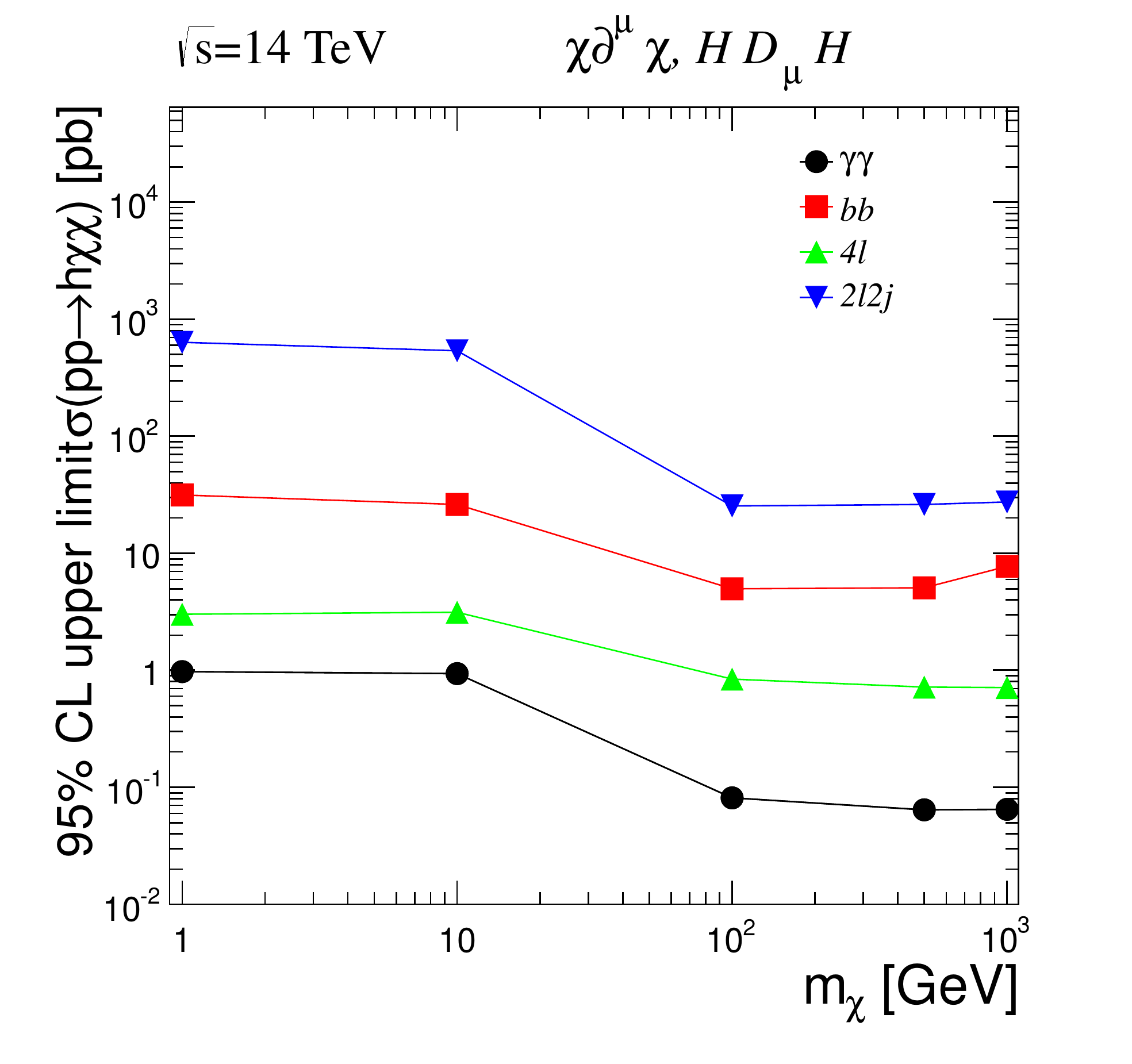}
\includegraphics[width=1.65in]{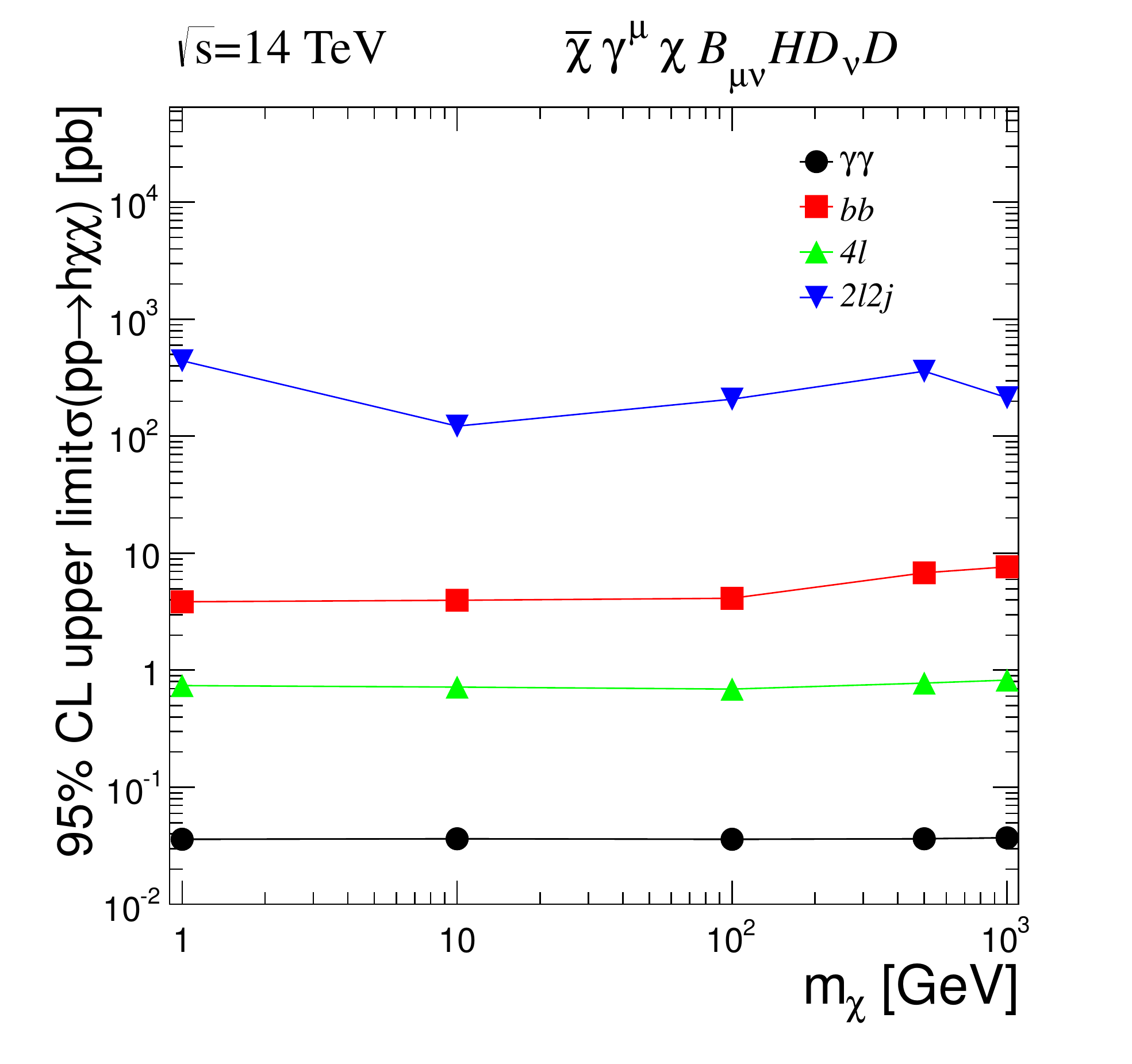}
\includegraphics[width=1.65in]{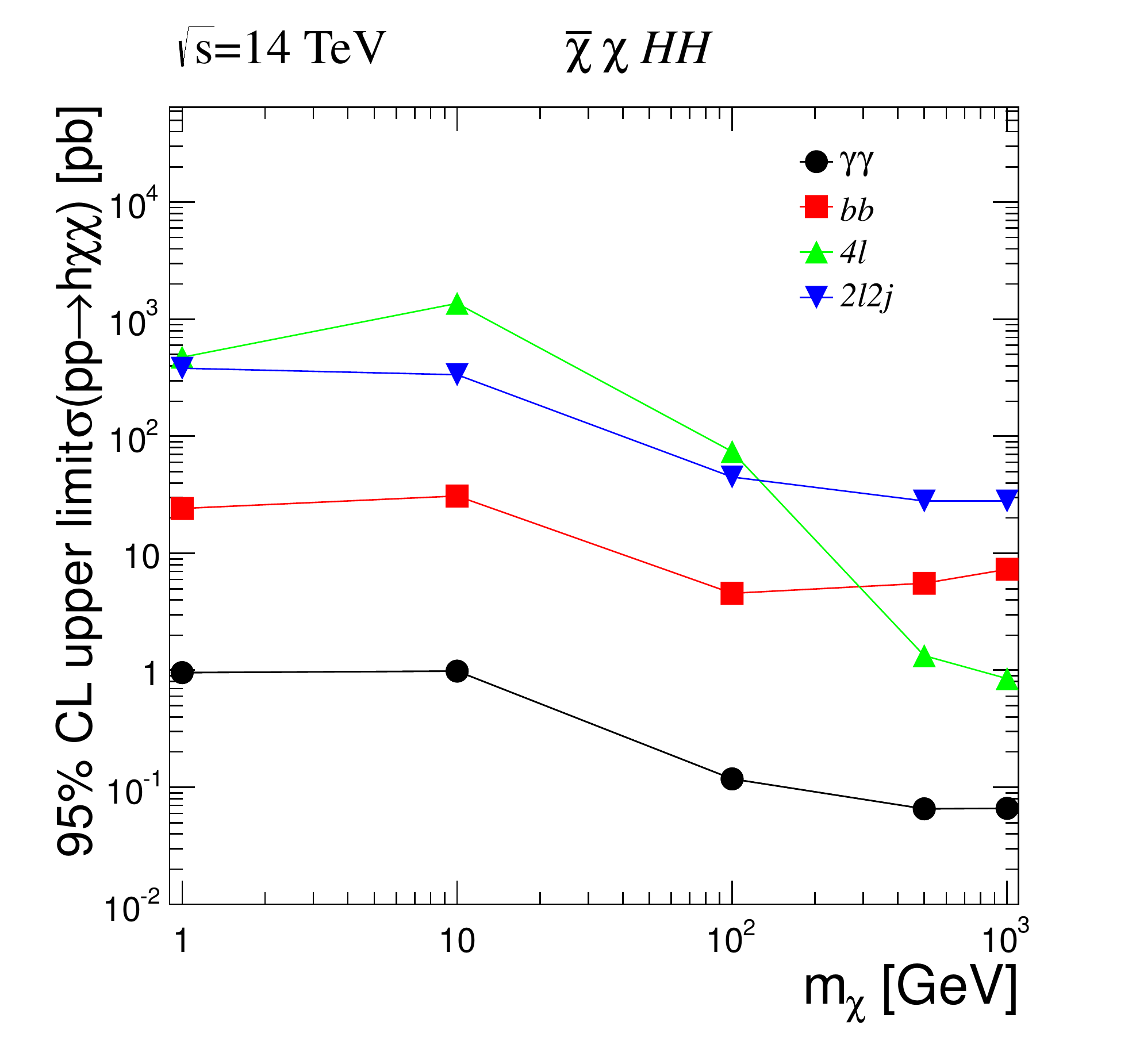}
\includegraphics[width=1.65in]{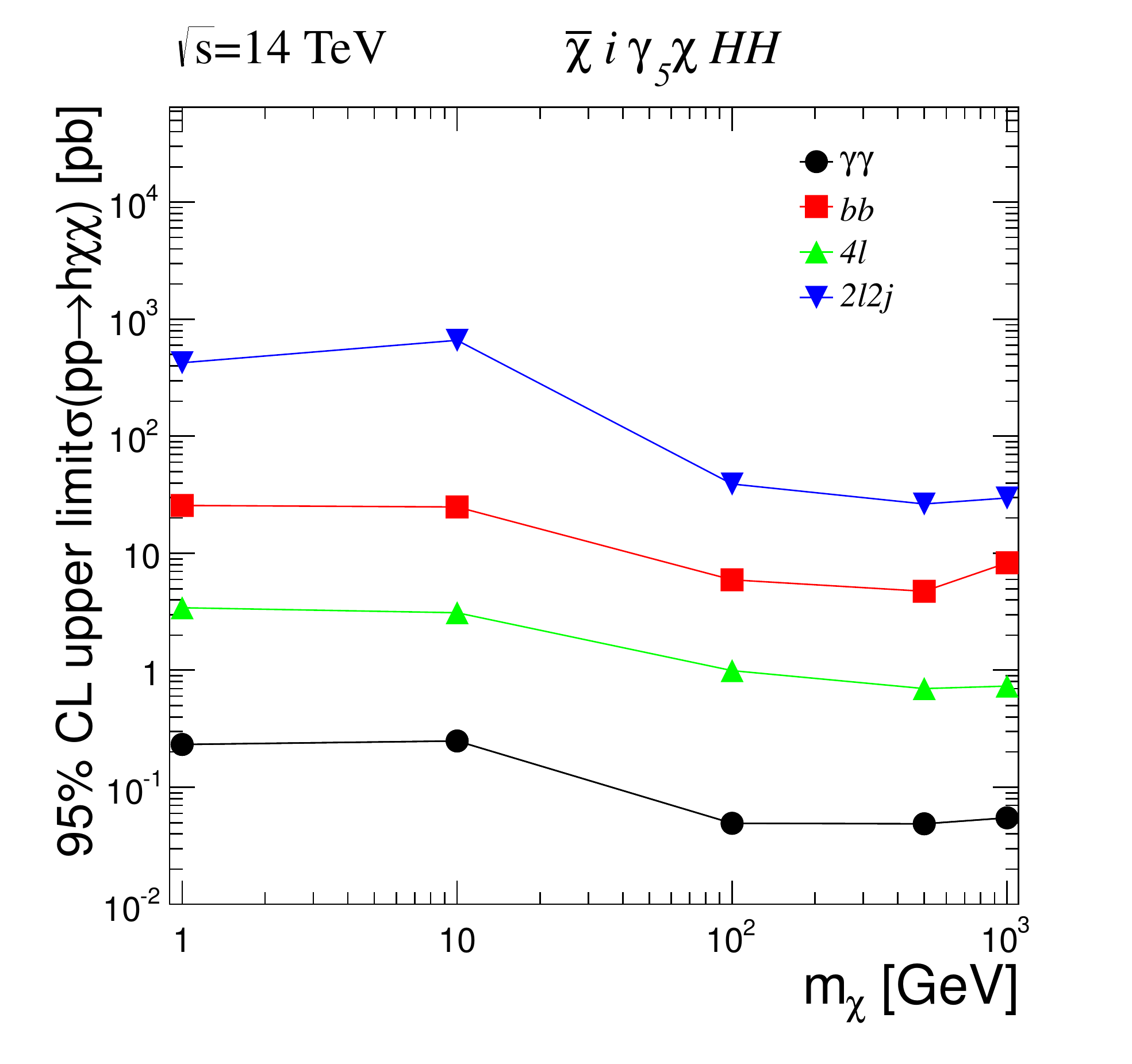}
\includegraphics[width=1.65in]{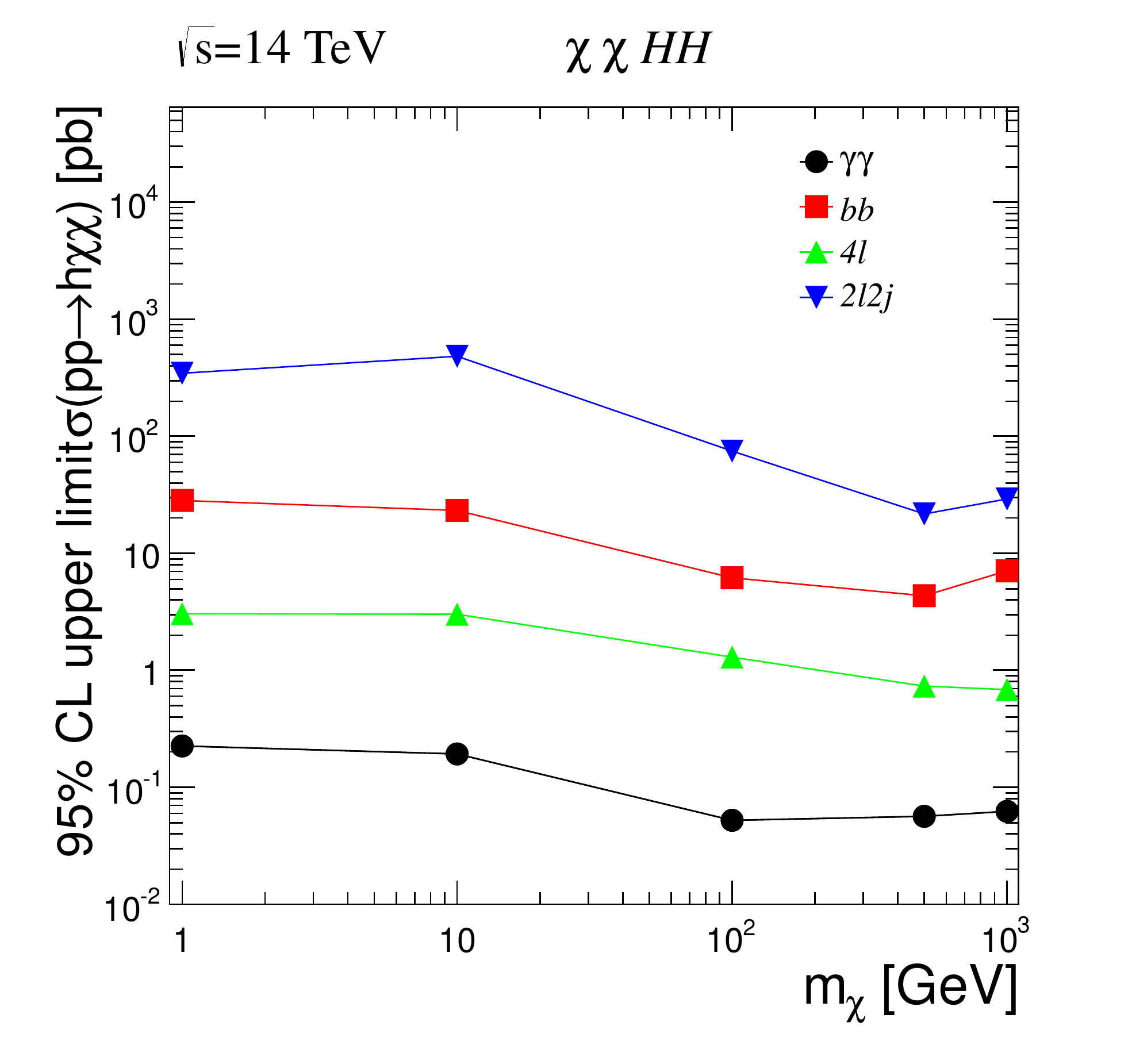}
\includegraphics[width=1.65in]{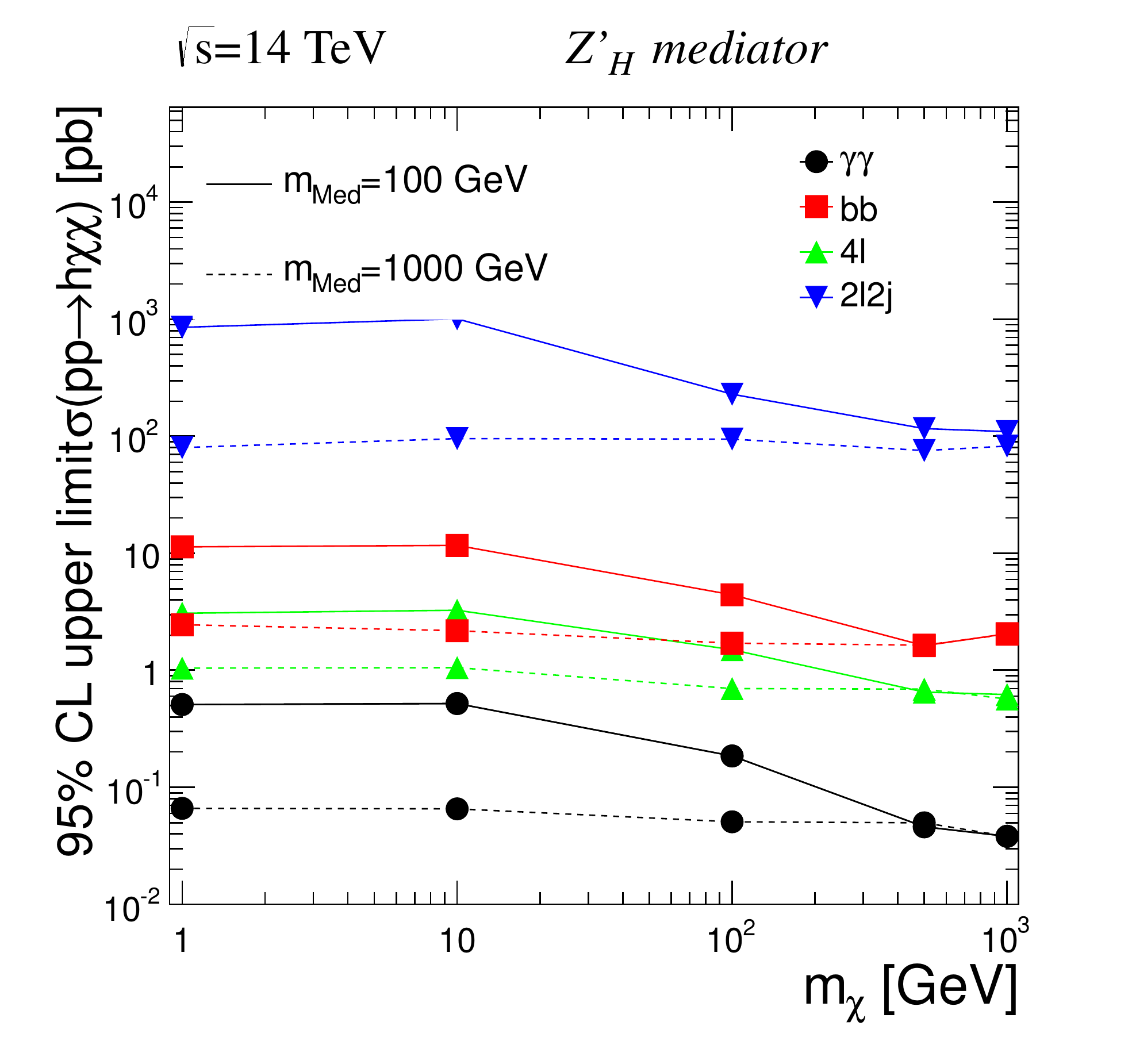}
\includegraphics[width=1.65in]{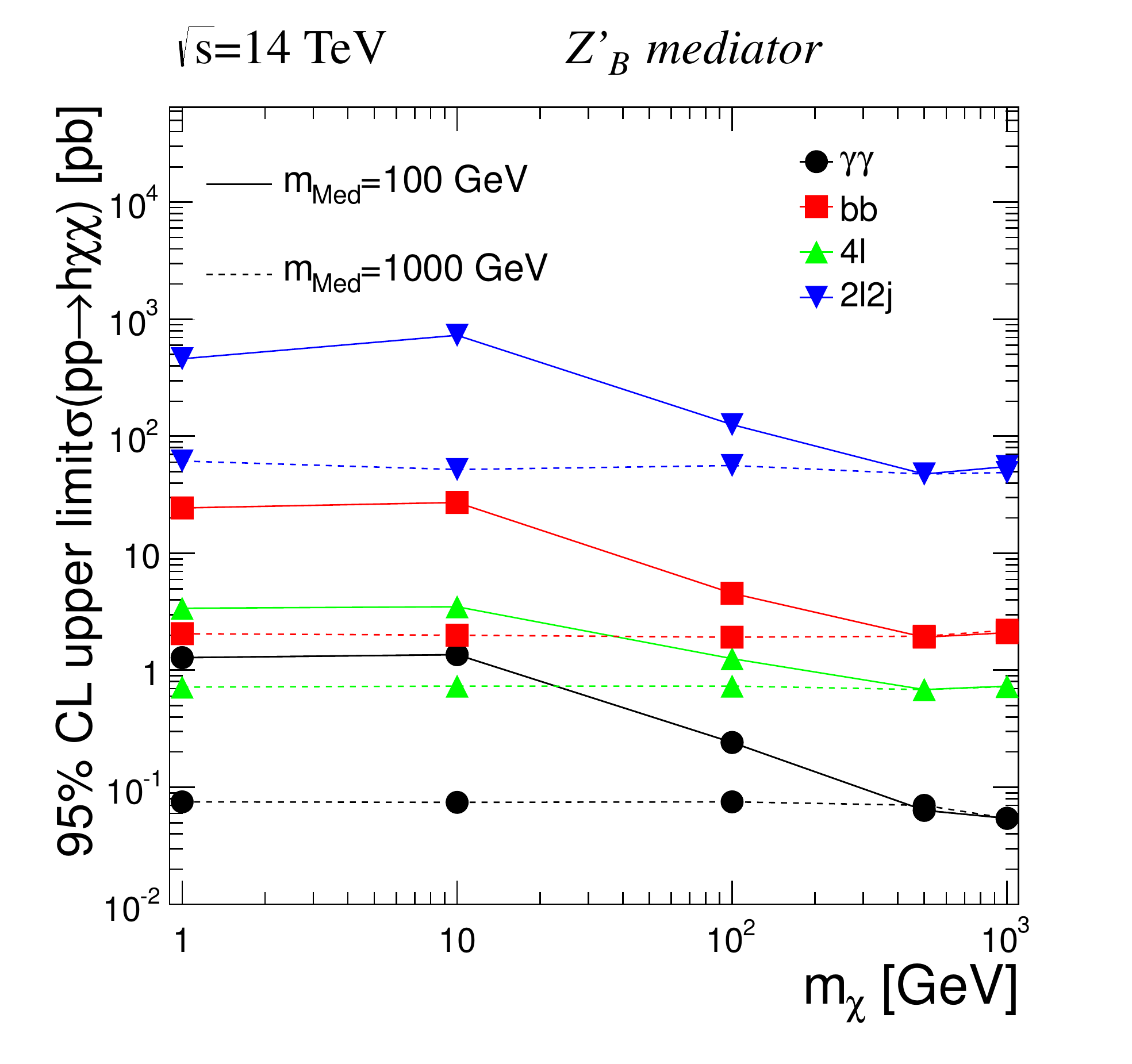}
\caption{ Upper limits on
  $\sigma(pp\rightarrow h\chi\bar{\chi})$ for $\sqrt{s}=14$ TeV in
  different decay modes and different models. For simplified models with explicit mediators, solid lines are for 100 GeV mediator, and dashed for 1000 GeV.}
\label{fig:all_lim14}
\end{figure}

\section{Discussion and results}
\label{sec:discuss}

For a range of different models and DM mass $m_\chi$, the LHC sensitivity to mono-Higgs production is approximately $100 \; {\rm fb} - 1 \; {\rm pb}$.  (More precise values are given in Figs.~\ref{fig:all_lim8} and \ref{fig:all_lim14}.)  In this section, we compare these projected sensitivities to the predicted cross sections for our benchmark theories (Fig.~\ref{fig:xs}) in order to constrain the parameter space of these scenarios.  We also consider other important constraints, such as invisible $h$ or $Z$ decays, as well as the recent bound on the spin-independent (SI) direct detection cross section from the LUX experiment~\cite{Akerib:2013tjd}.  The SI cross section for DM scattering on a nucleus $N$ with atomic and mass numbers $(Z,A)$ is
\be
\sigma_{\chi N}^{\rm SI} = \frac{\mu_{\chi N}^2}{\pi} \big( Z f_p + (A-Z) f_n \big)^2 \label{SIDD}
\ee
where $\mu_{\chi N}$ is the $\chi$-$N$ reduced mass and $f_{p,n}$ are the DM couplings to protons/neutrons.  We emphasize, however, that direct detection constraints can be avoided if DM is inelastic~\cite{TuckerSmith:2001hy}, i.e., if the complex state $\chi$ is split into real states $\chi_{1,2}$ with an $\mathcal{O}({\rm MeV})$ or larger mass splitting, with no change to the collider phenomenology provided it is much smaller than the typical parton energy.

Our results are shown in Figs.~\ref{fig:all_param_1}-\ref{fig:all_param_3}.  The ``$\gamma\gamma + \missET$'' contours show the LHC reach on our models at $\sqrt{s} = 8$ and $14$ TeV, based on 20 and 300 fb$^{-1}$ respectively, from mono-Higgs searches with $\gamma\gamma$ final states, which provides the stronger bound compared to $b \bar b$ and $ZZ^*$.  The limit contours shown exclude larger values of couplings and mixing angles, or smaller values of the effective operator mass scale $\Lambda$.

\begin{figure*}
\includegraphics[scale=0.46]{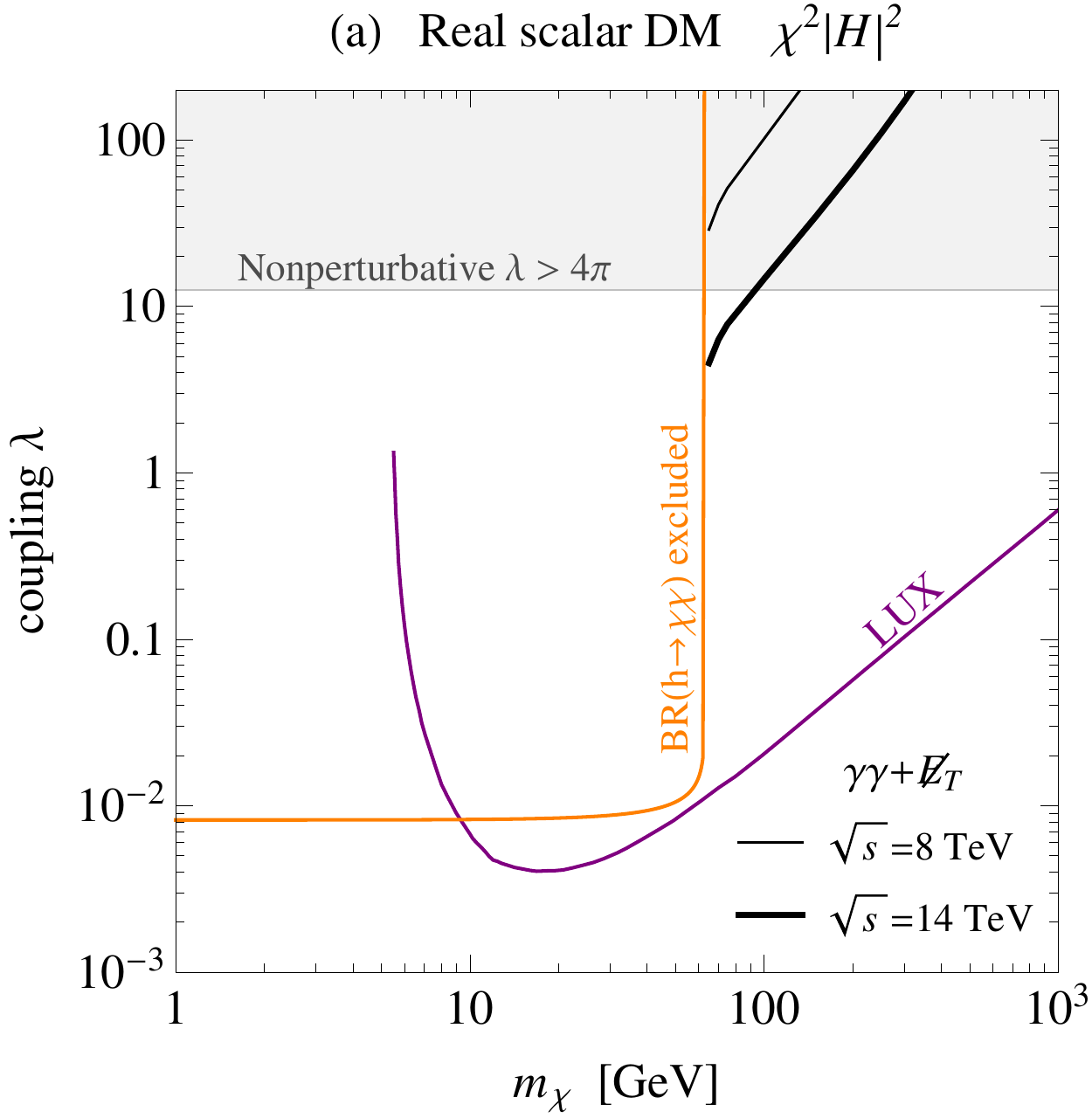}
\includegraphics[scale=0.46]{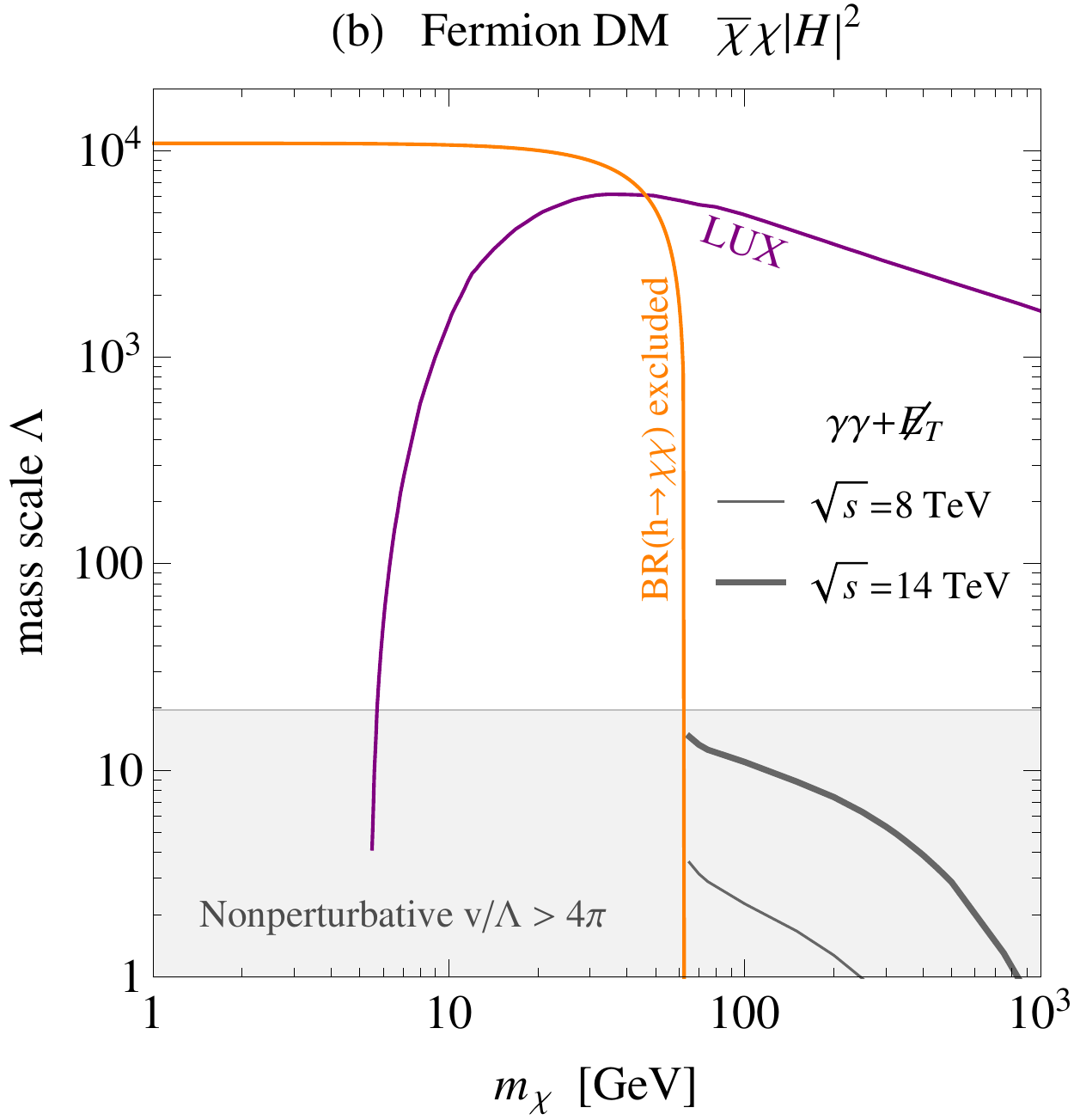}
\includegraphics[scale=0.46]{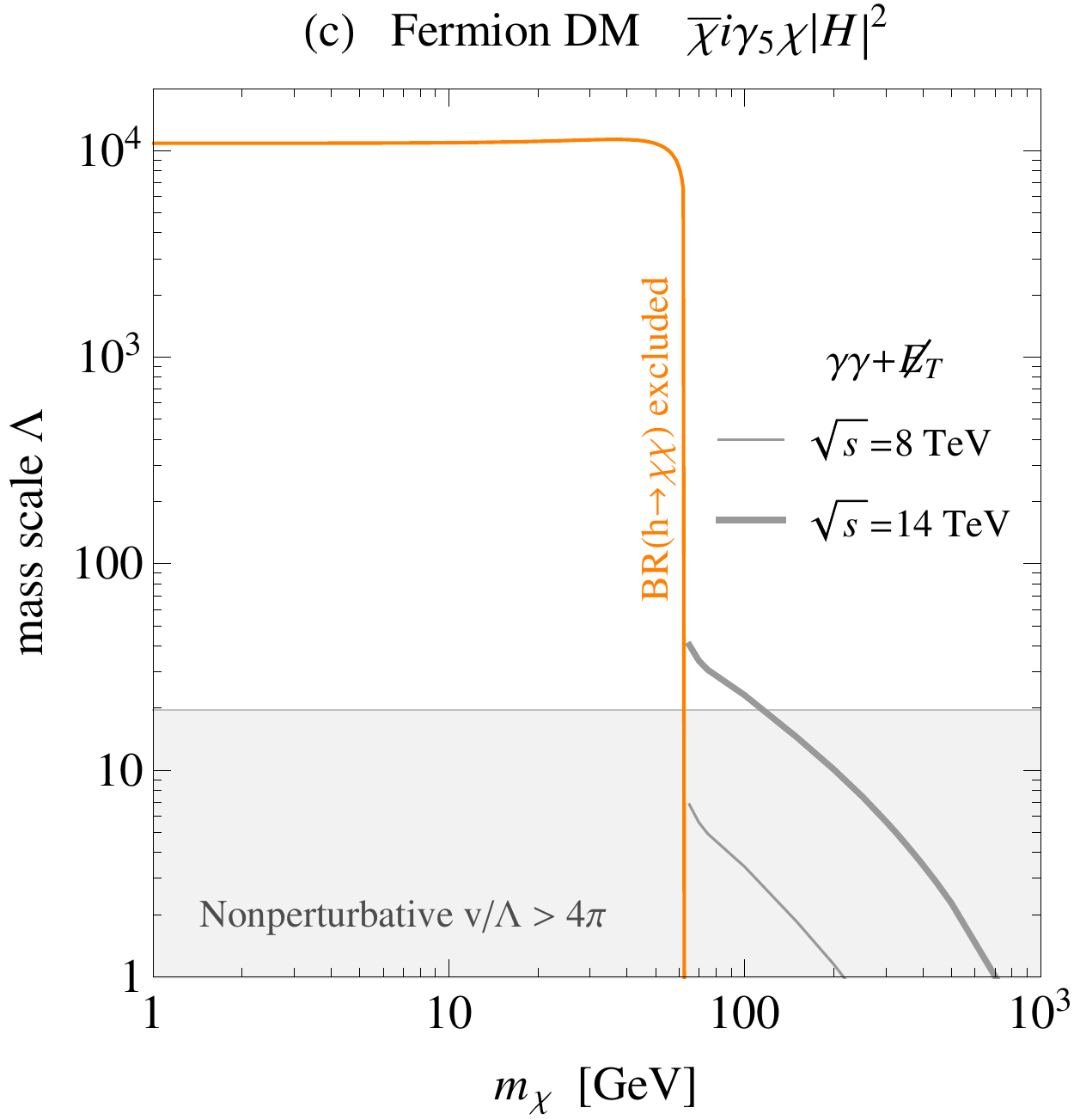}
\caption{Projected LHC mono-Higgs sensitivities at $\sqrt{s} = 8$ TeV ($20 \; {\rm fb}^{-1}$) and $14$ TeV ($300 \; {\rm fb}^{-1}$), with $\gamma\gamma + \missET$ final states, on Higgs portal effective operators.  All constraint contours exclude larger coupling $\lambda$ or smaller mass scale $\Lambda$.  Shaded region is excluded based on perturbativity arguments; orange contours denote limits from invisible $h$ decays; purple contours are exclusion limits from LUX.} \label{fig:all_param_1}
\end{figure*}

\subsection{Higgs portal effective operators}

The simplest models for coupling DM and the Higgs boson are the Higgs portal effective operators \eqref{eq:dim4} and \eqref{eq:dim5}.  For real scalar DM there is one operator $\chi^2 |H|^2$ with a dimensionless coupling $\lambda$, while for fermion DM there are two operators $\bar \chi \chi|H|^2$ and $\bar \chi i \gamma_5  \chi|H|^2$ suppressed by a mass scale $\Lambda$.  All three operators, which we consider separately, are qualitatively similar, and all three contribute to the invisible $h$ branching ratio ${\mathcal B}_{\rm inv}$ for $m_\chi < m_h/2$.  The LHC reach depends on whether $m_\chi$ is above or below $m_h/2$.

For $m_\chi < m_h/2$, mono-Higgs signals cannot be observed for these operators unless LHC sensitivities can be improved by a factor of $\sim 30$ over our estimates.  Actually, this is true for {\it any} value of ${\mathcal B}_{\rm inv}$, independently of whether one imposes a constraint on ${\mathcal B}_{\rm inv}$ or not.  Although $h\chi\chi$ production is enhanced as $\lambda$ becomes larger (or $\Lambda$ smaller), the {\it visible} branching ratios to $b\bar b$, $\gamma\gamma$, etc., become quenched as $\mathcal{B}_{\rm inv}$ becomes large, thereby suppressing the mono-Higgs signal.  The most favorable trade-off is for $\mathcal{B}_{\rm inv} \sim 50\%$, close to the present bound.  In this case, the dominant $h \chi\chi$ channel is resonant di-Higgs boson production (i.e. $hh$ produced on-shell) followed by an invisible decay $h \to \chi\chi$.  The $h \chi\chi$ cross section is bounded by the $hh$ cross section, 10 fb (34 fb) at 8 (14) TeV \cite{Dicus:1987ic,Dawson:1998py}, which is below the sensitivity limits we have found.  For recent LHC studies of di-Higgs cross sections, see for example \cite{Barger:2013jfa}.  As $\mathcal{B}_{\rm inv}$ becomes larger, resonant production saturates when ${\mathcal{B}}_{\rm inv} \sim 100\%$, while nonresonant production continues to grow as $\lambda^2$ or $1/\Lambda^2$.  However, the visible branching ratios fall as $\lambda^{-2}$ or $\Lambda^2$, compensating any enhancement in production.

On the other hand, for $m_\chi > m_h/2$, there is no bound on $\lambda$ or $\Lambda$ from $\mathcal{B}_{\rm inv}$.   Fig.~\ref{fig:xs} shows that our benchmark points with $\lambda = 1$ or $\Lambda = 100$ GeV are below the LHC sensitivity reach.  However, since $h\chi\chi$ is produced purely nonresonantly ($h$ cannot decay on-shell to $\chi\chi$), the cross section grows with $\lambda^2$ or $1/\Lambda^2$ with no suppression of visible decays.

In Fig.~\ref{fig:all_param_1}, we show how our LHC sensitivities map onto the parameter space of these scenarios.  The ``$\gamma\gamma + \missET$'' contours show the LHC reach from mono-Higgs searches with $\gamma\gamma$ final states.  These limits should be interpretted with care since such values push the boundaries imposed by perturbativity and validity of the effective field theory.  The shaded region is excluded based on perturbativity.   For scalar DM we require $\lambda < 4\pi$, while for fermion DM we require that the $h \bar \chi \chi$ Yukawa coupling $v/\Lambda$ be less than $4 \pi$.  As discussed above, there is no LHC sensitivity for $m_\chi < m_h/2$.  However, this region is strongly contrained by invisible Higgs decays, shown by the orange contour, taking $\mathcal{B}_{\rm inv} < 38\%$.  For direct detection, the SI cross section is given by Eq.~\eqref{SIDD} where
\be
f_{p,n} = \frac{m_{p,n}}{m_h^2} \big( 1 - \tfrac{7}{9} f_{TG} \big) \times \left\{ \begin{array}{cc}
\lambda/m_\chi \, ,   &  \chi^2 |H|^2 \\
1/\Lambda \, , &  \bar\chi \chi |H|^2
\end{array} \right. \, ,
\ee
taking $f_{TG} = 0.92$~\cite{Belanger:2013oya}.  The purple contour shows the exclusion limit from the LUX experiment.  On the other hand, the $\bar \chi i \gamma_5 \chi |H|^2$ operator leads to a velocity-suppressed SI cross section that is very weakly constrained.

\subsection{Other effective operators}

\begin{figure*}
\includegraphics[scale=0.46]{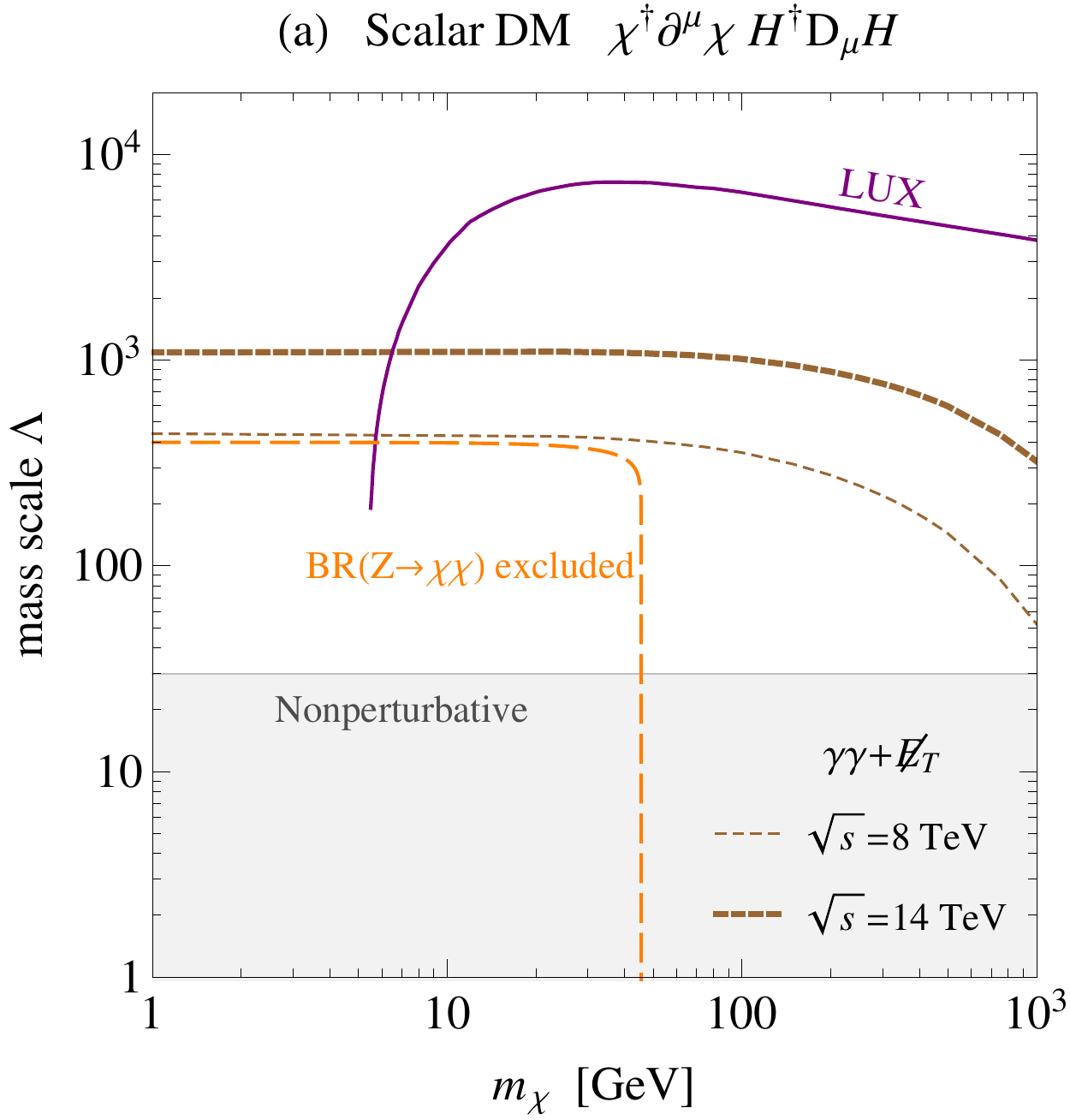}
\includegraphics[scale=0.46]{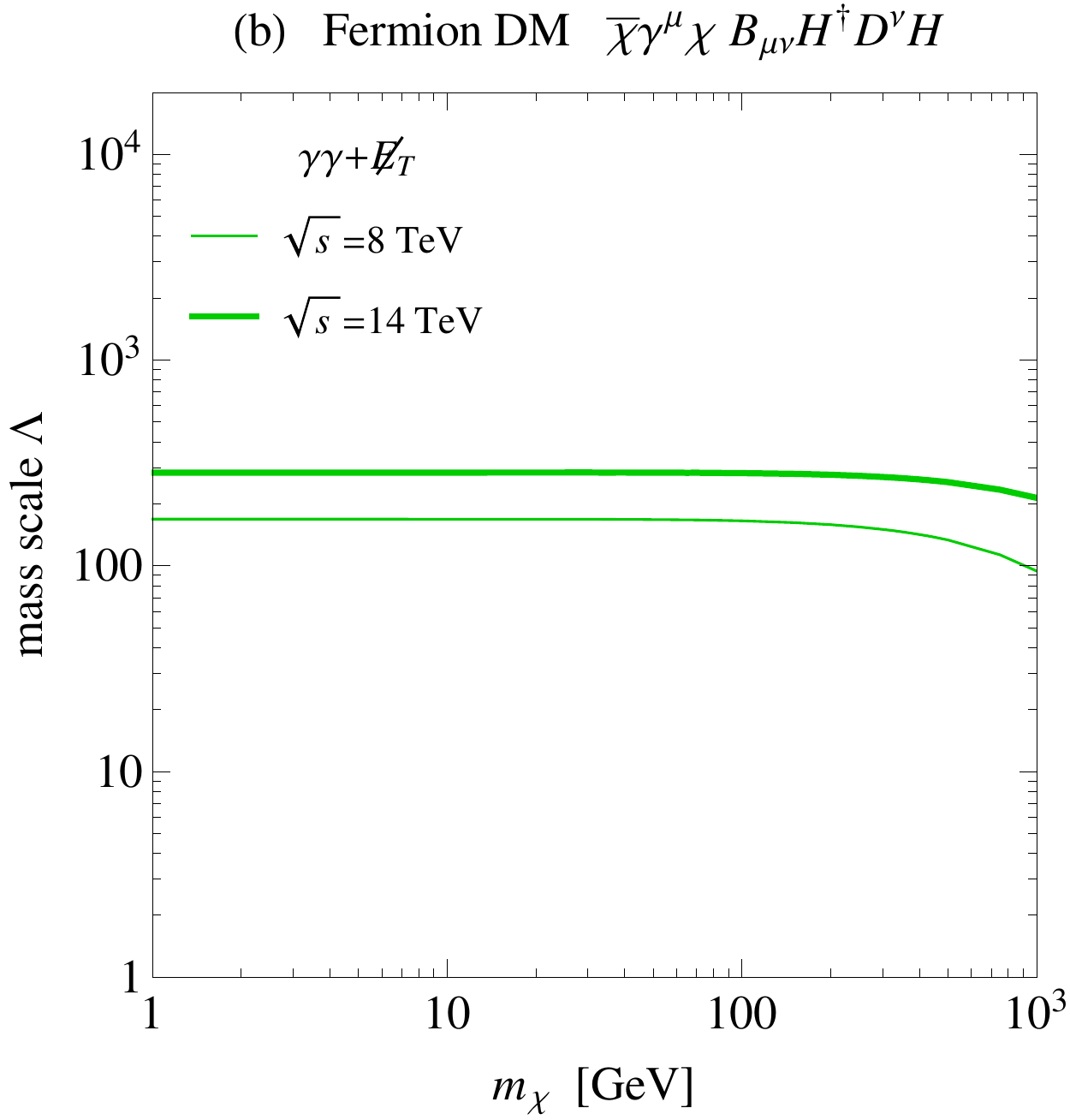}
\includegraphics[scale=0.46]{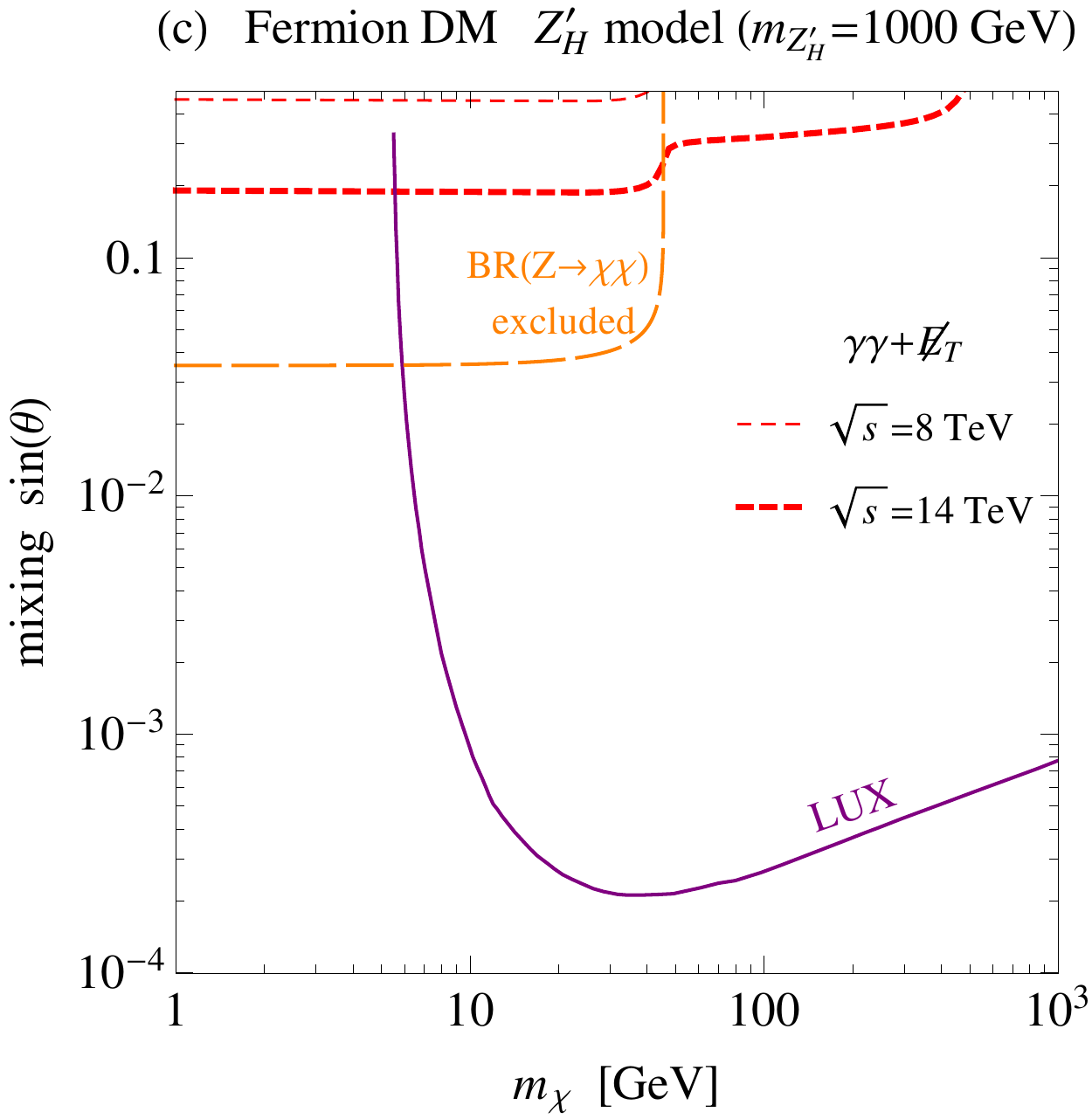}
\caption{Projected LHC mono-Higgs sensitivities at $\sqrt{s} = 8$ TeV ($20 \; {\rm fb}^{-1}$) and $14$ TeV ($300 \; {\rm fb}^{-1}$), with $\gamma\gamma + \missET$ final states, on effective operator models and simplified models.  All constraint contours exclude larger mixing angles or smaller mass scales $\Lambda$.  Shaded region is excluded based on perturbativity arguments or requiring $\sin\theta \le 1$; dashed orange contours denote limits from invisible $Z$ decays; purple contours are exclusion limits from LUX.} \label{fig:all_param_2}
\end{figure*}

Beyond the Higgs portal, we have studied mono-Higgs signals from two effective operators coupling DM to the electroweak degrees of freedom.  First, at dimension six, we have a scalar DM model with interaction \eqref{eq:O6Zscalar}, which generates an effective coupling of $\chi$ to the $Z$ boson that is $\mathcal{O}(v^2/\Lambda^2)$ as strong as a neutrino.  In Fig.~\ref{fig:all_param_2}(a), the brown dashed contours show the LHC mono-Higgs sensitivity with $\gamma\gamma$ final states.  For $m_\chi < m_Z/2$, the invisible $Z$ width measured at LEP constrains this operator, requiring $\Lambda \lesssim 400$ GeV for $\Gamma(Z \to \chi \chi^\dagger) \lesssim 3 \; {\rm MeV}$~\cite{Beringer:1900zz} (dashed orange contour).  There is no invisible Higgs decay.   It is interesting to note that mono-Higgs searches can be more powerful than the invisible $Z$ width bound.  However, the $Z$ coupling between DM and nucleons leads to a sizable cross section for direct detection.  The proton and neutron couplings are
\be
f_n = \frac{1}{4 \Lambda^2} \, , \quad f_p = - (1- 4 s_W^2) \frac{1}{4 \Lambda^2} \, .
\ee
The purple contour shows the current LUX bound, rescaled by the Xenon neutron fraction of $\approx 0.6$ since DM couples predominantly to neutrons only.  While the LUX bound is highly constraining, it may be evaded completely if DM is inelastic.  If the complex field $\chi$ is split into two real components $\chi_{1,2}$, operator \eqref{eq:O6Zscalar} becomes a transition interaction between $\chi_1$ and $\chi_2$, and the energetics of direct detection may be insufficient to excite a transition.  We also exclude the shaded region by perturbativity, where the effective $Z$ coupling to DM, given in \eqref{eq:hhbilin}, becomes larger than ${4\pi}$.

At dimension eight, there are many operators that are constrained neither by invisible decays nor direct detection.  As an example, we considered here operator \eqref{eq:O8} with fermionic DM.  The LHC sensitivities are shown in Fig.~\ref{fig:all_param_2}(b) by the green contours.  Direct detection signals, arising at one-loop order, are expected to be suppressed, especially compared to other potential operators generated from the same UV physics as operator \eqref{eq:O8}.

\subsection{Simplified models}

Beyond effective operators, we have described three simplified models for mono-Higgs signals with a new $s$-channel mediator particle that couples $\chi$ to SM particles.

\subsubsection{Hidden sector $Z^\prime$}

First, we consider a hidden sector $Z^\prime$, denoted $Z^\prime_H$, that couples to SM particles by mixing with the $Z$ boson.  The only parameters in this model are $m_{Z^\prime_H}$ ($Z^\prime_H$ mass), $g_\chi$ (DM-$Z^\prime_H$ coupling), and $\sin\theta$ ($Z$-$Z^\prime_H$ mixing angle).  Fig.~\ref{fig:all_param_2}(c) shows the LHC mono-Higgs sensitivity to this model, as a function of $\sin\theta$, for $m_{Z^\prime_H} = 1000$ GeV and $g_\chi = 1$.  The dashed orange contour shows the exclusion limit from the invisible $Z$ width if $m_\chi < m_Z/2$, requiring $\sin\theta \lesssim 0.03$ for $\Gamma(Z \to \chi \bar \chi) \lesssim 3 \; {\rm MeV}$~\cite{Beringer:1900zz}.  The $\rho$-parameter provides in principle a much stronger limit, at the level of $\sin\theta \lesssim 3 \times 10^{-3}$ for any $m_\chi$.  However, the quantitative details depend on doing a global fit to precision electroweak data, which is sensitive to other sources of new physics that may affect those observables.  Nevertheless, large values of $\sin\theta$ accessible to mono-Higgs searches have significant tension with precision electroweak observables.

For direct detection, the proton and neutron couplings entering Eq.~\eqref{SIDD} are
\be
f_n = \frac{g_2 g_\chi \sin 2\theta }{8 c_W m_Z^2} \left( 1 - \frac{m_Z^2}{m^2_{Z^\prime_H}} \right) \, , \quad f_p = - (1-4 s_W^2) f_n \, .
\ee
The LUX exclusion limits, rescaled by the Xenon neutron fraction $\approx 0.6$, are denoted by the purple contour.  These limits may be evaded for a $Z^\prime_H$ that couples to the axial vector current $\bar \chi \gamma^\mu \gamma_5 \chi$, instead of the vector current $\bar \chi \gamma^\mu \chi$ that we had assumed.  In that case, although the collider phenomenology would be identical, this model would contribute to the spin-dependent direction detection cross section only (at leading order in velocity), which is less constrained.  Alternately, $\chi$ may be inelastic if it has a Majorana mass term that splits the Dirac field $\chi$ into two Majorana fields $\chi_{1,2}$.  Eq.~\eqref{eq:LintZH} becomes a transition coupling between $\chi_1$ and $\chi_2$ that can be energetically forbidden in direct detection scattering.

\begin{figure*}
\includegraphics[scale=0.46]{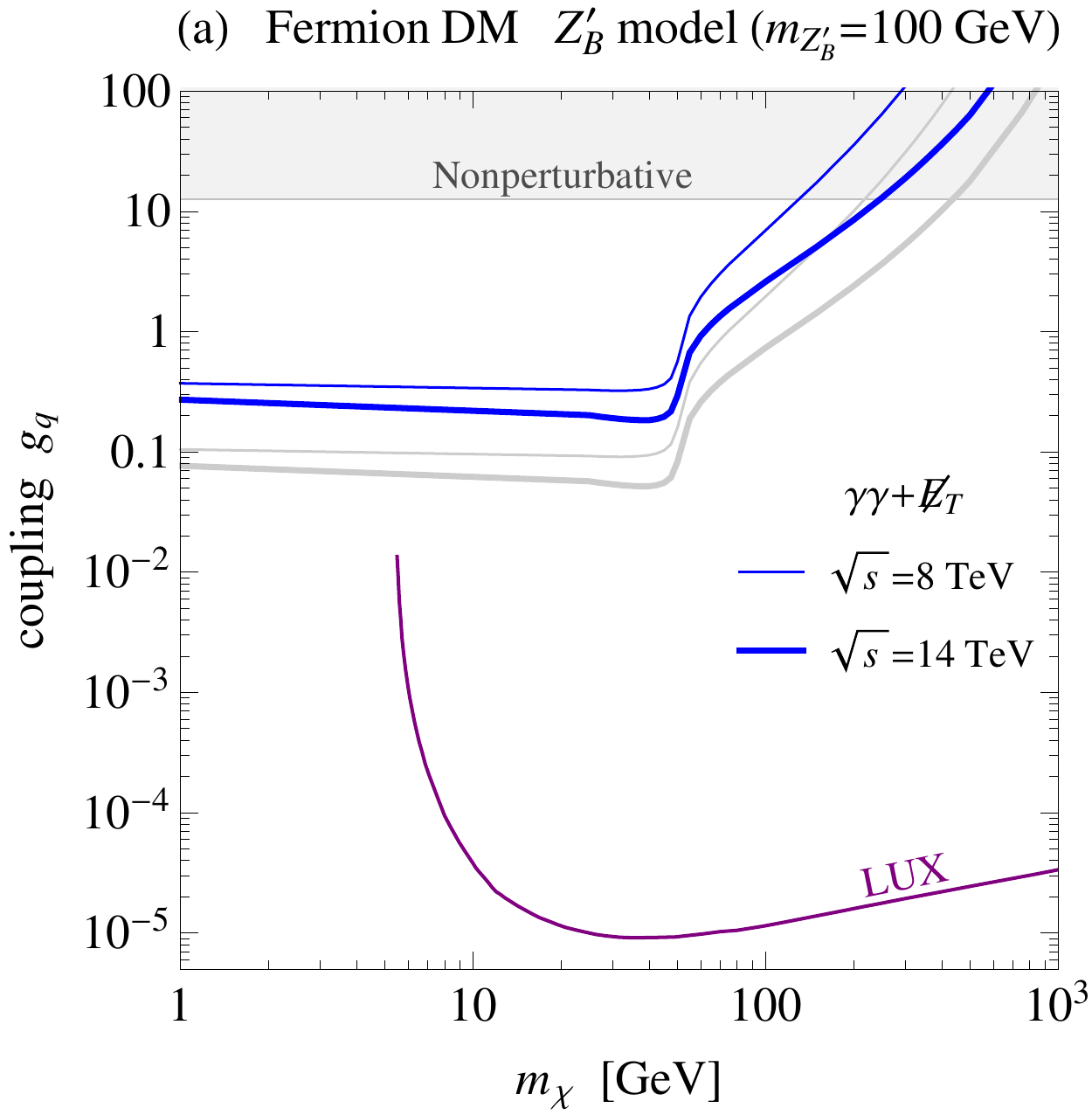}
\includegraphics[scale=0.46]{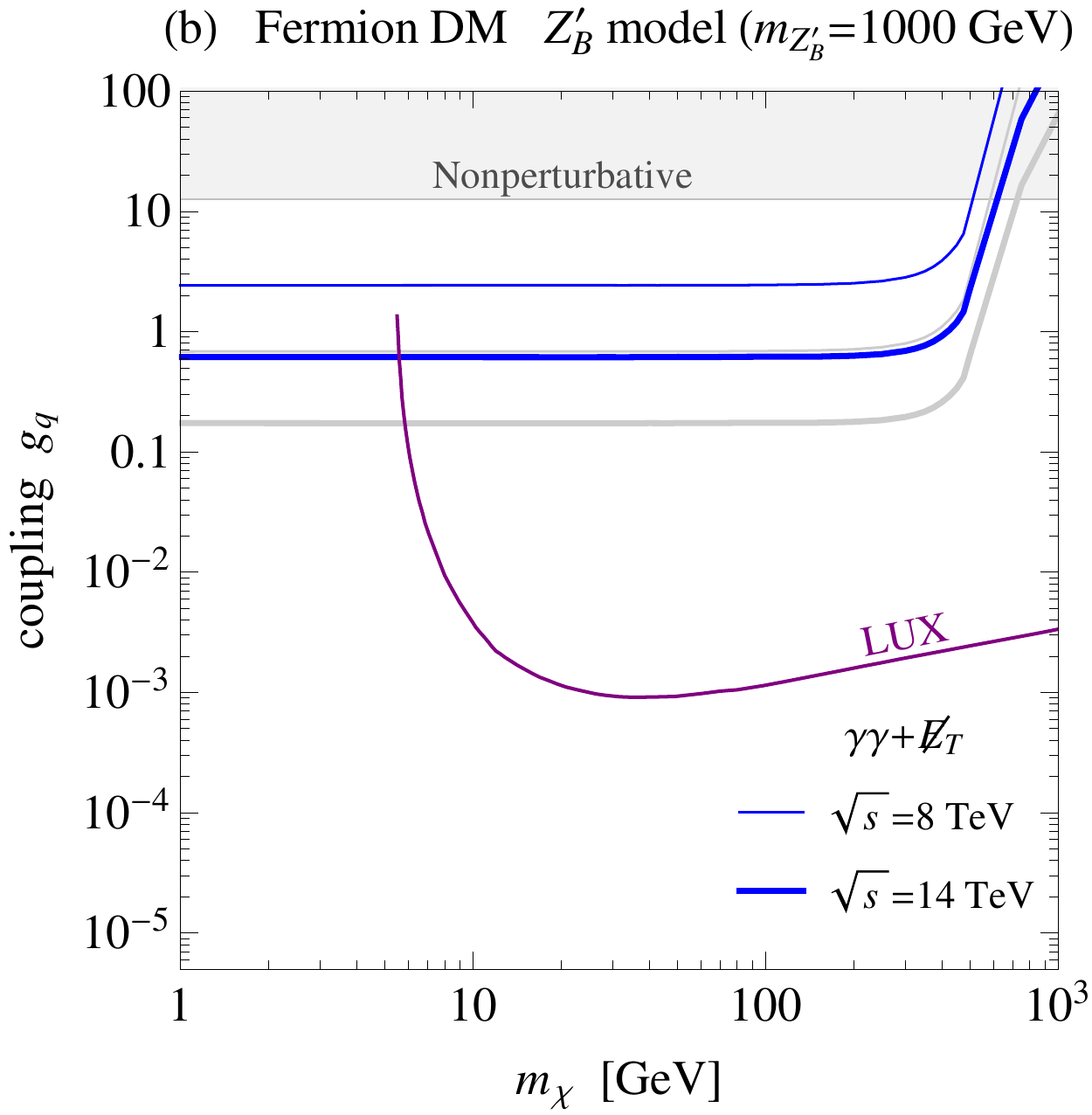}
\includegraphics[scale=0.46]{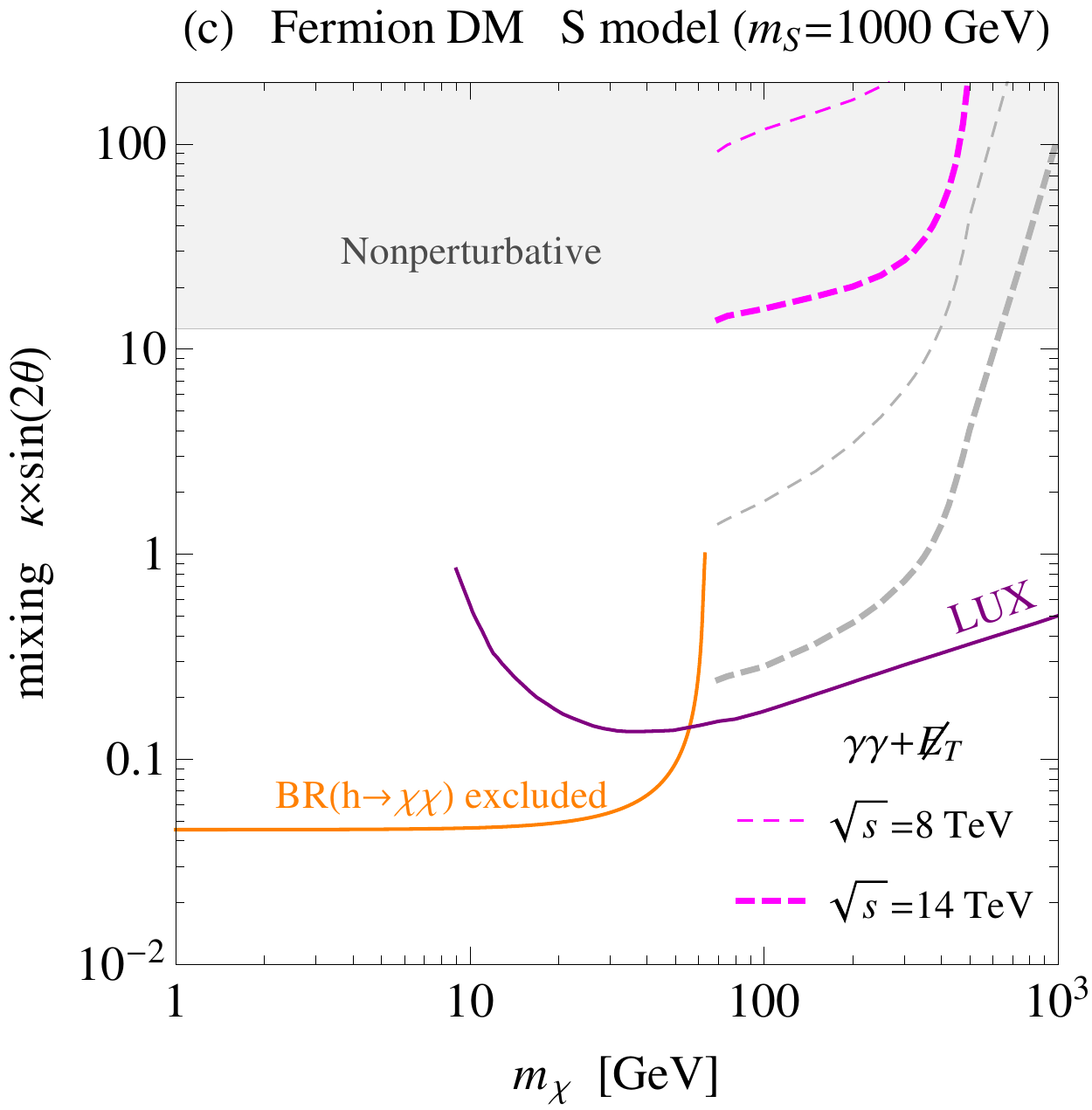}
\caption{Projected LHC mono-Higgs sensitivities at $\sqrt{s} = 8$ TeV ($20 \; {\rm fb}^{-1}$) and $14$ TeV ($300 \; {\rm fb}^{-1}$), with $\gamma\gamma + \missET$ final states, on simplified models.  All constraint contours exclude larger couplings or mixing angles. The light gray contours show the LHC reach if $g_{hZ^\prime Z^\prime}$ is pushed as large as allowed by perturbativity arguments.
 Shaded region is excluded based on perturbativity arguments or requiring $\sin\theta \le 1$; orange contour denotes limit from invisible $h$ decays; purple contours are exclusion limits from LUX.} \label{fig:all_param_3}
\end{figure*}

\subsubsection{Baryon-number $Z^\prime$}

Second, we consider a leptophobic $Z^\prime$, denoted $Z^\prime_B$, that couples to both baryon number and DM.  The dark blue contours in Figs.~\ref{fig:all_param_3}(a,b) show the LHC mono-Higgs sensitivities for our two $Z^\prime_B$ benchmarks ($m_{Z^\prime_B} = 100$ GeV and $m_{Z^\prime_B} = 1000$ GeV; see Table~\ref{tab:bench}) except we have allowed the quark-$Z^\prime_B$ coupling $g_q$ to vary.  The mono-Higgs cross section scales with $g_{h Z^\prime Z^\prime}^2$.  The light gray contours show the LHC reach if $g_{hZ^\prime Z^\prime}$ is pushed as large as allowed by perturbativity arguments.\footnote{We impose a bound $m_{Z^\prime_B}^2/v_B^2 < 4\pi$ based on perturbativity of the underlying $h_B^2 Z^\prime Z^\prime$ coupling in Eq.~\eqref{ZBmassterm} before mixing, which requires $g_{h Z^\prime Z^\prime} < \sqrt{4\pi} m_{Z^\prime_B} \sin\theta$.}
 There is no constraint from invisible $h$ or $Z$ decays.

The DM-nucleon couplings for SI direct detection are
\be
f_{p,n} = \frac{ 3 g_q g_\chi}{m_{Z^\prime_B}^2} \, .
\ee
For our benchmark scenarios, LUX strongly excludes the entire parameter region above kinematic threshold, thereby requiring extremely small values of $g_q$.  However, like the $Z^\prime_H$ model, this bound can be evaded by appealing to inelastic DM or an axial-vector DM interaction.  Alternately, DM particles below $\sim 5$ GeV are below LUX thresholds and are not excluded by invisible $h$ or $Z$ decays.

It is clear from Figs.~\ref{fig:all_param_3}(a,b) that mono-Higgs searches have sensitivity only for $m_\chi < m_{Z^\prime_B}/2$, where the $Z^\prime_B$ can be produced on-shell and then decays into $\chi \bar \chi$.  However, the $Z^\prime_B$ has a sizable coupling to quarks and can decay back into jets instead, giving a dijet resonance.  The constraints on a leptophobic $Z^\prime$ DM model from both mono-jet and dijet resonance searches were explored by Ref.~\cite{An:2012va} based on Tevatron and 7 TeV LHC studies with $\sim 1$ fb.  The constraints obtained therein are stronger than the our projected $h+\missET$ sensitivities.  Thus, mono-Higgs searches do not give a strong probe of the $Z^\prime_B$ model compared to other existing analyses.  (This conclusion would be strengthened by considering more recent searches compared to  Ref.~\cite{An:2012va}.)

\subsubsection{Scalar singlet $S$ model}

Lastly, we consider the scalar singlet model, where $S$ couples to SM particles by mixing with the Higgs boson.  We focus on the benchmark case with $m_S=1000$ GeV (see Table~\ref{tab:bench}).  For $m_\chi < m_h/2$, LHC sensitivities are insufficient to put any limit on this model, similar to the Higgs portal operators discussed above, due to the visible $\gamma\gamma$ signal becoming diluted by a large $h \to \chi \bar\chi$ branching ratio.  However, this mass range is strongly constrained by the invisible $h$ width.

For $m_\chi > m_h/2$, the dominant $h \chi\bar\chi$ channel is via Higgstrahlung from an intermediate $S$ propagator, shown in Fig.~\ref{fig:effhSxx}(b).  This process is proportional to $\sin^2(2\theta)$, where $\theta$ is the $h$-$S$ mixing angle in \eqref{hSrotation}.  To present our bounds, it is useful to introduce an extra scaling parameter $\kappa$, defined by $\sin(2\theta) \to \kappa \sin(2\theta)$, such that now the $h \chi\bar\chi$ cross section is proportional to $\kappa^2 \sin^2(2\theta)$.  The model we have discussed in Sec.~\ref{sec:models} is obtained with $\kappa=1$.  However, larger values of $\kappa$ may be obtained in more complicated models, e.g., with an additional Higgs doublet~\cite{Cotta:2013jna}.\footnote{Suppose we introduce an additional Higgs doublet $H^\prime$ coupled to quarks.  The CP-even neutral scalar couplings to fermions are
\be
\mathscr{L} \supset - y_\chi \bar\chi \chi S - \frac{m_q}{v} \bar q q( h + \kappa \, h^\prime) \, , \label{eq:2HDM}
\ee
where $h^\prime$ is an additional neutral Higgs state with couplings aligned with those of the SM Higgs boson $h$, up to a constant $\kappa$.  If $S$ mixes with $h^\prime$, as opposed to $h$, the diagram in Fig.~\ref{fig:effhSxx}(b) is enhanced by a factor $\kappa$.}
The magenta contours in Fig.~\ref{fig:all_param_3}(c) show the LHC sensitivities on $\kappa\times \sin(2\theta)$, with other model parameters fixed as in Table~\ref{tab:bench}.  The corresponding light gray contours show the enhanced LHC reach if we take coupling parameters $b = 4\pi$ and $y_\chi = 4\pi$ as large as perturbatively allowed.  For $\kappa=1$, since $\sin(2\theta)$ cannot be larger than unity, the LHC has sensititivity only for $\sqrt{s} = 14$ TeV and for values of $b,y_\chi$ near their perturbative limits.   Mono-Higgs signals may be more readily observable, however, in scalar extended models with $\kappa > 1$.  The shaded region is excluded if we require $\kappa < {4\pi}$ based on perturbativity of the top Yukawa coupling in Eq.~\eqref{eq:2HDM}.  The purple contour denotes the current LUX bound for the $\kappa=1$, with nucleon couplings
\be
f_{p,n} = \frac{y_\chi m_{p,n} \sin(2\theta)}{2 v m_h^2} \left( 1 - \frac{m_h^2}{m_S^2} \right) \big( 1 - \tfrac{7}{9} f_{TG} \big)\,.
\ee
However, these limits can be evaded if $S$ couples to $\bar\chi i \gamma_5 \chi$, rather than $\bar\chi\chi$ as we assumed, with little impact on the collider signatures.  The orange contour shows the limit from $\mathcal{B}_{\rm inv} < 38\%$ for $\kappa = 1$, although this bound can be weakened if $S$ mixes primarily with an additional Higgs boson, rather than $h$.

\section{Conclusions}

\label{sec:conclude}

Since the particle theory of DM is as yet unknown, it is worthwhile exploring all possible avenues for discovery.  In this work, we have studied a new DM signature to be explored at the LHC: missing energy from DM particles produced in association with a Higgs boson ($h + \missET$).  Coupling between DM and the Higgs boson is a generic feature of many weak-scale DM models.  While the $h$ invisible branching fraction $\mathcal{B}_{\rm inv}$ is a sensitive probe of Higgs boson couplings to light DM, mono-Higgs searches provide a complementary window into DM masses above $m_h/2$ or into models that otherwise do not lead to invisible $h$ decays.

We have considered several benchmark DM models for generating mono-Higgs signals at the LHC, including both EFT operators and simplified models with new $s$-channel mediators.  We performed a study of SM backgrounds to $h + \missET$ searches at $\sqrt{s} = 8$ TeV (20 fb$^{-1}$) and $14$ TeV (300 fb$^{-1}$) for four $h$ decay channels: $h \to b \bar b$, $\gamma\gamma$, and $ZZ^* \to 4\ell$, $\ell\ell j j$.  The $h \to b \bar b$ channel, despite having the largest branching ratio, does not give the best LHC sensitivity reach due to a large $t \bar t$ background.  The greatest reach for all our models is set by the $h \to \gamma\gamma$ channel.  Future experimental analyses may achieve reduced systematic uncertainties from data-driven background extrapolation or may find avenues for more aggressive background supporession than were plausible in the context of approximations made for these sensitivity studies.

The most promising scenarios for mono-Higgs signals are models where the effective coupling of DM to SM particles requires additional insertions of the Higgs field $H$.  One example is an effective coupling of DM to the $Z$ boson via higher dimensional operators.  We showed that, for scalar DM, LHC mono-Higgs searches at 14 TeV can set a limit $\Lambda \gtrsim$ TeV on the effective mass scale governing this coupling.  For light DM ($m_\chi < m_Z/2$), this constraint would be stronger than the invisible $Z$ width bound from LEP.  Another scenario is the case of scalar mediator models.  While LHC sensitivities to the minimal Higgs-mixing model we considered are insufficient to constrain this scenario, extended scalar models (e.g.,~\cite{He:2013suk}) offer a promising direction for mono-Higgs studies.

{\it Note added:} In the final stages of preparing this paper, Ref.~\cite{Petrov:2013nia} appeared, exploring similar mono-Higgs ideas within the framework of EFT operators with fermionic DM.  These authors reinterpreted a recent CMS search for $Z(\nu \bar \nu) + h(b \bar b)$ at 8 TeV with 19 ${\rm fb}^{-1}$~\cite{Chatrchyan:2013zna} in terms limits on $h \chi \chi$.  The new physics sensitivity adopted in Ref.~\cite{Petrov:2013nia} is consistent with our projected 8 TeV sensitivities with 20 ${\rm fb}^{-1}$ in the $b \bar b$ channel, although somewhat different cuts were adopted.  Ref.~\cite{Petrov:2013nia} did not consider other Higgs boson final states $ZZ^*$ and $\gamma\gamma$ studied herein, the latter of which is significantly more sensitive than $b\bar{b}$.  For EFT operators, Ref.~\cite{Petrov:2013nia} studied the dimension-5 operators given in \eqref{eq:dim5}, obtaining a lower bound $\Lambda \gtrsim$ GeV.  Such limits are not physically meaningful since not only is the EFT invalid at LHC energies, but more generally any perturbative analysis would fail since the $h \bar\chi \chi$ Yukawa coupling would be $v/\Lambda \sim 100$.  On the other hand, interesting limits were obtained for a variety of other operators.  Ref.~\cite{Petrov:2013nia} considered dimension-6 operators given in \eqref{eq:O6Zferm} --- whereas we considered a similar operator but for scalar DM, given in \eqref{eq:O6Zscalar} --- as well as dimension-7 and -8 operators involving quarks and gluons.  Although we did not consider the latter operators here, they arise in the low-energy limit of the simplified models we have studied.

\vspace{0.5cm}
\section*{Acknowledgements}
We acknowledge useful conversations with Andy Haas, Tongyan Lin, Bjorn Penning, Ning Zhou, and Kathryn Zurek.
DW, MM and CS are supported by grants from the Department of Energy
Office of Science.  ST is supported by the DOE under contract DE-SC0007859 and NASA Astrophysics Theory Grant NNX11AI17G.  ST and DW thank the support and hospitality of the Aspen Center for Physics and NSF Grant No.~1066293.

\bibliography{monoh}

\clearpage
\appendix

\end{document}